\documentclass[%
 reprint,
superscriptaddress,
 amsmath,amssymb,
 aps,
]{revtex4-2}

\usepackage{float}
\usepackage{graphicx}
\usepackage{dcolumn}
\usepackage{bm}
\usepackage{booktabs} 
\usepackage{graphicx} 
\usepackage[font=small,labelfont=bf]{caption} 
\usepackage{lipsum} 
\usepackage{booktabs}
\usepackage{geometry}
\usepackage{rotating}
\usepackage{hyperref}

\usepackage{amsmath,amssymb,lmodern}
\usepackage[autostyle=true,czech=quotes]{csquotes}
\usepackage{caption}
\usepackage{lipsum}
\usepackage[export]{adjustbox}
\usepackage{subfig}
\usepackage{hyperref}
\usepackage{amsfonts} 
\usepackage{amssymb} 
\usepackage{bbm}     
\usepackage{subfig}
\usepackage{hyperref}
\usepackage{amsfonts} 
\usepackage{amssymb} 
\usepackage{bbm}     
\usepackage{algorithm}
\usepackage{algpseudocode}
\usepackage{physics}
\usepackage{float}
\usepackage{booktabs}
\usepackage{tikz}
\usetikzlibrary{quantikz2}
\usepackage{tikz}
\usepackage{chronosys}
\usepackage{multirow}
\usepackage{booktabs}
\usepackage{multirow} 
\usepackage{siunitx}  
\usepackage{placeins}
\usepackage{framed}
\begin{document}

\preprint{APS/123-QED}

\title{Quantum Machine Learning for Predicting Anastomotic Leak: A Clinical Study}

\author{Vojtěch Novák}
\email{vojtech.novak.st1@vsb.cz}
\affiliation{
 Department of Computer Science, Faculty of Electrical Engineering and Computer science, VSB-Technical University of Ostrava, Ostrava, Czech Republic
}
\affiliation{
 IT4Innovations National Supercomputing Center, VSB - Technical University of Ostrava, 708 00 Ostrava, Czech Republic
}
\author{Ivan Zelinka}%
\affiliation{
 Dpt. of Informatics and Statistics, Marine Research Institute, Klaipeda University, Lithuania
}
\affiliation{
 IT4Innovations National Supercomputing Center, VSB - Technical University of Ostrava, 708 00 Ostrava, Czech Republic
}
\author{Lenka Přibylová}
\affiliation{
Department of Applied Mathematics, Faculty of Electrical Engineering and Computer science, VSB-Technical University of Ostrava, Ostrava, Czech Republic
}

\author{Lubomír Martínek}
\affiliation{Department of Surgical Studies, Faculty of Medicine, University
of Ostrava, Ostrava, Czech Republic}
\affiliation{Department of Surgery, University Hospital Ostrava, Ostrava, Czech Republic}

\author{Vladimír Benčurik}
\affiliation{Surgical Department, Hospital AGEL Novy Jicin, Novy Jicin, Czech Republic}

\begin{abstract}

Anastomotic leak (AL) is a life-threatening complication following colorectal surgery, and its accurate prediction remains a significant clinical challenge. This study explores the potential of Quantum Neural Networks (QNNs) for AL prediction, presenting a rigorous benchmark against hyperparameter-tuned classical models including logistic regression, multilayer perceptrons, and boosting algorithms. Using a clinical dataset of 200 patients and four key predictors identified through statistical analysis, we evaluated QNNs with ZZFeatureMap encoding and EfficientSU2 and RealAmplitudes ansätze simulated under realistic hardware noise models.
Our framework emphasizes robustness, with performance metrics averaged over 10 independent optimization runs using multiple algorithms. The EfficientSU2-BFGS combination achieved the highest mean AUC of $0.7966 \pm 0.0237$, while RealAmplitudes with CMA-ES excelled in Average Precision ($0.5041 \pm 0.1214$), critical for imbalanced medical datasets. We establish a direct link between optimizer convergence and model efficacy, where effective variational parameter optimization translates to improved classification metrics. Interpretability analysis suggests QNNs may capture complex, non-linear feature relationships not evident in classical linear models. This work highlights QNNs' potential as screening tools while underscoring the need for model selection based on specific clinical goals, pending validation on larger datasets.

\end{abstract}

\maketitle

\section{Introduction}

Machine Learning (ML) has emerged as a cornerstone of artificial intelligence, permeating diverse domains including computer vision, image recognition, natural language processing, healthcare, and beyond \cite{carleo2018constructing}, \cite{liu2019unitary}. Concurrently, the realm of Quantum Computing (QC) has witnessed rapid expansion in recent years, despite its current limitations stemming from Noise Intermediate-Scale Quantum (NISQ) devices \cite{lau2022nisq}. However, there exists a tantalizing prospect wherein QC could potentially outstrip classical computers in select ML applications \cite{biamonte2017quantum}.

Quantum Machine Learning (QML) \cite{ciliberto2018quantum} represents an intersection of quantum physics and ML, heralding a new era of interdisciplinary research \cite{schuld2019quantum}. Harnessing QML methodologies not only enhances performance but also accelerates data processing on QC platforms \cite{martin2022quantum}, \cite{wiebe2016quantum}. However, the efficacy of modern machine learning is fundamentally bounded by the polynomial computing time \cite{somma2003quantum}, necessitating the simplification of quantum algorithms to yield reliable results. Within the domain of QML, four distinct approaches emerge based on the nature of data and the processing device, be it classical or quantum \cite{schuld2015introduction}.

Supervised learning tasks, such as those undertaken by Support Vector Machines (SVM) \cite{sierra2020dementia}, have undergone extensive exploration across various datasets, showcasing superior performance particularly with kernel-based methodologies \cite{schuld2021supervised}. Among quantum paradigms, the Variational Quantum Classifier (VQC) has garnered significant attention, particularly for classification tasks on NISQ devices \cite{havlicek2019supervised}. Notably, a plenty of approaches exists for categorizing well-established supervised QML algorithms, including QSVM and VQC. Further strides have been made in this arena with the advent of quantum-inspired neural networks \cite{farhi2018classification}, coupled with innovative applications such as hybridized low-depth VQC classification methods integrated with simple error-mitigation strategies \cite{moll2018quantum}. Additionally, pre-processing techniques like Principle Component Analysis (PCA) \cite{lloyd2014quantum} have been leveraged, culminating in marked enhancements in categorization performance.

While prior QML studies have explored classification on benchmark datasets \cite{Caro2022, Kumar2022, Kharsa2023, Jerbi2021, Abohashima2020}, its application to complex, noisy, and highly imbalanced real-world clinical problems like AL prediction remains nascent. This study establishes its novelty in three key areas: It presents one of the first applications of QNNs to AL prediction, a problem space dominated by classical statistical methods. It introduces a uniquely robust comparative framework, where QNNs simulated under realistic noise are benchmarked against hyperparameter-optimized classical models, with QNN performance validated across 10 independent runs to ensure statistical reliability. It moves beyond performance metrics to investigate QNN interpretability through experimental perturbation-based methods, offering a direct comparison to the feature importance derived from traditional clinical models like logistic regression. This work thus provides a practical blueprint for evaluating and interpreting QNNs in a challenging, high-stakes medical domain.

Anastomotic leak (AL) is a serious and potentially life-threatening complication arising from surgical procedures involving anastomosis, such as bowel resection, where two ends of a bowel are surgically connected. AL occurs when the connection fails to heal, resulting in leakage of contents into the abdominal cavity. In the case of bowel surgery, this can lead to peritonitis, sepsis, and other severe outcomes. Numerous risk factors, including smoking, malnutrition, immunosuppression, and prolonged operation times, have been associated with an increased likelihood of AL. Despite advancements in surgical techniques and perioperative care, accurately predicting and managing the risk of AL remains a critical challenge.

In this context, leveraging statistical and machine learning methodologies to identify key risk factors and develop predictive models is crucial for improving patient outcomes. ML offers the potential to process large datasets and uncover complex patterns that traditional statistical approaches might miss. By integrating these techniques, clinicians can better stratify patient risk, tailor surgical and postoperative care, and ultimately reduce the incidence and severity of AL. The incorporation of ML and, potentially, QML for such predictive modeling represents an innovative step toward addressing this pressing medical issue.

A recent study \cite{benvcurik2021intraoperative} analyzed the impact of intraoperative fluorescence angiography with indocyanine green (ICG) on reducing anastomotic leakage in mini-invasive low anterior resections with total mesorectal excision for rectal cancer. The authors reported a significant reduction in AL rates (19 \% in the control group vs. 9 \% in the ICG group, p=0.042) and identified diabetes and the use of a transanal drain as significant risk factors. Building upon this data, our approach integrates predictive modeling to evaluate these and additional clinically relevant risk factors, combining statistical analysis with expert medical insights to enhance prediction accuracy and inform clinical decision-making.

\section{Problem Description}
\label{sec:problem_desc}
Anastomotic leak (AL) is a severe complication following colorectal surgeries, particularly low rectal resections with total mesorectal excision for cancer. When treating malignant or large benign tumors in the colorectal area, the affected section of the intestine must be removed, and the two remaining ends are surgically reconnected. If the anastomotic site does not heal properly, it can rupture, leading to AL. This complication is associated with significant morbidity and mortality, occurring in approximately 14\% of cases in our dataset and contributing to up to 40\% of surgery-related deaths. Identifying risk factors for AL and developing predictive models are essential for improving patient outcomes.

Our study utilizes data collected from the Surgical Department of Hospital Nový Jičín a.s. between 2015 and 2016, comprising 200 patients (28 with AL, 172 without AL). The dataset includes 76 explanatory variables, categorized into two main groups: intraoperative techniques and patient history variables. The primary goal of this study was to identify the statistical significance of intraoperative techniques such as NoCoil, ACSP, PERFB, and ICG, and to identify potential risk factors for AL occurrence. Furthermore, we aimed to establish if we could predict anastomotic leak from the significant factors associated with its occurrence. We will particularly focus on smoking status (Smoking), Diabetes Mellitus (DM), preservation of the left colic artery (ACSP), and the use of a transanal drain (NoCoil), as these were found to be statistically significant in a later chapter. We will delve more into the ACSP factor rather than ICG, given that the effectiveness of ICG has already been thoroughly assessed and studied on data used in this study \cite{benvcurik2021intraoperative} and elsewhere \cite{blanco2018intraoperative}. This predictive analysis will be conducted using both classical machine learning methods and a novel quantum-enhanced approach.

Approval for human experiments: This study was approved by the institutional review board of Hospital Nový Jičín a.s., and all experiments were performed in accordance with the relevant guidelines and regulations. Informed consent was obtained from all participants and/or their legal guardians. We have obtained confirmation from the doctors and surgeons who conducted the study that we are permitted to use this data for our research.

This study specifically evaluates the performance of classical and quantum classifiers on raw clinical data without modifications, preserving the natural distribution of cases. This approach ensures ecological validity, providing a realistic assessment of classifier performance in clinical settings. By avoiding techniques like SMOTE or stratified sampling, we maintain the dataset’s inherent imbalance, which reflects the real-world prevalence of AL. This allows for a direct and fair comparison between classical and quantum approaches under identical, real-world conditions.

To mitigate overfitting risks, we reduced the number of variables through statistical analysis, including goodness-of-fit tests and risk ratios, and selected features based on medical importance in consultation with clinicians. This feature selection process not only enhances model interpretability but also ensures that the models are trained on clinically relevant predictors, reducing the risk of overfitting despite the small sample size.

Our analysis incorporates several key \textbf{intraoperative techniques and surgical variables} that are critical to understanding the surgical procedure and its immediate impact on the patient. These variables include:

\begin{itemize}
\item \textbf{NoCoil}: A transanal drain used at the end of the surgery to maintain decompression and improve healing conditions at the anastomotic site (Yes/No).
\item \textbf{ICG}: Intraoperative fluorescence angiography using indocyanine green to assess tissue perfusion (Yes/No).
\item \textbf{ACSP}: Preservation of the left colic artery (LCA), which may improve perfusion at the anastomotic site (Yes/No).
\item \textbf{PERFB}: Assessment of bowel perfusion using fluorescence angiography (Good/Poor).
\end{itemize}

In addition to surgical factors, \textbf{patient history and other intraoperative variables} provide crucial context regarding a patient's overall health and the conditions leading up to and during the surgery. These include:

\begin{itemize}
\item \textbf{CRP}: C-reactive protein, a marker of systemic inflammation (mg/L).
\item \textbf{HT}: Hypertension, characterized by elevated blood pressure (Yes/No).
\item \textbf{ICHS}: Ischemic heart disease (Yes/No).
\item \textbf{KOAG}: Coagulopathy, indicating abnormal blood clotting (Yes/No).
\item \textbf{DM}: Diabetes mellitus, a known risk factor for impaired wound healing (Yes/No).
\item \textbf{Smoking}: Smoking status (Yes/No).
\item \textbf{Sex}: Gender (Male/Female).
\item \textbf{Age}: Age (years).
\item \textbf{BMI}: Body mass index (kg/m$^2$).
\item \textbf{HB}: Hemoglobin level (g/L).
\item \textbf{KORT}: Corticosteroid therapy, which may impact immune response and healing (Yes/No).
\item \textbf{ASA}: Preoperative physical status classification (2, $>$2).
\item \textbf{LN}: Number of lymph nodes removed.
\end{itemize}

Among these factors, the preservation of the left colic artery (ACSP variable) plays a crucial role in maintaining blood supply to the descending and sigmoid colon. During colorectal surgery, preserving the LCA is often considered beneficial for ensuring adequate perfusion (PERFB) of the anastomotic site, thereby promoting healing and reducing the risk of AL \cite{kato2019impact}. However, in some cases, its preservation may not be feasible due to anatomical constraints or oncological requirements.

Another factor influencing AL risk is the use of a transanal drain. NoCoil refers to a silicone-based tube, typically 1.5 to 2 cm in diameter, inserted into the rectum at the end of the surgery to decompress the rectal stump and protect the anastomosis. This device facilitates passive evacuation of gas and residual stool, reducing intraluminal pressure and helping to minimize the risk of leakage \cite{zhao2021transanal}.

Diabetes Mellitus and Smoking have been consistently identified as significant risk factors for AL \cite{axtell2025anastomotic}, primarily due to their effects on vascular health and immune response. Similarly, smoking has been shown to impair tissue oxygenation and angiogenesis, further contributing to poor anastomotic healing.

Recent advancements in intraoperative monitoring, such as fluorescence angiography with indocyanine green (ICG), offer a potential method for reducing AL incidence by providing real-time visualization of bowel perfusion. Studies indicate that the use of ICG during surgery may improve anastomotic outcomes by allowing surgeons to identify poorly perfused areas and adjust their technique accordingly \cite{trastulli2021indocyanine}.

Risk factors such as diabetes, smoking, and the absence of transanal drain have been consistently linked to higher rates of AL in surgical literature. A systematic review and meta-analysis by He et al. (2023) confirmed that diabetes and smoking significantly increase the risk of AL following colorectal surgery, alongside other factors such as male sex and intraoperative blood transfusion \cite{He2023}. Similarly, Awad et al. (2021) highlighted that intraoperative complications, prolonged surgery duration, and inadequate tissue perfusion contribute to higher postoperative AL rates \cite{Awad2021}.

Novel intraoperative techniques, such as fluorescence angiography (FA), have shown promise in reducing AL incidence. Lin et al. (2021) conducted a meta-analysis demonstrating that the use of intraoperative indocyanine green (ICG) fluorescence angiography significantly reduces AL rates in colorectal resections by improving the assessment of bowel perfusion \cite{Lin2021}. More recent work by Balamurugan et al. (2025) further supports this, emphasizing that FA leads to better anastomotic outcomes, though additional large-scale studies are needed to validate its routine clinical application \cite{Balamurugan2025}.

While existing studies emphasize biological and clinical interventions, recent efforts have explored predictive modeling for AL. Venn et al. (2023) reviewed preoperative and intraoperative scoring systems designed to predict AL occurrence, revealing that although current models provide some predictive power, they remain limited by variability in surgical techniques and patient-specific factors \cite{Venn2023}. Our work introduces a computational perspective by leveraging advanced machine learning and quantum computing to enhance AL prediction models, potentially improving clinical decision-making and patient outcomes.

\section{Predictive Modeling Approaches}
Our study compares classical and quantum-enhanced approaches to predictive modeling of AL:

\begin{enumerate}
\item \textbf{Classical Methods:}
We employed a diverse set of classical machine learning models, including logistic regression (LR), linear discriminant analysis (LDA), support vector machines (SVM), multilayer perceptron (MLP), and nearest neighbors (NN). Additionally, we utilized ensemble methods such as gradient boosting machines (GBM) and AdaBoost to enhance predictive performance. The implementation of these classical models was carried out using the \texttt{scipy} \cite{2020SciPy-NMeth} module in Python. The effectiveness of these models was evaluated using Receiver Operating Characteristic (ROC) curves, with a primary focus on sensitivity (true positive rate) to prioritize the identification of patients at high risk of AL.

\item \textbf{Quantum Neural Networks:}
A novel quantum-enhanced approach using Variational Quantum Circuits (VQCs) is applied to the dataset. These circuits leverage the principles of quantum mechanics to explore high-dimensional feature spaces, potentially improving the predictive power for imbalanced datasets. The VQC models were implemented using the \texttt{qiskit} \cite{qiskit2024} library.
\end{enumerate} 

Model performance is evaluated using the following metrics:  

\begin{itemize}  
    \item \textbf{ROC Curves and AUC:} The Area Under the Curve (AUC) of the Receiver Operating Characteristic (ROC) curve measures a model’s ability to discriminate between patients with and without the complication. AUC values closer to 1 indicate better performance, as the model correctly distinguishes between positive and negative cases more often.  

    \item \textbf{Sensitivity and Specificity:} Sensitivity (true positive rate) reflects the model’s ability to correctly identify patients who experience the complication, ensuring high-risk individuals are not missed. Specificity (true negative rate) measures how well the model correctly identifies those who do not develop the complication, reducing false positives.  

    \item \textbf{Predictive Power, Calibration, and Classification Accuracy:} We assess overall model performance using various statistical measures:  
    \begin{itemize}  
        \item \textit{Efron's R$^2$, McKelvey \& Zavoina R$^2$, and Count R$^2$} evaluate how well the model explains variance in the outcome.  
        \item \textit{Brier Score and Log Loss} assess model calibration, measuring how well predicted probabilities align with actual outcomes.  
        \item \textit{F1 Score} balances precision and recall, particularly relevant for imbalanced datasets.  
    \end{itemize}  
\end{itemize}  

Additionally, \textbf{Positive Predictive Value (PPV) and Negative Predictive Value (NPV)} provide practical insights into prediction reliability. PPV represents the probability that a positive prediction corresponds to an actual occurrence of the complication, while NPV indicates the probability that a negative prediction correctly identifies a complication-free patient. These metrics are particularly useful in clinical decision-making, where understanding the likelihood of correct predictions is crucial.  

To rigorously evaluate and compare the performance of our predictive models, we established a comprehensive evaluation framework. This framework encompasses cross-validation to assess robustness, ROC curve analysis for threshold optimization, and a comparative analysis of classification metrics at fixed sensitivity levels.  These methodologies are designed to provide a multifaceted understanding of each model’s strengths and limitations in the context of rare post-surgical complication prediction.

To ensure the reliability and generalizability of our findings, we employed cross-validation (CV). This technique is crucial for evaluating how well models generalize to unseen data and for mitigating the risk of overfitting.  By systematically training and testing models on different partitions of the dataset, cross-validation provides a more robust estimate of predictive performance than a single train-test split.

\section{Variational Quantum Algorithms in the NISQ Era}

Quantum Machine Learning (QML) in the NISQ era offers new computational possibilities, working within the constraints of contemporary quantum hardware that typically provides 50 to 1000 noisy qubits. Deep quantum circuits remain challenging to execute due to noise and decoherence \cite{preskill2018quantum}. The introduction of parameterized quantum circuits in 2014 enabled significant QML advancements, leading to influential algorithms like the Quantum Approximate Optimization Algorithm (QAOA) \cite{farhi2014quantum} and the Variational Quantum Eigensolver (VQE) \cite{peruzzo2014variational}.

These variational quantum algorithms (VQAs) use parameterized quantum circuits (PQCs) optimized through classical techniques. This hybrid approach combines quantum and classical computing, with the quantum processor serving as a parameterized model refined through iterative measurements. VQAs provide a robust framework for diverse problems through hybrid quantum-classical optimization loops \cite{cerezo2021variational}.

In QML contexts, PQCs are termed "quantum neural networks" due to their machine learning applications \cite{schuld2014quest} \cite{abbas2021power}. These models leverage quantum systems' representational capabilities for supervised learning and generative modeling \cite{benedetti2019parameterized}.

Current research focuses on improving VQA performance and scalability \cite{caro2022generalization, larocca2022group, meyer2023exploiting}. Whether QML algorithms can outperform classical counterparts remains an open question \cite{liu2021rigorous} \cite{schreiber2023classical} \cite{jager2023universal}. Nevertheless, quantum circuit design has progressed considerably, with architectures including hardware-efficient circuits \cite{kandala2017hardware}, the quantum alternating operator ansatz \cite{hadfield2019quantum}, and dissipative quantum neural networks (dQNNs) \cite{beer2020training}. Data re-uploading techniques have also shown promise for encoding classical data into quantum models \cite{heimann2022learning}.

Hybrid quantum-classical approaches have demonstrated effectiveness across multiple domains, including quantum chemistry, combinatorial optimization, and machine learning. VQE approximates ground states of molecular electronic Hamiltonians, providing quantum alternatives for certain molecular systems. QAOA addresses combinatorial problems like MaxCut and classical Ising models. These applications demonstrate the versatility of hybrid algorithms in utilizing current quantum hardware capabilities.

VQAs face challenges including noise-induced barren plateaus and stochastic quantum measurements that complicate optimization. However, advancing quantum hardware and software increasingly demonstrate VQAs' practical potential through experimental results \cite{benedetti2019parameterized}. This evolving field shows promise for real-world applications in material science, drug discovery, and artificial intelligence.

\section*{Quantum Machine Learning in the Biomedical Domain}

Biotechnology, an interdisciplinary field rooted in engineering, physics, biology, and chemistry, applies scientific and engineering principles to biological processes \cite{bud2001history, van2006oecd}. In healthcare, biomedical engineering plays a crucial role in disease prediction and classification, encompassing four primary domains: 	\textbf{Omics}, 	\textbf{biomedical imaging}, 	\textbf{Biosignals}, and 	\textbf{medical healthcare records}. Quantum Machine Learning (QML) is increasingly being integrated into these domains, enhancing data analysis and accelerating advancements in disease detection and treatment.

Omics research, involving genomic sequences, gene expression, and protein structure prediction, benefits significantly from QML techniques. These approaches aid in disease prevention and personalized treatments by analyzing genetic data. Researchers have utilized QML models like Quantum Moth Flame Optimization Algorithm (QMFOA) and Quantum Ising approaches for classifying small gene subsets \cite{dabba2021hybridization, sergioli2021quantum, li2021quantum}. Such applications are particularly relevant in cancer detection, where quantum algorithms enhance classification accuracy and efficiency.

Biomedical imaging—encompassing cytopathology, histopathology, ultrasound, X-rays, CT, MRI, PET, and fNIRS—has seen advancements with QML, enabling faster and more precise disease classification. QML has been applied to COVID-19 diagnosis using X-ray images \cite{sengupta2021quantum, acar2021covid, amin2021quantum, elshafeiy2021approach} and cancer detection \cite{iliyasu2017quantum}. Hybrid quantum-classical convolutional neural networks (QCCNN) show promising results in classifying malignant lesions in CT scans, demonstrating performance comparable to classical models \cite{matic2022quantumclassical}.

Biosignals, such as EEG, ECG, and EMG, provide critical insights into neurological and physiological conditions. QML techniques, including Quantum Variational Classifiers (VQC) and Quantum Annealing, have been applied to EEG signals for cognitive response prediction \cite{aishwarya2020quantum}. Additionally, Quantum Recurrent Neural Networks (QRNN) have been employed in brain-computer interface applications, enhancing signal processing capabilities \cite{gandhi2013quantum}.

Medical Healthcare Records (MHRs) integrate clinical test results, imaging data, biosignals, and treatment histories. QML models such as QSVM, VQC, and HQMLP have demonstrated effectiveness in classifying diseases like diabetes and ischemic heart disease (IHD) \cite{maheshwari2020machine, gupta2022comparative, sierra2021diabetes, maheshwari2023quantum}. Studies have shown that optimized quantum support vector machines (OQSVM) and hybrid quantum neural networks achieve high classification accuracies for IHD, highlighting QML's potential in real-time healthcare applications. Alzheimer’s disease diagnosis, particularly through handwriting analysis, is another emerging area where QML is being explored for early detection.

QML is rapidly transforming biomedical research by offering powerful tools for complex data analysis. As quantum hardware advances, these techniques are expected to further enhance disease diagnosis, treatment optimization, and healthcare decision-making.

\section{Simulating Quantum Neural Networks with Noise}

To simulate quantum computing behavior as realistically as possible, we implemented a comprehensive noise model that closely approximates the characteristics of real quantum hardware. Our simulation framework utilizes Qiskit's AerSimulator with a custom noise model that incorporates several key aspects of quantum decoherence and gate errors, providing a realistic testbed for evaluating quantum neural network performance under practical conditions.

The noise model implementation focuses on depolarizing errors, which are among the most significant sources of noise in current quantum hardware. This choice was deliberate as depolarizing noise represents a key paradigm in quantum error modeling that approximates the combined effects of various physical error mechanisms. In actual quantum computers, errors arise from multiple sources including thermal fluctuations, electromagnetic interference, and imperfect control pulses. These varied sources manifest as a combination of bit-flip ($X$), phase-flip ($Z$), and bit-phase-flip ($Y$) errors, which is precisely what the depolarizing channel models.

We selected a gate error probability of $p_{gate} = 0.05$ based on reported error rates from IBM's Manila quantum processor and similar NISQ (Noisy Intermediate-Scale Quantum) devices, where single-qubit gate fidelities typically range from 99.5\% to 99.9\%, corresponding to error rates of 0.001 to 0.05. Our choice represents a conservative upper bound that accounts for the performance degradation observed in quantum processors under extended operation. This value also aligns with published benchmarks from quantum hardware manufacturers and recent literature on quantum error characterization \cite{Zhou2018, Cross2019}.

We modeled these errors using Kraus operators, which provide a complete description of quantum operations including noise effects. The noise model includes single-qubit gate errors with a probability $p_{gate} = 0.05$, distributed equally among $X$, $Y$, and $Z$ errors. The quantum channel effects are represented by Kraus operators with the identity operation ($I$) occurring with probability $1 - p_{gate}$, while Pauli $X$, $Y$, and $Z$ operations each occur with probability $p_{gate}/3$. This equal distribution of error types reflects the equiprobable nature of different error mechanisms in depolarizing channels, which is consistent with the quantum information theory principle that noise in quantum systems tends toward maximum entropy in the absence of specific environmental biases. These error channels were applied to all single-qubit gates across all qubits in the system, ensuring comprehensive noise modeling throughout the quantum neural network.

The simulation was configured to use 1024 shots per circuit execution, providing a balance between statistical significance and computational efficiency. This shot count is comparable to typical experimental runs on actual quantum hardware, where 1000-2000 shots are commonly used to overcome the probabilistic nature of quantum measurements while respecting hardware time limitations and queue constraints. Circuit execution was managed through a custom NoiseSimulator class that handles the integration of the noise model with the quantum circuit execution pipeline.

\subsection{Data Encoding and Feature Mapping}
For encoding classical data into quantum states, we employed the ZZFeatureMap, which is particularly well-suited for machine learning tasks due to its ability to create non-linear feature spaces. The ZZFeatureMap implements a second-order feature map that encodes classical data into quantum states through a series of single-qubit rotations and two-qubit ZZ operations, enabling the quantum neural network to process classical information in a quantum computational framework.

The encoding process begins with an initial rotation layer where each qubit $i$ receives a rotation $R_x(x[i])$ based on the input feature $x[i]$. This is followed by an entangling layer that applies ZZ operations between pairs of qubits, creating quantum correlations that encode feature interactions. A second rotation layer applies another set of single-qubit rotations, and this process can be repeated for the specified number of repetitions to increase the encoding depth.

Mathematically, the ZZFeatureMap implements the unitary transformation:
\begin{equation}
U(\mathbf{x}) = \exp\left(i \sum_{i} x_i Z_i\right) \exp\left(i \sum_{i<j} x_i x_j Z_i Z_j\right)
\end{equation}
where $x_i$ represents the $i$-th feature of the input data, $Z_i$ denotes the Pauli Z operator on qubit $i$, and the second term creates entanglement between qubits based on pairwise products of features. This formulation allows the quantum circuit to capture both linear and quadratic feature combinations, potentially revealing hidden patterns in the data that might not be accessible through classical linear transformations.

The feature mapping circuit for 4 features with single repetition exhibits a depth of 17 with a total of 26 individual quantum gates, comprising 12 CNOT gates for entanglement generation, 10 parameterized phase gates for feature encoding, and 4 Hadamard gates for superposition initialization. This gate composition reflects the circuit's emphasis on creating quantum correlations through the CNOT operations while encoding classical feature information through the parameterized phase rotations. The relatively moderate depth of 17 time steps ensures that the feature encoding process remains feasible on near-term quantum devices while still providing sufficient complexity to create meaningful quantum feature spaces for the 4-dimensional input data. Circuit depth represents a critical performance metric that directly impacts practical implementation, as it refers to the number of sequential time steps required for execution, with lower depth generally implying shorter execution time and reduced exposure to decoherence effects that can degrade quantum states over time.

\begin{figure*}
    \centering
    \includegraphics[width=1\linewidth]{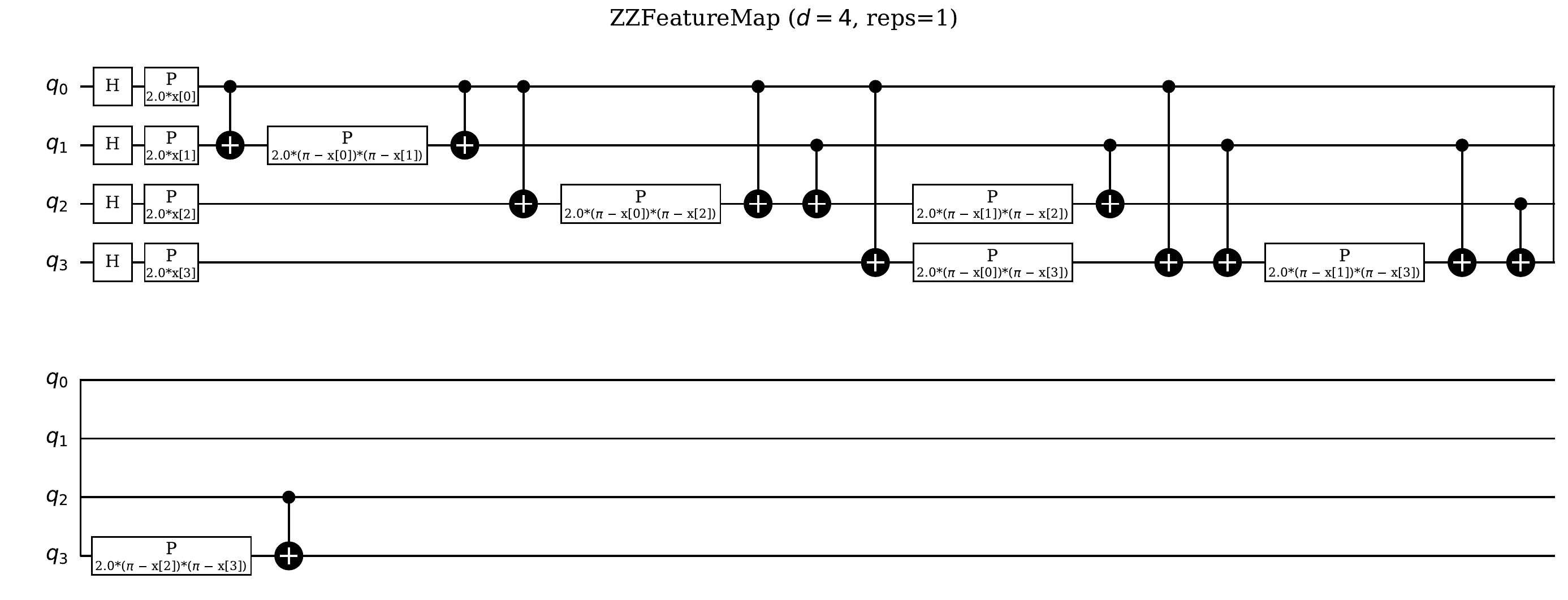}
    \caption{The ZZFeatureMap applies Hadamard gates to initialize superposition, followed by entangling CNOT gates and parameterized phase gates \( P(2.0\phi(x_i, x_j)) \) to encode feature interactions. This combination enables quantum interference, where correlations between input data are mapped into the entangled quantum state.}
    \label{fig:ZZfeature}
\end{figure*}

In our implementation, we configured the ZZFeatureMap with 4 feature dimensions and a single repetition of the encoding circuit. This configuration maintains the dimensionality of the input space while creating non-linear feature combinations through quantum operations. The quantum circuit representing the ZZFeatureMap is shown in Figure~\ref{fig:ZZfeature}, which illustrates how Hadamard gates initialize superposition, followed by entangling CNOT gates and parameterized phase gates $P(2.0\phi(x_i, x_j))$ to encode feature interactions. This combination enables quantum interference, where correlations between input data are mapped into the entangled quantum state, providing the foundation for the subsequent variational quantum processing.

For instance, Havlíček et al. (2019) \cite{havlicek2019supervised} demonstrated the potential of quantum-enhanced feature spaces, prominently featuring the ZZFeatureMap, to make classically challenging datasets amenable to linear classification. Furthermore, its empirical effectiveness has been observed in various studies utilizing quantum kernel methods and Variational Quantum Classifiers \cite{vasques2023application}.

\subsection{Variational Ansätze Design}
Following the data encoding stage, the quantum neural network employs variational ansätze to process the encoded information through parameterized quantum circuits. The selection of appropriate ansätze involves balancing expressive power with computational efficiency, as more complex circuits can represent richer quantum states but require increased computational resources and may be more susceptible to noise effects. We investigated two prominent ansätze that offer different trade-offs between these competing requirements: the Real Amplitudes ansatz and the EfficientSU2 ansatz.

\begin{figure*}[htpb]
    \centering
    \includegraphics[width=1\linewidth]{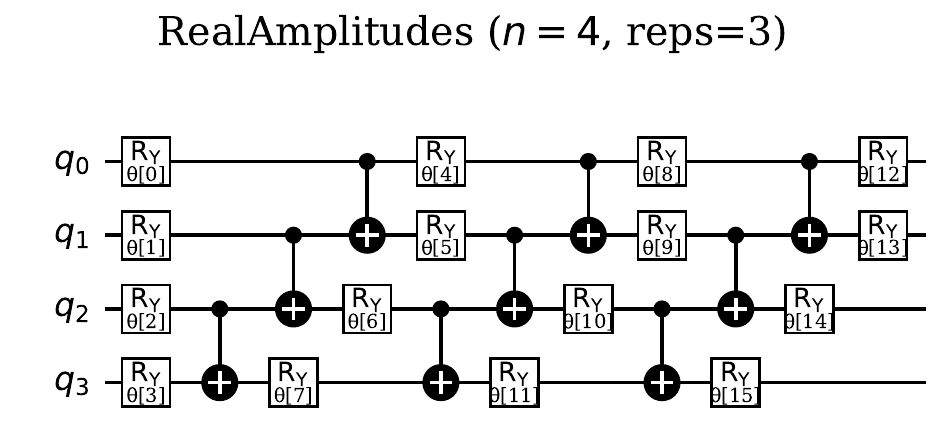}
    \caption{Quantum circuit diagram of the Real Amplitudes ansatz, featuring a layered structure with alternating single-qubit $R_y$ rotations and multi-qubit entangling gates (e.g., CNOT). This ansatz generates quantum states with real amplitudes, simplifying interpretation while maintaining expressiveness for complex data. Its gate efficiency reduces computational costs, making it a practical choice for variational quantum circuits.}
    \label{Realamp}
\end{figure*}

The Real Amplitudes ansatz represents a well-established choice for variational quantum classifiers due to its simplicity and computational efficiency. This ansatz utilizes a layered structure consisting of alternating single-qubit rotations ($R_y$ gates) and multi-qubit entangling gates (typically CNOT gates), as illustrated in Figure~\ref{Realamp}. The layered architecture allows for systematic construction of expressive quantum states while maintaining a relatively shallow circuit depth, which is crucial for maintaining coherence in noisy quantum environments.

The Real Amplitudes ansatz with 4 qubits and 3 repetitions demonstrates favorable circuit characteristics with a depth of 11 and a total of 25 quantum gates, reflecting its design philosophy of computational efficiency while maintaining expressive capability. The gate composition emphasizes rotation operations for parameterization while utilizing entangling gates judiciously to create necessary quantum correlations. This relatively shallow depth of 11 time steps, achieved through the efficient layering of the 3 repetitions, minimizes the exposure to decoherence effects, making it particularly suitable for near-term quantum implementations where coherence times are limited. When combined with the ZZFeatureMap, the complete Real Amplitudes quantum neural network achieves a total circuit depth of 28 with 51 gates, representing an efficient balance between expressiveness and practical implementability on current quantum hardware.

A defining characteristic of the Real Amplitudes ansatz is its generation of quantum states with only real amplitudes, which simplifies the interpretation and analysis of the circuit's output since all basis state probabilities are represented by real numbers. This property also reduces the parameter space that must be explored during optimization, potentially leading to more efficient training procedures. Additionally, the Real Amplitudes ansatz requires a relatively low number of gates compared to more intricate ansätze, translating to lower computational costs for circuit implementation and reduced susceptibility to accumulated gate errors in noisy environments.

However, the simplicity that makes the Real Amplitudes ansatz computationally attractive also imposes limitations on its expressive capabilities. Compared to more complex ansätze, it may have reduced ability to represent highly intricate quantum states, which could potentially limit its performance for tasks demanding the most sophisticated quantum computations. Furthermore, the restriction to real amplitudes may prevent the circuit from fully exploiting quantum interference effects that require complex phases.

\begin{figure*}[htpb]
    \centering
    \includegraphics[width=1\linewidth]{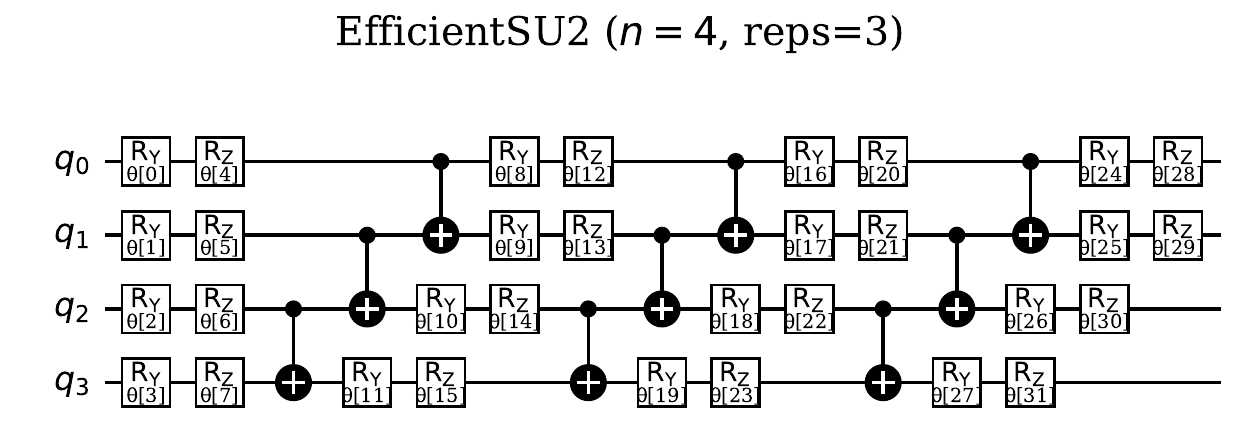}
    \caption{Quantum circuit diagram of the EfficientSU2 ansatz used in the quantum neural networks. The circuit features a layered structure with parametrized single-qubit rotations around two axes (typically $R_x$ and $R_y$), followed by entangling gates to create correlations between qubits.}
    \label{SU2}
\end{figure*}

In contrast, the EfficientSU2 ansatz offers enhanced flexibility through a richer parameterization scheme. This ansatz retains the beneficial layered structure but utilizes single-qubit rotations around two axes (typically $R_x$ and $R_y$) followed by entangling gates, as shown in Figure~\ref{SU2}. The additional rotation axis provides access to a broader range of single-qubit unitaries, enabling the circuit to explore a more extensive portion of the quantum state space and potentially capture more complex data patterns.

The EfficientSU2 ansatz with 4 qubits and 3 repetitions exhibits increased circuit complexity with a depth of 15 and a total of 41 quantum gates, reflecting its enhanced expressiveness through additional parameterized rotations across multiple axes. The gate distribution includes rotations around both x and y axes within each repetition layer, providing greater flexibility in quantum state preparation and potentially enabling more sophisticated pattern recognition capabilities. However, this increased depth of 15 time steps compared to the Real Amplitudes' 11 steps represents a trade-off between expressiveness and susceptibility to noise, requiring careful consideration of the coherence properties of the target quantum hardware. The complete EfficientSU2-based quantum neural network, when combined with the ZZFeatureMap, results in a total circuit depth of 32 with 67 gates, demonstrating the computational overhead associated with enhanced quantum expressiveness.

The increased expressive power of the EfficientSU2 ansatz stems from its ability to generate quantum states with complex amplitudes and phases, allowing for more sophisticated quantum interference patterns. This enhanced expressiveness can translate to superior performance for certain classification tasks, particularly those involving complex, non-linearly separable data structures. The additional degrees of freedom provided by the dual-axis rotations enable the ansatz to adapt more flexibly to diverse data characteristics during the optimization process.

However, this enhanced capability comes with computational trade-offs. The EfficientSU2 ansatz typically requires more gates than the Real Amplitudes ansatz, leading to increased circuit depth and higher computational burden for both circuit implementation and optimization. The larger parameter space also presents optimization challenges, as the increased dimensionality can lead to more complex loss landscapes with additional local minima, potentially making the training process more difficult and requiring more sophisticated optimization strategies.

The choice between these ansätze ultimately depends on the specific requirements of the machine learning task, the available computational resources, and the noise characteristics of the target quantum hardware. In our comparative study, both ansätze were evaluated under identical noisy conditions to assess their relative performance and robustness, providing insights into the practical trade-offs between circuit complexity and classification accuracy in realistic quantum computing environments.

\section{Parameter Optimization}

To optimize the parameters of our quantum neural networks, we employed a diverse set of optimization algorithms, each offering distinct advantages for navigating the complex parameter landscapes inherent in variational quantum circuits. The selection encompasses both gradient-based and gradient-free methods, providing robustness against the challenges posed by noisy quantum environments and barren plateaus.

\begin{table}[htpb]
\centering
\begin{tabular}{@{}ll@{}}
\toprule
\textbf{Optimizer} & \textbf{Category}                        \\ \midrule
BFGS               & Gradient-Based Method                    \\
SLSQP              & Gradient-Based Method                    \\
CMA-ES             & Metaheuristic Method                   \\
COBYLA             & Gradient-Free Method                     \\
SPSA               & Gradient-Free Method                     \\ \bottomrule
\end{tabular}
\caption{Classification of Optimizers by Principle}
\label{tab:optimizer_classification}
\end{table}

BFGS is a quasi-Newton optimization method that approximates the inverse Hessian matrix using gradient information to achieve rapid convergence for smooth objective functions \cite{liu1989limited}. In quantum circuit optimization, BFGS can effectively navigate parameter spaces when gradients are accessible, though its performance may degrade in the presence of sampling noise and measurement uncertainties typical in quantum hardware.

CMA-ES represents a derivative-free evolutionary strategy that maintains a population of candidate solutions while adapting a multivariate normal distribution to guide the search process \cite{hansen2003reducing}. This metaheuristic approach proves particularly valuable for quantum optimization as it naturally handles the non-convex, multimodal landscapes often encountered in variational circuits and remains robust to noise inherent in quantum measurements \cite{bonet2023performance, illesova2025numerical, novak2025optimization}.

COBYLA employs a simplex-based approach to construct linear approximations of the objective function, making it well-suited for noisy optimization environments \cite{powell1994direct}. Its derivative-free nature allows it to handle the stochastic fluctuations common in quantum circuit evaluation, though convergence may be slower compared to gradient-based methods in noiseless scenarios.

SLSQP combines sequential quadratic programming with constraint handling capabilities, constructing quadratic approximations to iteratively refine solutions \cite{kraft1988software}. While effective for smooth, differentiable problems, its reliance on gradient information can be challenging in quantum settings where parameter shift rules or finite differences must be employed for gradient estimation.

SPSA offers a highly efficient gradient-free approach that estimates gradients using only two function evaluations per iteration, regardless of parameter dimensionality \cite{spall1992multivariate}. This efficiency makes it particularly attractive for quantum optimization where circuit evaluations are computationally expensive, and its inherent robustness to noise aligns well with the stochastic nature of quantum measurements and decoherence effects.

The optimization algorithms considered have the following hyperparameters and initialization settings, assuming \( m \) parameters where applicable: CMA-ES uses an initial standard deviation \(\sigma_0 = 0.15\), a population size of \(\lceil 4 + 3 \log(m) \rceil\), a parent fraction \(\mu = 0.5\), a mean update factor \(c_{\text{mean}} = 1.0\), and a damping factor of 1.0; SPSA is configured with a maximum of 75 iterations , an allowed increase of \(10^{-3}\), blocking enabled, and a termination checker with \(N=10\); BFGS, specifically L-BFGS-B, is initialized with a maximum of 75 iterations  and a function tolerance of \(10^{-4}\); SLSQP is set with a maximum of 75 iterations  and a function tolerance of \(10^{-4}\); COBYLA is configured with a maximum of 75 iterations.
\section{Feature Importance Analysis for Quantum Neural Network}

Interpreting the predictive mechanisms of machine learning models is paramount for their application in critical domains such as clinical risk assessment. While classical linear models like logistic regression offer straightforward interpretability through their model coefficients, the complex, high-dimensional nature of quantum neural networks (QNNs) presents a significant challenge. In our work, the association between clinical risk factors---diabetes mellitus (DM), smoking, prior abdominal surgery (ACSP), and neoadjuvant therapy (NoCoil)---and post-surgical anastomotic leak was modeled using a Variational Quantum Classifier (VQC) in Qiskit. This QNN was constructed with a \texttt{ZZFeatureMap} to encode the four-dimensional feature vector into an entangled quantum state, and an \texttt{EfficientSU2} ansatz to create a trainable parameterized quantum circuit. Optimization was performed via a Covariance Matrix Adaptation Evolution Strategy (CMAES).

Unlike a logistic regression model, where the feature importance is directly quantified by the coefficients $\beta_i$ in the equation 
\[
\ln\left(\frac{p}{1-p}\right) = \beta_0 + \sum_i \beta_i x_i,
\]
the trained parameters of a QNN lack a direct, human-interpretable meaning. Consequently, to understand our model's decision-making process, we explored two experimental, perturbation-based feature importance techniques. These methods aim to approximate a feature's influence by observing the model's behavior when its inputs are systematically altered.

Our first experimental approach was a permutation-based analysis, a model-agnostic technique that assesses the global impact of a feature on the model's overall performance. This method first establishes a baseline predictive accuracy, measured by the Area Under the Curve (AUC) on the test set. Then, for each feature, its values are randomly shuffled across all samples, effectively breaking any learned relationship between that feature and the outcome. The model's AUC is re-evaluated on this perturbed data, a process repeated one hundred times for statistical robustness. The average decrease in AUC is then taken as the importance score for that feature, with a larger drop indicating a greater reliance of the model on that variable. While this method captures complex, non-linear interactions as reflected in the final performance metric, its stochastic nature can lead to variance in the results, and the shuffling process may create unrealistic data instances that disrupt the very feature correlations the \texttt{ZZFeatureMap} is designed to learn.

To address these limitations, we implemented a second experimental technique: a gradient-based importance analysis. This method provides a more direct and stable probe of the model's sensitivity to each feature. Instead of the large-scale disruption of shuffling, it applies a minute, controlled perturbation ($\epsilon$) to a single feature at a time and measures the immediate impact on the model's output probability. The importance is quantified as the average absolute change in the predicted probability across the entire test set. This value serves as a numerical approximation of the gradient of the model's output with respect to its input, effectively measuring how much the model's prediction changes for a small change in a given feature. This approach is deterministic and better preserves the intricate, learned correlations between features, offering a more nuanced view of a feature's local influence on the classification decision.

In contrasting these approaches, the permutation method provides a holistic score of a feature's contribution to the final AUC, whereas the gradient method reveals the model's internal sensitivity to variations in the feature's value. Both quantum-inspired techniques stand apart from logistic regression by their ability to account for the non-linear relationships and entanglement-based correlations captured by the QNN. While the classical approach provides a clear, linear, but potentially incomplete picture, our experimental methods represent a necessary step toward unpacking the ``black box'' of quantum classifiers, allowing us to build confidence in their predictions and gain deeper insights from the complex patterns they uncover. The development of such interpretability tools is a vital research direction for the advancement and practical deployment of quantum machine learning in scientific discovery and beyond.

\section{Statistical Analysis of Patient Data}
\label{sec:statistical}
Initial statistical analysis of patient data was performed to assess the significance of the selected explanatory variables. In Table \ref{tab_lecebne_metody}, the structure of the patient cohort is presented based on the occurrence of leaks and the treatment methods used. In Table \ref{anamneza_pacienta}, the structure of the patient cohort is presented based on the occurrence of leaks and variables describing the patient's medical history.

\begin{table*}[htpb]
\caption{Structure of patients based on leak occurrence and treatment methods}
\centering
  \begin{tabular}{l l l l l l l}
 Variable & Category & Leak Occurred & No Leak & Total & $ \widehat{RR} $ & p-value \\
    &     & N = 28 (14\%)    & N = 172 (86\%) & N = 200 & (95\% CI) & ($ \chi^2 $ test)\\
\hline
NoCoil & No & 25 (17\%)  &  120 (83\%) & 145 & 3.16 & \textbf{0.032}  \\
& Yes  &  3  (5\%)   &    52 (95\%) & 55 & (0.99; 10.05)  \\ 
\hline 
ICG & No  &  19  (19\%)  &   81 (81\%) & 100 & 2.11 & \textbf{0.042}  \\
& Yes  &  9 (9\%)   &   91 (91\%) & 100 & (1.00; 4.44)  \\ 
\hline 
ACSP & No &  23  (17\%)  & 112 (83\%) & 135 & 2.21 & 0.074  \\
& Yes  &  5 (8\%)   &   60 (92\%) & 65 & (0.88; 5.56) \\ 
\hline
PERFB & Bad & 3 (20\%) & 12 (80\%) & 15 & 1.48  & 0.757  \\
 & Good & 25 (14\%)  &   160 (86\%) & 185  & (0.50; 4.34) \\ 
\hline
\label{tab_lecebne_metody}
  \end{tabular}
\end{table*}

The analysis shows that the occurrence of leaks is significantly associated with the insertion of a special rectal tube (variable NoCoil). As shown in Table \ref{tab_lecebne_metody}, the insertion of a special rectal tube reduces the occurrence of leaks by approximately 3.16 times. The 95\% confidence interval for the relative risk and the test of independence confirm that this measure has a statistically significant effect on the occurrence of leaks. Another treatment method showing a statistically significant association with leak occurrence is the use of intraoperative fluorescence imaging (ICG), which reduces the occurrence of leaks by 2.11 times. For the other treatment methods studied, no statistically significant effect on leak occurrence was demonstrated, as shown in Table \ref{tab_lecebne_metody}.

\begin{table*}[htpb]
\caption{Structure of patients based on leak occurrence and patient medical history}
\centering
  \begin{tabular}{l l l l l l l}
Variable & Category & Leak Occurred & No Leak & Total & $ \widehat{RR} $ & p-value \\
    &     & N = 28 (14\%)    & N = 172 (86\%) & N = 200 & (95\% CI) & ($ \chi^2 $ test)\\
\hline
DM & Yes & 9 (25\%) & 27 (75\%) & 36 & 2.16  & \textbf{0.036} \\
& No  &  19 (12\%) & 145 (88\%) & 164 & (1.06; 4.37) \\ 
\hline 
Smoking & Yes &  9  (26\%)  & 25 (74\%) & 34 & 2.31 & \textbf{0.042}  \\
& No  &  19 (11\%)   &   147 (89\%) & 166 & (1.15; 4.67) \\ 
\hline
HT & Yes & 19 (17\%)  &   95 (83\%) & 114 & 1.59 & 0.211  \\
& No  &  9 (10\%) & 77 (90\%) & 86 & (0.76; 3.34) \\ 
\hline
ASA  & $>2$ & 13 (18\%)  &   59 (82\%) & 72  & 1.54 & 0.215  \\
& 2  &  15 (12\%) & 113 (88\%) & 128 & (0.78; 3.05)   \\ 
\hline
CORT & Yes  &  2 (29\%) & 5 (71\%) & 193 & 2.12 & 0.564  \\
 & No & 26 (13\%)  &   167 (87\%) & 7 & (0.62; 7.22)  \\ 
\hline
Sex & Female  &  10  (14\%)  & 60 (86\%) & 70 & 1.03 & 0.932  \\
& Male  &  18 (14\%)   &   112 (86\%) & 130 & (0.50; 2.11)  \\ 
\hline
COAG & No  &  27 (14\%) & 164 (86\%) & 191 & 1.27 & $>0.999$   \\
& Yes & 1 (11\%)  &   8 (89\%) & 9  & (0.19; 8.34) \\ 
\hline
ICHS & Yes & 3 (14\%)  &   18 (86\%) & 21 & 0.98 & $>0.999$  \\
& No  &  25 (14\%) & 154 (86\%) & 179 & (0.32; 2.96)  \\ 
\hline
\label{anamneza_pacienta}
  \end{tabular}
\end{table*}

From Table \ref{anamneza_pacienta} it is evident that the patient's medical history most significantly influencing leak occurrence is diabetes mellitus. The presence of diabetes increases the occurrence of leaks by 2.16 times, and the 95\% confidence interval for the relative risk, along with the test of independence, confirms that this condition has a statistically significant effect on leak occurrence. Another factor from the patient's medical history with a statistically significant association with leak occurrence is smoking. Smokers have a 2.31 times higher occurrence of leaks compared to non-smokers. Among smokers and patients with diabetes, leaks occur in approximately 25\% of cases. Elevated blood pressure, ASA classification, and corticosteroid use also show notable effects in some patients, as indicated by the relative risk intervals in Table \ref{anamneza_pacienta}, but the influence of these variables on leak occurrence is not statistically significant. For other categorical variables in the patient's medical history, no statistically significant association with leak occurrence was demonstrated.

For quantitative variables, the description includes the median along with the interquartile range (lower quartile–upper quartile), the point and 95\% interval estimate of the difference in medians, and the corresponding Mann-Whitney test, which allows us to determine whether the influence of a given variable is statistically significant.

The description of the quantitative variables related to the patient's medical history, depending on the occurrence of leaks, the point and 95\% interval estimates of median differences (median for patients with a leak – median for patients without a leak), and the corresponding results of the Mann-Whitney test can be found in Table \ref{leak_spoj_tab}.

\begin{table*}[htpb] 
\caption{Association of Quantitative Variables with Leak Occurrence} 
\centering 
\begin{tabular}{l l l l l } 
Variable & Leak Occurred & No Leak & Median Difference & p-value \\
\hline CRP & 108 (78-131) & 62 (34-94) & 46 (25; 79) & \textbf{$<$0.001} \\
BMI & 28 (24-31) & 26 (24-29) & 2 (-1; 3) & ~0.348 \\
LN & 13 (6-16) & 13 (8-17) & 0 (-4; 2) & ~0.487 \\
Age & 65 (60-67) & 65 (58-71) & 0 (-4; 2) & ~0.503 \\
HB & 115 (103-121) & 116 (107-123) & -1 (-7; 4) &~0.525 \\
\hline 
\label{leak_spoj_tab} 
\end{tabular} 
\end{table*}

A statistically significant difference in C-reactive protein (CRP) median levels between patients with and without a leak is observed because a leak represents a type of inflammation, and CRP is a well-established inflammatory marker. As Sproston and Ashworth highlight in "Role of C-reactive protein at sites of inflammation and infection" \cite{sproston2018role}, CRP's primary function is its rapid increase in response to inflammatory stimuli. From a medical perspective, CRP should therefore not be considered among potential risk factors that could predict the occurrence of a leak beforehand; instead, it is a consequence marker, reflecting the body's inflammatory response after a leak has developed.

The timing of the C-reactive protein (CRP) measurement is crucial for its diagnostic value.
For none of the other continuous variables was a statistically significant association with leak occurrence observed, see Table \ref{leak_spoj_tab}.

The final predictive model includes NoCoil, ACSP, DM, and Smoking. In clinical practice, when AL is suspected, multiple treatment methods, particularly NoCoil and ICG, are often applied together, leading to overlapping effects that reduce ICG’s independent predictive value in the presence of other variables (Table\ref{SignificantVariablesModel}). To optimize the model, Akaike’s Information Criterion (AIC) was applied for feature reduction, resulting in model S2 (Table\ref{ReducedModelAIC}), which excluded ICG to improve the AIC score. ACSP, despite its marginal individual significance ($p=0.074$), was retained for its more stable and independent contribution to the model, making it a preferable choice over ICG.

As shown in Table~\ref{tab:patient_features}, these variables demonstrate strong associations with anastomotic leak (AL) ($\chi^2$ test).

\begin{table*}[htpb]
    \centering
    \caption{Summary of selected patient features and their association with AL}
    \label{tab:patient_features}
    \begin{tabular}{l l l l l l l}
        \toprule
        Variable & Category & Leak Occurred & No Leak & Total & $\widehat{RR}$ & p-value \\
        \midrule
        NoCoil & No & 25 (17\%) & 120 (83\%) & 145 & 3.16 & \textbf{0.032} \\
               & Yes & 3 (5\%) & 52 (95\%) & 55 & (0.99; 10.05) \\
        ACSP   & No & 23 (17\%) & 112 (83\%) & 135 & 2.21 & 0.074 \\
               & Yes & 5 (8\%) & 60 (92\%) & 65 & (0.88; 5.56) \\
        DM     & Yes & 9 (25\%) & 27 (75\%) & 36 & 2.16 & \textbf{0.036} \\
               & No & 19 (12\%) & 145 (88\%) & 164 & (1.06; 4.37) \\
        Smoking  & Yes & 9 (26\%) & 25 (74\%) & 34 & 2.31 & \textbf{0.042} \\
               & No & 19 (11\%) & 147 (89\%) & 166 & (1.15; 4.67) \\
        \bottomrule
    \end{tabular}
\end{table*}

The figure clearly illustrates the protective effect of NoCoil usage, where patients who received the special rectal tube had anastomotic leak occurrence in only 5\% of cases compared to 17\% in patients without NoCoil. Similarly, the figure demonstrates the beneficial impact of intraoperative fluorescence imaging (ICG), with leak rates of 9\% in the ICG group versus 19\% in patients who did not receive this intervention.

Regarding patient risk factors, the figure highlights that diabetes mellitus substantially increases leak risk, with 25\% of diabetic patients experiencing anastomotic leak compared to only 12\% of non-diabetic patients. The smoking status shows an even more pronounced effect, with smokers having a 26\% leak occurrence rate versus 11\% in non-smokers. The figure also shows that while ACSP (Acute Care Surgery Program) appears to have a protective trend with 8\% leak occurrence compared to 17\% without ACSP, this difference did not reach statistical significance ($p=0.074$).

\begin{figure*}[htpb]
    \centering
    \includegraphics[width=0.48\linewidth]{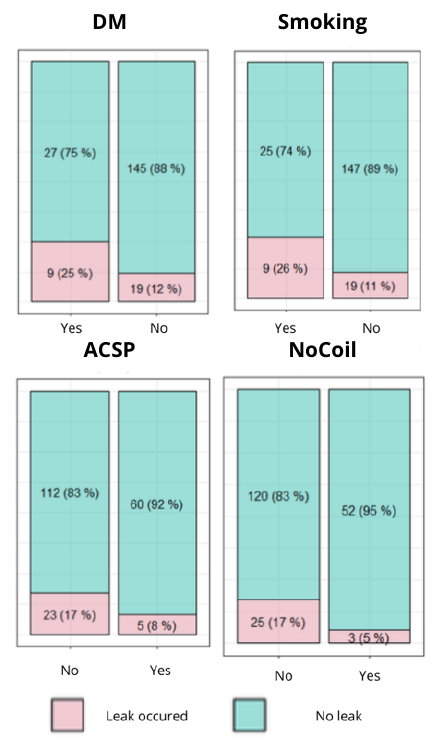}
    \caption{Risk factors of Anastomotic leak}
    \label{fig:riskfactors}
\end{figure*}

These visual comparisons in Figure~\ref{fig:riskfactors} effectively demonstrate the substantial clinical impact of both modifiable factors (smoking cessation, NoCoil usage, ICG implementation) and non-modifiable risk factors (diabetes mellitus) on anastomotic leak outcomes, supporting the statistical findings presented in the accompanying tables.

\section{Consolidated Logistic Regression Models}
\label{sec:consolidated_models}

Building on the initial statistical analysis (Section~\ref{sec:statistical}), which identified significant predictors of anastomotic leak (AL) such as NoCoil, ACSP, DM, and Smoking, we now construct consolidated logistic regression models to account for inter-variable relationships. To enhance model interpretability and reduce overfitting, we performed feature reduction from the initial set of 15 variables to focus on the most impactful predictors of AL.

In the logistic regression models presented in Tables~\ref{ComprehensiveModel1}, \ref{ReducedModelAIC}, \ref{ReducedModelBIC}, and \ref{SignificantVariablesModel}, the p-values for each predictor are derived from the Wald test, which evaluates the null hypothesis that a regression coefficient ($\beta$) is zero, indicating no effect on the log-odds of anastomotic leak (AL). The test statistic, $z = \hat{\beta}/\text{SE}(\hat{\beta})$, uses the estimated coefficient and its standard error, with low p-values (e.g., $<$ 0.05) indicating significant predictors like NoCoil or Smoking. Unlike the Wald test, which focuses on individual coefficients, the chi-squared test (used in likelihood ratio tests) assesses the overall significance of multiple predictors or model fit by comparing nested models, offering greater robustness for smaller samples.

The comprehensive model (S1) includes all 15 explanatory variables, as shown in Table~\ref{ComprehensiveModel1}. Unlike individual models, S1 captures interactions between variables, providing a more holistic view of their combined effects on AL.

\begin{table}[htpb]
\caption{Comprehensive Model (S1) with All Variables}
\centering
\begin{tabular}{l l l}
Variable & Coefficient Estimate & p-value \\
\hline
Constant & 6.729 & 0.061 \\
Age & -0.048 & 0.114 \\
Sex & -0.391 & 0.475 \\
BMI & 0.014 & 0.790 \\
ASA & -0.546 & 0.348 \\
Hypertension & -0.484 & 0.412 \\
ICHS & 0.098 & 0.906 \\
DM & -0.966 & \textbf{0.010} \\
Smoking & -1.380 & \textbf{0.014} \\
CORT & 0.103 & 0.923 \\
COAG & -0.746 & 0.539 \\
Hemoglobin & -0.006 & 0.752 \\
ICG & -0.577 & 0.378 \\
PERFB & -1.283 & 0.161 \\
ACSP & -0.764 & 0.249 \\
NoCoil & -1.601 & 0.055 \\
\hline
\end{tabular}
\label{ComprehensiveModel1}
\end{table}

To address non-significant variables, we applied Akaike’s Information Criterion (AIC) \cite{akaike2011akaike} for feature reduction, resulting in model S2 (Table~\ref{ReducedModelAIC}). This model retains DM, Smoking, NoCoil, and ACSP, which were significant or near-significant in individual analyses (Table~\ref{tab:patient_features}).

\begin{table}[htpb]
\caption{Reduced Model (S2) Based on AIC}
\centering
\begin{tabular}{l l l}
Variable & Coefficient Estimate & p-value \\
\hline
Constant & 1.436 & 0.073 \\
DM & -1.149 & \textbf{0.021} \\
Smoking & -1.429 & \textbf{0.005} \\
ACSP & -0.952 & 0.092 \\
NoCoil & -1.610 & \textbf{0.020} \\
\hline
\end{tabular}
\label{ReducedModelAIC}
\end{table}

ICG, initially significant (p = 0.042, Table~\ref{tab_lecebne_metody}), was excluded due to strong correlations with NoCoil, which reduces its independent predictive power. ACSP, while not significant at the 5\% level (p = 0.092), is a better predictor than ICG in this context. At a 10\% significance level, all S2 variables would be significant, but ACSP’s marginal significance suggests it could be excluded.

Further reduction using Bayesian Information Criterion (BIC) \cite{vrieze2012model}, which penalizes model complexity more stringently, yields model S3 (Table~\ref{ReducedModelBIC}). This model excludes ACSP, retaining DM, Smoking, and NoCoil, consistent with removing non-significant variables manually.

\begin{table}[htpb]
\caption{Reduced Model (S3) Based on BIC}
\centering
\begin{tabular}{l l l}
Variable & Coefficient Estimate & p-value \\
\hline
Constant & 0.319 & 0.592 \\
DM & -1.147 & \textbf{0.017} \\
Smoking & -1.261 & \textbf{0.010} \\
NoCoil & -1.439 & \textbf{0.027} \\
\hline
\end{tabular}
\label{ReducedModelBIC}
\end{table}

The significance of reduced models S2 and S3 was evaluated using likelihood ratio tests with a $\chi^2$ test, comparing them to the null model ($l_0$) and S1 (Table~\ref{LikelihoodRatioTests}).

\begin{table}[htpb]
\caption{Likelihood Ratio Tests for Reduced Models S2 and S3 (DoF: Degrees of Freedom)}
\centering
\begin{tabular}{l l l l}
Model & DoF & Log-Likelihood & p-value ($\chi^2$ test) \\
\hline
S2 & 5 & -70 & \\
$l_0$ & 1 & -81 & $<$\textbf{0.001} \\
\hline
S3 & 4 & -73 & \\
$l_0$ & 1 & -81 & \textbf{0.001} \\
\hline
S1 & 16 & -67 & \\
S2 & 5 & -70 & 0.879 \\
\hline
S1 & 16 & -67 & \\
S3 & 4 & -73 & 0.551 \\
\hline
\end{tabular}
\label{LikelihoodRatioTests}
\end{table}

Both S2 and S3 are significantly different from the null model but not from S1, indicating that they retain predictive power with fewer variables.

Model S4 includes variables significant at p $<$ 0.05 from individual analyses (DM, Smoking, NoCoil, ICG; Table~\ref{tab:patient_features} and Table~\ref{tab_lecebne_metody}). However, NoCoil and ICG lose significance in S4 (Table~\ref{SignificantVariablesModel}), likely due to their correlated use in practice, where multiple treatment methods are applied when AL is suspected.

\begin{table}[htpb]
\caption{Model (S4) with Significant Individual Variables}
\centering
\begin{tabular}{l l l}
Variable & Coefficient Estimate & p-value \\
\hline
Constant & 0.587 & 0.361 \\
DM & -1.128 & \textbf{0.021} \\
NoCoil & -1.051 & 0.150 \\
ICG & -0.633 & 0.220 \\
Smoking & -1.376 & \textbf{0.007} \\
\hline
\end{tabular}
\label{SignificantVariablesModel}
\end{table}

Model quality was assessed using determination coefficients (Table~\ref{DeterminationCoefficients}) and classification metrics \cite{mcfadden1987regression, smith2013comparison, ansin2015evaluation} (Table~\ref{ClassificationSummary}).

\begin{table}[htpb]
\caption{Determination Coefficients of Models}
\centering
\begin{tabular}{l l l l}
Model & McFadden’s & Cox-Snell’s & Nagelkerke’s \\
\hline
S1 & 0.171 & 0.129 & 0.233 \\
S2 & 0.134 & 0.103 & 0.185 \\
S3 & 0.098 & 0.077 & 0.138 \\
S4 & 0.108 & 0.084 & 0.151 \\
\hline
\end{tabular}
\label{DeterminationCoefficients}
\end{table}

\begin{table}[htpb]
\caption{Summary of Classification Performance}
\centering
\begin{tabular}{l l l l l}
Metric & Model S1 & Model S2 & Model S3 & Model S4 \\
\hline
Accuracy & 75\% & 69\% & 78\% & 78\% \\
Sensitivity & 57\% & 57\% & 29\% & 29\% \\
Specificity & 77\% & 71\% & 85\% & 85\% \\
PPV & 27\% & 22\% & 22\% & 22\% \\
NPV & 93\% & 92\% & 89\% & 89\% \\
\hline
\end{tabular}
\label{ClassificationSummary}
\end{table}

Feature reduction was performed to simplify the model by eliminating redundant or non-significant variables from the initial 15 predictors. This process mitigates overfitting, enhances interpretability, and focuses on key predictors (DM, Smoking, NoCoil, and potentially ACSP) that drive AL risk. By using AIC and BIC, we systematically identified a parsimonious model that maintains predictive power while accounting for inter-variable relationships, which individual models failed to capture.

\section{Results}

Building on the statistical insights from Section \ref{sec:statistical}, where NoCoil, ACSP, DM, and Smoking emerged as key predictors of anastomotic leakage, we now evaluate the performance of Quantum Neural Networks (QNNs) for this clinical classification task. The analysis begins by optimizing variational parameters in quantum circuits under realistic noise conditions, comparing the convergence and stability of ansätze like Real Amplitudes (RA) and EfficientSU2 (ESU2) when paired with classical optimizers such as BFGS and CMA-ES. These optimized QNNs are then rigorously benchmarked against classical machine learning models—including Logistic Regression and Multi-Layer Perceptrons—using metrics tailored to clinical needs, such as sensitivity-driven thresholds and probability calibration.

Beyond raw performance, we dissect the interpretability of QNNs through two novel feature importance methods (permutation- and gradient-based), contrasting their non-linear insights with the odds ratios from classical logistic regression. This reveals surprising divergences: while logistic regression highlights smoking and NoCoil as dominant risk factors, the QNN's quantum correlations emphasize DM and ACSP in ways that suggest entangled feature interactions. The clinical implications of these findings are carefully weighed, balancing the QNN's superior screening potential (with its high negative predictive value) against the well-calibrated probabilities of classical models like MLPs. Together, these results map a path toward hybrid quantum-classical approaches for postoperative risk stratification, where ansatz selection and optimizer choice become critical levers for aligning model behavior with clinical priorities.

\subsection{Optimization of Variational Parameters in Quantum Neural Network Ansätze}

This subsection focuses on the mean convergence plots of various optimization algorithms when applied to both the Real Amplitudes (RA) and EfficientSU2 (ESU2) ansätze in a quantum neural network. The results presented here are averaged over 10 independent runs for each optimizer-ansatz pair. It is crucial to note that this optimization was performed in a noisy environment, incorporating sampling noise (1024 shots) and realistic modeled errors, including quantum decoherence and gate errors, to simulate practical quantum computing conditions. Individual convergence traces for each run can be found in Appendix \ref{sec:runs}.

\begin{figure*}
    \centering
    \includegraphics[width=1\linewidth]{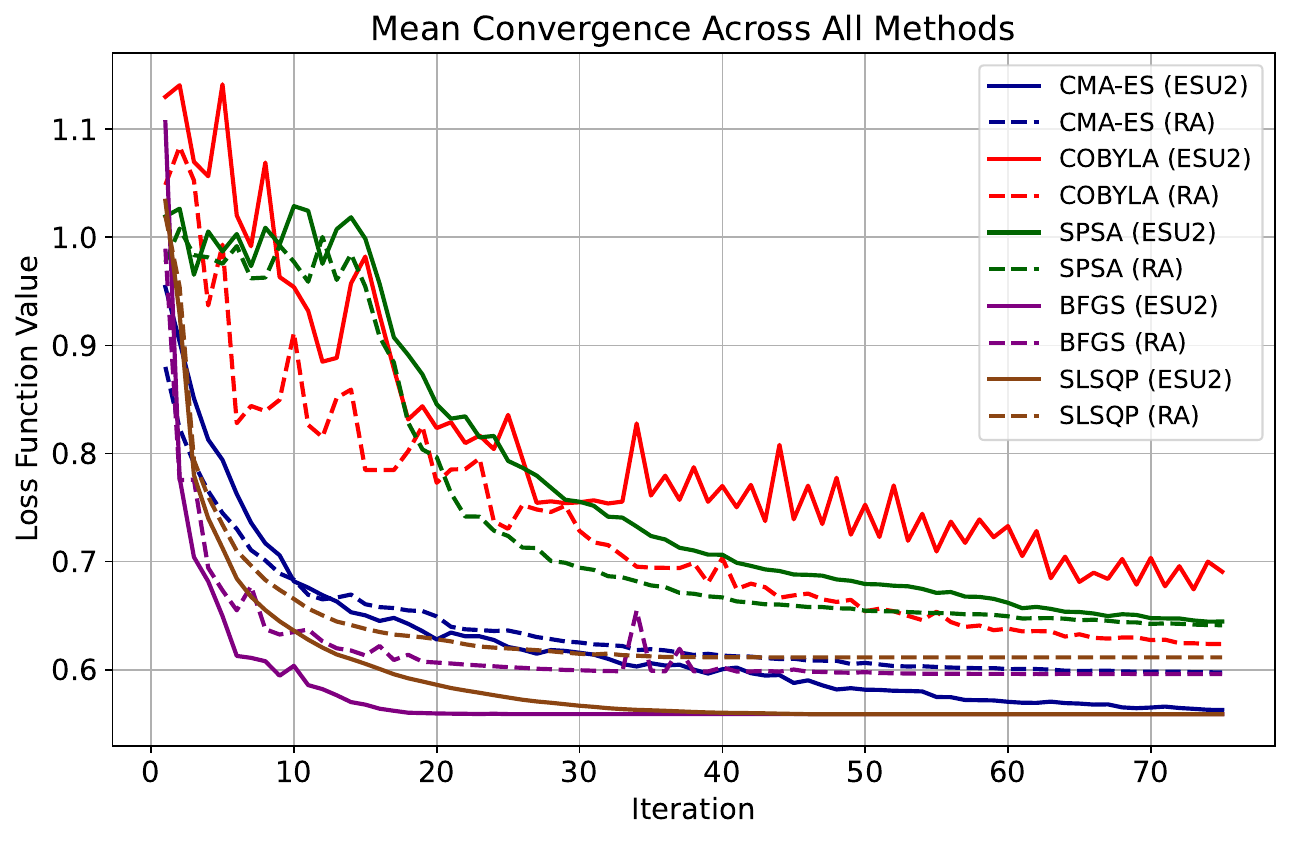}
    \caption{Mean convergence of loss function values over 10 runs for Real Amplitudes (RA) and EfficientSU2 (ESU2) ansätze, optimized using different algorithms. The optimization includes sampling noise (1024 shots) and modeled quantum decoherence and gate errors.}
    \label{fig:mean-convergence-all-methods}
\end{figure*}

The mean convergence plots, shown in Figure \ref{fig:mean-convergence-all-methods}, illustrate the performance of different optimizers in minimizing the loss function. The summarized results for the mean final error and their standard deviations over the 10 runs are presented in Table \ref{tab:optimization-results}.

\begin{table}[htpb]
\centering
\caption{Summarized Mean Final Error and Standard Deviation for Different Optimizers and Ansätze (Mean over 10 Runs)}
\label{tab:optimization-results}
\begin{tabular}{|l|l|c|c|}
\hline
\textbf{Optimizer} & \textbf{Ansatz} & \textbf{Mean Error} & \textbf{Std Dev} \\
\hline
BFGS & ESU2 & 0.558914 & 0.008642 \\
BFGS & RA & 0.596073 & 0.010018 \\
CMA-ES & ESU2 & 0.562686 & 0.010454 \\
CMA-ES & RA & 0.597349 & 0.011816 \\
COBYLA & ESU2 & 0.690626 & 0.02874 \\
COBYLA & RA & 0.623657 & 0.015775 \\
SLSQP & ESU2 & 0.558703 & 0.011492 \\
SLSQP & RA & 0.611398 & 0.018218 \\
SPSA & ESU2 & 0.644543 & 0.035807 \\
SPSA & RA & 0.641119 & 0.027142 \\
\hline
\end{tabular}
\end{table}

As observed from Table \ref{tab:optimization-results}, the SLSQP optimizer achieved the lowest mean final error for the ESU2 ansatz ($0.558703 \pm 0.011492$), closely followed by BFGS with ESU2 ($0.558914 \pm 0.008642$) and CMA-ES with ESU2 ($0.562686 \pm 0.010454$). These three optimizers, BFGS, CMA-ES, and SLSQP, consistently demonstrated superior performance across both ansätze, converging to lower loss function values with relatively small standard deviations. This indicates robust optimization capabilities even when operating within a noisy quantum environment that includes sampling noise (1024 shots) and modeled errors such as quantum decoherence and gate errors.

Comparing the two ansätze, the ESU2 ansatz generally yielded slightly better (lower) mean final errors when paired with these top-performing optimizers compared to the RA ansatz. For instance, the lowest error overall was achieved by SLSQP with ESU2. In contrast, COBYLA showed higher final errors, particularly with the ESU2 ansatz ($0.690626 \pm 0.02874$), although it performed comparatively better with the RA ansatz ($0.623657 \pm 0.015775$). SPSA exhibited weaker performance than other tested optimizers, characterized by significantly higher mean final errors (e.g., $0.644543 \pm 0.035807$ for ESU2) and larger standard deviations, suggesting less stable and less effective convergence in the presence of noise. This comprehensive analysis underscores the critical importance of judiciously selecting an appropriate optimizer for the training of quantum neural networks, especially under realistic noisy conditions, where some optimizers prove far more resilient and effective than others.

\subsection{QNN Performance Analysis Across Multiple Metrics}

\begin{figure*}[htpb]
    \centering
    \includegraphics[width=1\linewidth]{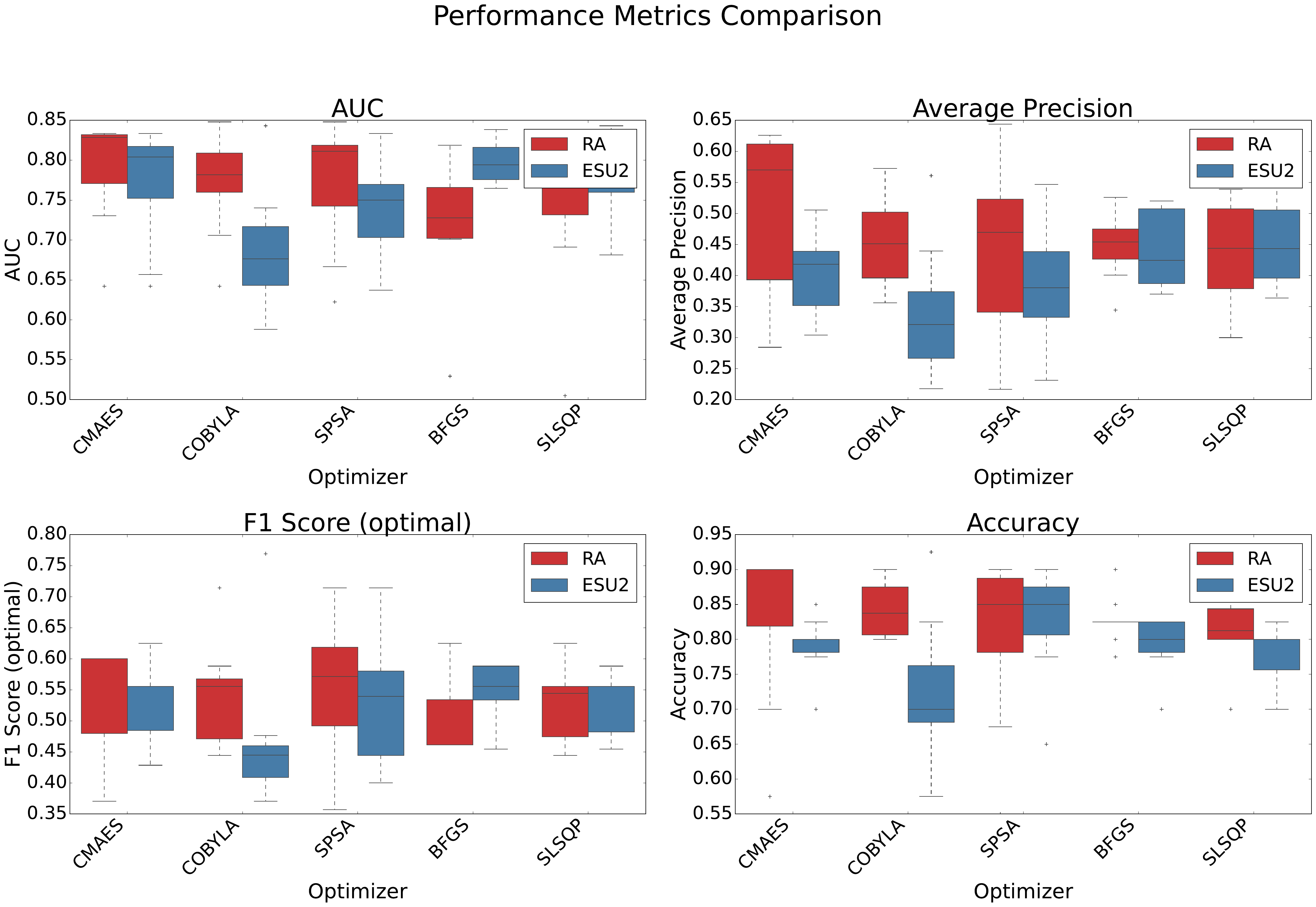}
    \caption{Comparison of performance metrics (AUC, Average Precision, F1 Score, and Accuracy) for QNN models with Real Amplitudes and Efficient SU(2) ansätze across different optimizers. Box plots show the distribution of metrics over 10 independent runs.}
    \label{fig:performance_metrics}
\end{figure*}

\begin{figure*}[htpb]
    \centering
    \includegraphics[width=1\linewidth]{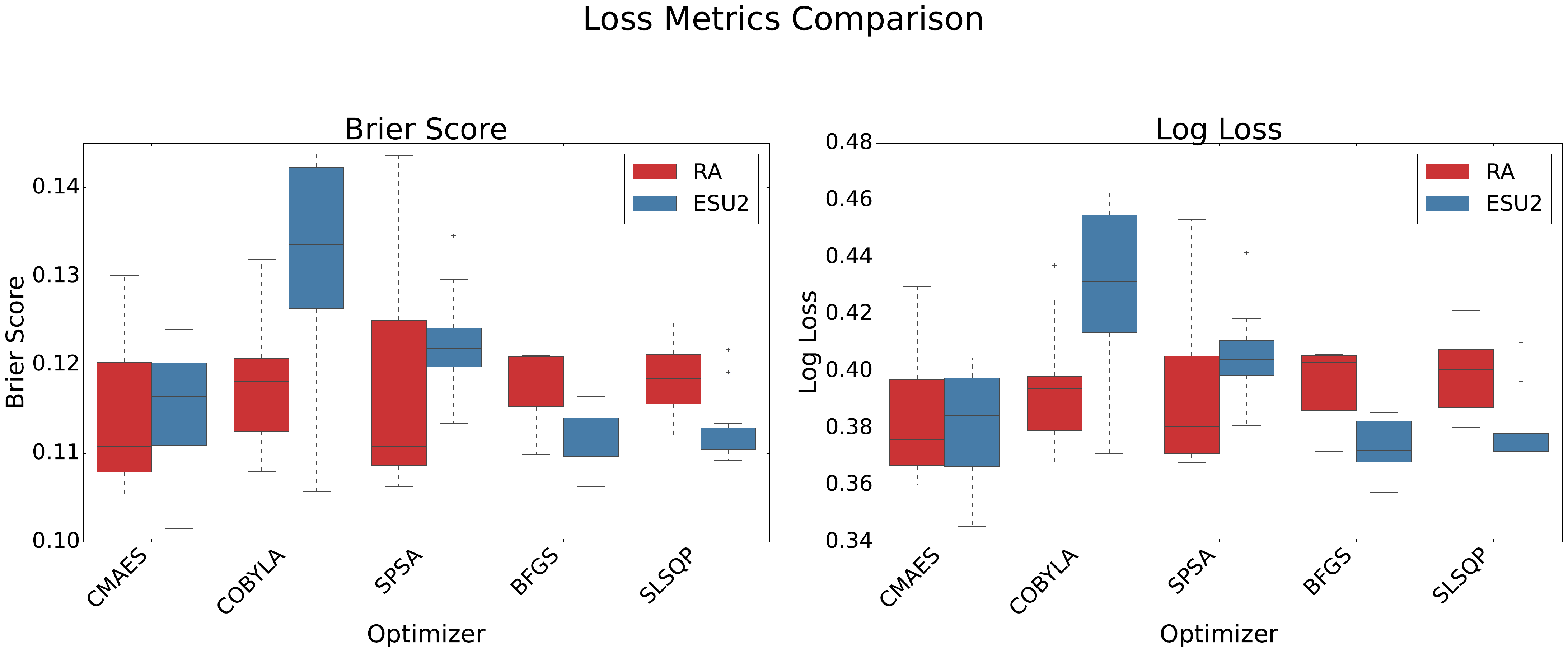}
    \caption{Comparison of loss metrics (Brier Score and Log Loss) for QNN models with Real Amplitudes and Efficient SU(2) ansätze across different optimizers. Box plots show the distribution of metrics over 10 independent runs.}
    \label{fig:loss_metrics}
\end{figure*}

\begin{figure*}[htpb]
    \centering
    \includegraphics[width=1\linewidth]{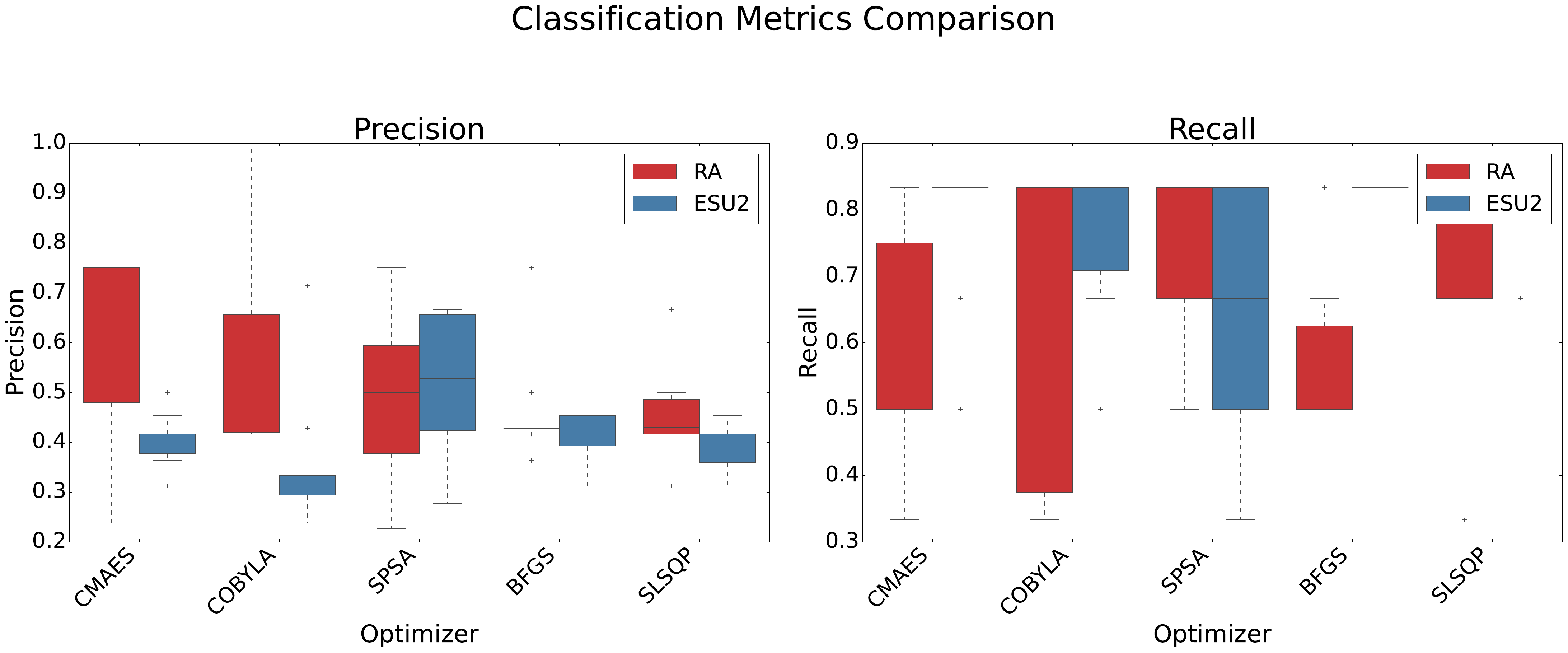}
    \caption{Comparison of classification metrics (Precision and Recall) for QNN models with Real Amplitudes and Efficient SU(2) ansätze across different optimizers. Box plots show the distribution of metrics over 10 independent runs.}
    \label{fig:classification_metrics}
\end{figure*}

\begin{figure*}[htpb]
    \centering
    \includegraphics[width=1\linewidth]{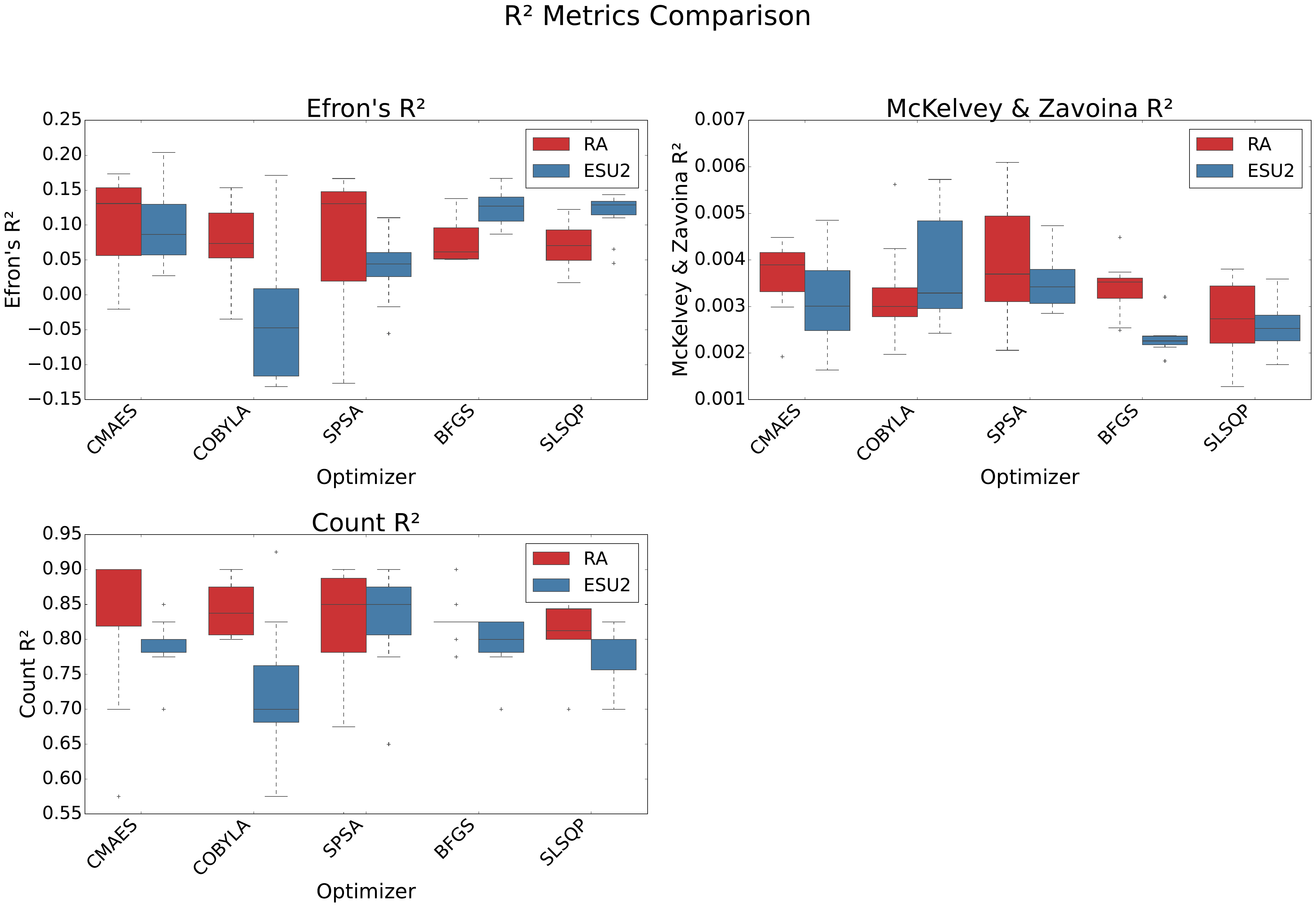}
    \caption{Comparison of pseudo-$R^2$ scores (Efron's, McKelvey \& Zavoina's, and Count) for QNN models with Real Amplitudes and Efficient SU(2) ansätze across different optimizers. Box plots show the distribution of metrics over 10 independent runs.}
    \label{fig:r2_metrics}
\end{figure*}

We evaluated two ansätze—Real Amplitudes (RA) and Efficient SU(2) (ESU2)—paired with six optimization algorithms: Broyden–Fletcher–Goldfarb–Shanno (BFGS), Covariance Matrix Adaptation Evolution Strategy (CMA-ES), Constrained Optimization by Linear Approximation (COBYLA), Simultaneous Perturbation Stochastic Approximation (SPSA), Sequential Least Squares Programming (SLSQP), and one additional optimizer. To account for the stochastic nature of quantum circuit sampling and parameter training, we conducted 10 independent runs for each optimizer-ansatz pair, ensuring statistical robustness. Performance, loss, classification, and pseudo-$R^2$ metrics are visualized in Figures~\ref{fig:performance_metrics}--\ref{fig:r2_metrics}, using box plots to illustrate metric distributions over these runs.

The overall performance of QNN models is summarized in Figure~\ref{fig:performance_metrics}, which displays the Area Under the Receiver Operating Characteristic Curve (AUC), Average Precision, F1 Score, and Accuracy. The ESU2-BFGS combination achieved the highest mean AUC of $0.7966 \pm 0.0237$, indicating strong discriminative power in distinguishing patients with and without post-surgical complications. For Average Precision, critical for imbalanced medical datasets, the RA-CMA-ES model excelled with a score of $0.5041 \pm 0.1214$. The RA-SPSA configuration provided the best F1 Score ($0.5567 \pm 0.1030$), balancing precision and recall effectively. The highest accuracy was achieved by RA-COBYLA ($0.8450 \pm 0.0384$), though accuracy alone can be misleading in imbalanced datasets. These results highlight that different ansatz-optimizer pairs excel in specific performance aspects, underscoring the importance of task-specific configuration selection.

Figure~\ref{fig:loss_metrics} presents the Brier Score and Log Loss, which assess predictive accuracy and probability calibration. Lower values indicate better-calibrated predictions. The ESU2-BFGS model achieved the lowest Brier Score ($0.1116 \pm 0.0029$) and Log Loss ($0.3737 \pm 0.0087$), with small standard deviations indicating stable convergence and reliable probability estimates. In contrast, gradient-free optimizers like CMA-ES and COBYLA exhibited higher variance in loss metrics, reflecting their susceptibility to the stochasticity of quantum optimization landscapes.

The precision-recall trade-off, crucial for medical applications, is illustrated in Figure~\ref{fig:classification_metrics}. The RA-CMA-ES model achieved the highest mean precision ($0.6134 \pm 0.1962$), making it suitable for minimizing false positives, a priority in scenarios where misclassification carries significant consequences. Conversely, the ESU2-BFGS model demonstrated perfect and consistent recall ($0.8333 \pm 0.0000$), ensuring all positive cases were identified, a critical feature for medical diagnostics where missing complications can be detrimental. The RA ansatz generally showed higher precision but with greater variance, while ESU2 maintained consistency in recall.

The goodness-of-fit of the models is evaluated using pseudo-$R^2$ metrics, shown in Figure~\ref{fig:r2_metrics}. The ESU2-BFGS model achieved the highest Efron's $R^2$ ($0.1247 \pm 0.0230$), indicating a better fit to the data distribution. The RA-COBYLA model led in Count $R^2$, consistent with its high accuracy. However, McKelvey and Zavoina's $R^2$ values were low across all models, with RA-SPSA performing best ($0.0039 \pm 0.0012$), suggesting limited explained variance, typical for complex medical classification tasks with inherent noise.

\begin{framed}
\textbf{Connection to Variational Parameter Optimization:} The effective optimization of variational parameters in the parameterized quantum gates of the ansätze directly influences all downstream performance metrics. The superior convergence achieved by optimizers like BFGS, CMA-ES, and SLSQP with the ESU2 ansatz (as shown in Table \ref{tab:optimization-results}) translates into better calibrated probabilities, improved classification accuracy, and more reliable clinical predictions. This demonstrates the critical pathway from quantum circuit optimization to practical medical applications.
\end{framed}

The comprehensive QNN analysis reveals the complex interplay between ansatz and optimizer in quantum neural network performance. The ESU2-BFGS combination consistently excelled in AUC, loss metrics, and recall, making it ideal for applications requiring reliable class separation and probability calibration. The RA ansatz, particularly with CMA-ES, demonstrated superior precision and robustness, as evidenced by smaller interquartile ranges and fewer outliers in box plots, making it suitable for minimizing false positives. CMA-ES's evolutionary approach proved particularly robust against the stochasticity of quantum optimization, including shot noise and gate errors, offering stable performance across runs. This reliability is critical for practical deployment in medical settings.

No single configuration dominated across all metrics, highlighting the need for task-specific optimization. For instance, ESU2-BFGS is preferable for high recall and calibration, while RA-CMA-ES suits precision-critical tasks. The observed performance variations emphasize the necessity of comprehensive empirical evaluation in quantum machine learning, particularly for medical classification, where both accuracy and reliability are paramount.

A key finding of our analysis is the superior probability calibration of classical models like MLP compared to the QNNs. This weaker calibration in QNNs may stem from several factors inherent to the variational quantum approach: the stochasticity of shot-noise in the measurement process, the complexity of the optimization landscapes which can lead to convergence in local minima, and the impact of simulated hardware noise which can impair the fine-tuning of output probabilities. While QNNs excelled at the classification task of separating high-risk from low-risk patients, their raw output probabilities are less reliable. Future work should investigate post-hoc calibration techniques, such as Platt scaling or isotonic regression, which could be applied to QNN outputs to improve their reliability without sacrificing their classification strength. Developing hybrid quantum-classical training methodologies that explicitly optimize for a calibration-aware loss function is another promising research direction.

\subsection{Comparative Evaluation Against Classical Machine Learning Models}

To evaluate the performance of QNNs against established classical machine learning models for predicting rare post-surgical complications, specifically anastomotic leakage, we conducted a comprehensive assessment using multiple performance metrics. The classical models— Logistic Regression (LR), AdaBoost, Linear Discriminant Analysis (LDA), Gaussian Naive Bayes (GNB), and Multi-Layer Perceptron (MLP)—were optimized with fine-tuned hyperparameters to utilize their full potential (hyperparameters described in detail in Section \ref{sec:hyperparams}). This ensures a fair comparison by leveraging the optimal configuration of each classical approach.

Anastomotic leakage, occurring in approximately 14\% of patients, is a severe complication where high sensitivity (recall) is critical to ensure timely identification and intervention. To enable a fair comparison, we optimized the decision threshold for each model to achieve the highest possible sensitivity while using an F-beta score to balance precision and recall, prioritizing the detection of true positives. This approach reflects the clinical importance of minimizing false negatives, as missing an anastomotic leak can lead to life-threatening consequences.

For QNNs, which incorporate variational quantum circuits with Real Amplitudes (RA) or Efficient SU2 (ESU2) ansätze and are optimized using CMAES, COBYLA, SPSA, BFGS, or SLSQP, the reported metrics represent the mean performance over 10 runs due to their stochastic nature. In contrast, classical models are evaluated based on a single run, as most classical models are deterministic, and for the MLP, a fixed random seed was used to ensure reproducibility.

Table~\ref{tab:model_comparison} presents a detailed comparison of model performance at a fixed sensitivity of 83\%, achieved by adjusting thresholds to maximize sensitivity while optimizing the F-beta score. The table includes both threshold-dependent metrics (Count R², F1 Score, Specificity, PPV, NPV, Accuracy) and probability calibration metrics (Efron's R², McKelvey \& Zavoina R², Brier Score, Log Loss).

Among the classical models, MLP demonstrates the highest Efron's R² (0.191) and the lowest Brier Score (0.103) and Log Loss (0.359), indicating strong calibration and discriminative power, though its F1 Score (21\%) is the lowest due to lower precision. GNB achieves the highest Count R² (0.500) and F1 Score (33\%) among classical models, with a Specificity of 44\% and PPV of 21\%, reflecting a balanced performance. LR, LDA, and AdaBoost show moderate performance, with Count R² values ranging from 0.400 to 0.450 and F1 Scores from 29\% to 31\%.

QNNs generally outperform classical models across most metrics. QNN - COBYLA - RA achieves the highest Count R² (0.845) and Accuracy (85\%), with a strong F1 Score (54\%) and Specificity (58\%). QNN - BFGS - ESU2 shows the highest Efron's R² (0.125) among QNNs and a competitive Brier Score (0.112) and Log Loss (0.374), alongside a high Specificity (66\%) and PPV (32\%). QNN - CMAES - RA and QNN - SPSA - ESU2 also perform well, with Count R² values of 0.835 and 0.830, respectively, and F1 Scores of 54\% and 53\%. Notably, QNN - CMAES - RA achieves the highest NPV (96\%), indicating excellent performance in ruling out non-cases. However, QNN - COBYLA - ESU2 underperforms, with a negative Efron's R² (-0.034) and the highest Brier Score (0.132) and Log Loss (0.430), suggesting poorer calibration for this configuration.

\begin{table*}[htbp]
    \centering
    \caption{Performance Comparison of Predictive Models at Fixed Sensitivity of 83\%, with QNN Metrics Averaged Over 10 Runs and Classical Model Metrics from a Single Run}
    \label{tab:model_comparison}
    \begin{adjustbox}{width=\textwidth}
        \begin{tabular}{lcccccc}
            \toprule
            \multirow{2}{*}{Model} & \multicolumn{6}{c}{Model performance metrics at fixed value of sensitivity = 83\%} \\
            \cmidrule(lr){2-7}
            & Efron's R² & McKelvey \& Zavoina R² & Count R² & Brier Score & Log Loss & F1 Score \\
            \midrule
            LR & 0.072 & 0.001 & 0.400 & 0.118 & 0.390 & 29\% \\
            AdaBoost & -0.054 & 0.001 & 0.450 & 0.134 & 0.443 & 31\% \\
            LDA & 0.095 & 0.002 & 0.400 & 0.115 & 0.381 & 29\% \\
            GNB & 0.087 & 0.007 & 0.500 & 0.116 & 0.393 & 33\% \\
            MLP & 0.191 & 0.007 & 0.425 & 0.103 & 0.359 & 21\% \\
            QNN - CMAES - RA & 0.103 & 0.004 & 0.835 & 0.114 & 0.384 & 54\% \\
            QNN - CMAES - ESU2 & 0.099 & 0.003 & 0.793 & 0.115 & 0.381 & 53\% \\
            QNN - COBYLA - RA & 0.070 & 0.003 & 0.845 & 0.119 & 0.395 & 54\% \\
            QNN - COBYLA - ESU2 & -0.034 & 0.004 & 0.728 & 0.132 & 0.430 & 47\% \\
            QNN - SPSA - RA & 0.082 & 0.004 & 0.805 & 0.117 & 0.392 & 56\% \\
            QNN - SPSA - ESU2 & 0.040 & 0.004 & 0.830 & 0.122 & 0.406 & 53\% \\
            QNN - BFGS - RA & 0.077 & 0.003 & 0.828 & 0.118 & 0.395 & 50\% \\
            QNN - BFGS - ESU2 & 0.125 & 0.002 & 0.788 & 0.112 & 0.374 & 55\% \\
            QNN - SLSQP - RA & 0.071 & 0.003 & 0.815 & 0.119 & 0.399 & 53\% \\
            QNN - SLSQP - ESU2 & 0.115 & 0.003 & 0.780 & 0.113 & 0.379 & 53\% \\
            \midrule
            \multicolumn{7}{c}{Predictive power and predictive validity metrics at sensitivity = 83\%} \\
            \midrule
            GNB & Threshold: 0.4103 & Specificity: 44\% & PPV: 21\% & NPV: 94\% & Accuracy: 50\% \\
            LR & Threshold: 0.1565 & Specificity: 32\% & PPV: 18\% & NPV: 92\% & Accuracy: 40\% \\
            AdaBoost & Threshold: 0.2800 & Specificity: 38\% & PPV: 19\% & NPV: 93\% & Accuracy: 45\% \\
            MLP & Threshold: 0.1333 & Specificity: 38\% & PPV: 19\% & NPV: 93\% & Accuracy: 45\% \\
            LDA & Threshold: 0.1348 & Specificity: 32\% & PPV: 18\% & NPV: 92\% & Accuracy: 40\% \\
            QNN - CMAES - RA & Threshold: 0.29 & Specificity: 64\% & PPV: 29\% & NPV: 96\% & Accuracy: 84\% \\
            QNN - CMAES - ESU2 & Threshold: 0.28 & Specificity: 58\% & PPV: 29\% & NPV: 85\% & Accuracy: 79\% \\
            QNN - COBYLA - RA & Threshold: 0.26 & Specificity: 58\% & PPV: 26\% & NPV: 95\% & Accuracy: 85\% \\
            QNN - COBYLA - ESU2 & Threshold: 0.23 & Specificity: 49\% & PPV: 23\% & NPV: 93\% & Accuracy: 73\% \\
            QNN - SPSA - RA & Threshold: 0.26 & Specificity: 58\% & PPV: 26\% & NPV: 95\% & Accuracy: 81\% \\
            QNN - SPSA - ESU2 & Threshold: 0.24 & Specificity: 53\% & PPV: 24\% & NPV: 95\% & Accuracy: 83\% \\
            QNN - BFGS - RA & Threshold: 0.21 & Specificity: 35\% & PPV: 21\% & NPV: 60\% & Accuracy: 83\% \\
            QNN - BFGS - ESU2 & Threshold: 0.32 & Specificity: 66\% & PPV: 32\% & NPV: 96\% & Accuracy: 79\% \\
            QNN - SLSQP - RA & Threshold: 0.22 & Specificity: 47\% & PPV: 22\% & NPV: 90\% & Accuracy: 82\% \\
            QNN - SLSQP - ESU2 & Threshold: 0.34 & Specificity: 65\% & PPV: 34\% & NPV: 86\% & Accuracy: 78\% \\
            \bottomrule
        \end{tabular}
    \end{adjustbox}
\end{table*}

The results highlight the potential of QNNs, particularly with RA and ESU2 ansätze optimized by COBYLA, CMAES, or BFGS, to outperform classical models in detecting anastomotic leakage at high sensitivity. The superior Count R², Accuracy, and NPV of QNNs suggest they are effective at both identifying true positives and minimizing false positives, critical for clinical applications. However, the variability in QNN performance across optimizers and ansätze underscores the importance of careful configuration selection.

The results in Table~\ref{tab:model_comparison}, evaluated at a fixed sensitivity of 83\% with F-beta optimized thresholds to prioritize anastomotic leakage detection, reveal distinct performance profiles for QNNs and classical models. QNNs generally show stronger threshold-dependent metrics (Count R², F1 Score, Specificity, PPV, NPV, Accuracy), particularly with certain ansatz-optimizer combinations, making them effective for identifying true positives and minimizing false positives in clinical screening. However, their probability calibration metrics (Efron's R², Brier Score, Log Loss) vary, with some configurations exhibiting weaker calibration. Classical models, particularly MLP, excel in calibration, producing reliable probability estimates suitable for risk stratification, but often at the cost of lower classification performance due to increased false positives at high sensitivity.

These differences likely arise from architectural and optimization distinctions. QNNs' variational quantum circuits may leverage complex feature interactions, potentially via entanglement, to enhance classification, but simulated noise (e.g., depolarizing errors) and stochasticity from multiple runs can impair probability calibration. Conversely, classical models like MLP, optimized for log loss minimization, prioritize well-calibrated probabilities, though lower decision thresholds at 83\% sensitivity reduce their specificity and PPV. 

Clinically, QNNs are promising for screening anastomotic leakage due to robust classification performance, while classical models like MLP are better suited for precise risk assessment. The findings provide valuable insights into the practical implementation of QNNs, guiding the selection of ansatz-optimizer pairs to achieve the desired balance of performance, robustness, and reliability in post-surgical complication detection. Future work could explore post-hoc calibration techniques for QNNs or hybrid quantum-classical approaches to combine the classification strength of quantum approaches with the calibration reliability of classical methods.

\subsection{Clinical Implications}

The findings of this comparative analysis bear significant implications for the application of quantum machine learning in biomedical contexts, particularly for critical tasks like the early detection of rare post-surgical complications.  The QNN’s demonstrated strength in threshold-based classification, particularly its superior performance in metrics pertinent to high sensitivity, suggests its potential as a powerful tool for screening and identifying patients at risk of anastomotic leakage.  

In settings where minimizing missed cases (false negatives) is paramount—a defining characteristic of rare, severe complication detection—the QNN’s classification profile offers a compelling advantage. However, the generalizability of these results is constrained by the study’s limited sample size (N=200, with 28 AL events), which inherently reduces statistical power and increases overfitting risk. While rigorous feature selection and cross-validation mitigate this, external validation in larger, multi-center cohorts is essential before clinical deployment.

In contrast, the MLP’s proficiency in probability calibration positions it as a valuable asset in scenarios where accurate risk stratification and reliable probability estimates are central.  For example, in clinical decision-making processes that rely on nuanced risk assessments to guide personalized interventions, the well-calibrated probabilities provided by MLP could be highly beneficial.

The trade-off we identified has direct clinical relevance. A high QNN score, indicating a high risk of AL, could serve as an effective clinical decision support alert. For example, it could automatically trigger a recommendation for closer post-operative monitoring, a delayed oral diet, or a follow-up CT scan for at-risk patients. The high NPV of the top-performing QNNs (up to 96\%) is particularly valuable, providing strong confidence in identifying patients who are unlikely to develop a leak. Conversely, the well-calibrated probability from an MLP could be integrated into a broader patient risk score, helping clinicians balance the risk of AL against other surgical risks when planning the intervention itself. Even a modest improvement in NPV from 93\% (MLP) to 96\% (QNN) means that for every 100 patients predicted to be low-risk, three fewer will have been incorrectly classified, representing a meaningful gain in patient safety.

\subsection{Feature Importance}

The feature importance methods developed for the quantum neural network (QNN) are highly experimental, as interpreting feature contributions in quantum models is an emerging area of research. While logistic regression provides a well-established, interpretable framework for understanding feature effects through odds ratios, the QNN feature importance analyses (permutation- and gradient-based) are novel and less validated, serving as an exploratory approach to uncover complex patterns in the quantum model’s decision-making process.
The logistic regression model developed to predict the probability of anastomotic leak is expressed as:
\begin{align*}
\ln\left(\frac{p}{1-p}\right) &= 1.436 - 1.149 \cdot \text{DM} \\
&\quad - 1.429 \cdot \text{Smoking} \\
&\quad - 0.952 \cdot \text{ACSP} \\
&\quad - 1.610 \cdot \text{NoCoil},
\end{align*}
where $p$ represents the probability of an anastomotic leak, and the variables DM (Diabetes Mellitus), Smoking, ACSP (use of a specific treatment method), and NoCoil (non-use of a rectal tube) are dichotomous predictors. The point estimates of the odds ratios (OR) with their corresponding 95
{\footnotesize
\begin{align*}
\text{OR(DM)} &= e^{-1.149} = 3.16 , (95\% , \text{CI}: 1.17, 8.32), \\
\text{OR(Smoking)} &= e^{-1.429} = 4.18 , (95\% , \text{CI}: 1.51, 11.48), \\
\text{OR(ACSP)} &= e^{-0.952} = 2.59 , (95\% , \text{CI}: 0.92, 8.75), \\
\text{OR(NoCoil)} &= e^{-1.610} = 5.00 , (95\% , \text{CI}: 1.47, 23.82).
\end{align*}
}
For dichotomous variables, the odds ratio compares the odds of the outcome (anastomotic leak) in the presence versus the absence of the risk factor. In this model, the riskier category for DM and Smoking is coded as 1 (indicating the presence of Diabetes Mellitus or Smoking, respectively), while for ACSP and NoCoil, the riskier category is coded as 0 (indicating the absence of these treatment methods). Thus, the results indicate that patients with diabetes have approximately 3.16 times higher odds of experiencing an anastomotic leak compared to those without diabetes. Similarly, smokers have approximately 4.18 times higher odds of an anastomotic leak compared to non-smokers. The absence of the ACSP treatment method increases the odds of a leak by approximately 2.59 times, though the confidence interval suggests some uncertainty (0.92, 8.75). Notably, patients without a rectal tube (NoCoil = 0) have approximately 5.00 times higher odds of an anastomotic leak compared to those with the tube, underscoring the protective effect of this intervention. The probability of an anastomotic leak can be calculated using the equation:
{\small
\begin{align*}
P(\text{LEAK} = 1) &= \frac{\exp\left(1.436 - 1.149 \cdot \text{DM}\right.}{1 + \exp\left(1.436 - 1.149 \cdot \text{DM}\right.} \\
&\quad \frac{\left. - 1.429 \cdot \text{Smoking}\right.}{\left. - 1.429 \cdot \text{Smoking}\right.} \\
&\quad \frac{\left. - 0.952 \cdot \text{ACSP} - 1.610 \cdot \text{NoCoil}\right)}{\left. - 0.952 \cdot \text{ACSP} - 1.610 \cdot \text{NoCoil}\right)}
\end{align*}
}
These findings highlight the significant impact of smoking and the non-use of a rectal tube as key risk factors for anastomotic leak, with odds ratios indicating strong associations. The presence of diabetes also substantially increases the risk, while the effect of ACSP, though suggestive of increased risk when absent, warrants further investigation due to the wide confidence interval. Clinically, these results emphasize the importance of smoking cessation prior to surgery and the potential benefit of using a rectal tube to reduce the risk of anastomotic leak. The model provides a robust framework for identifying high-risk patients and tailoring preventive strategies, such as enhanced perioperative monitoring or targeted interventions, to mitigate this serious complication.

\begin{figure}
\centering
\includegraphics[width=1\linewidth]{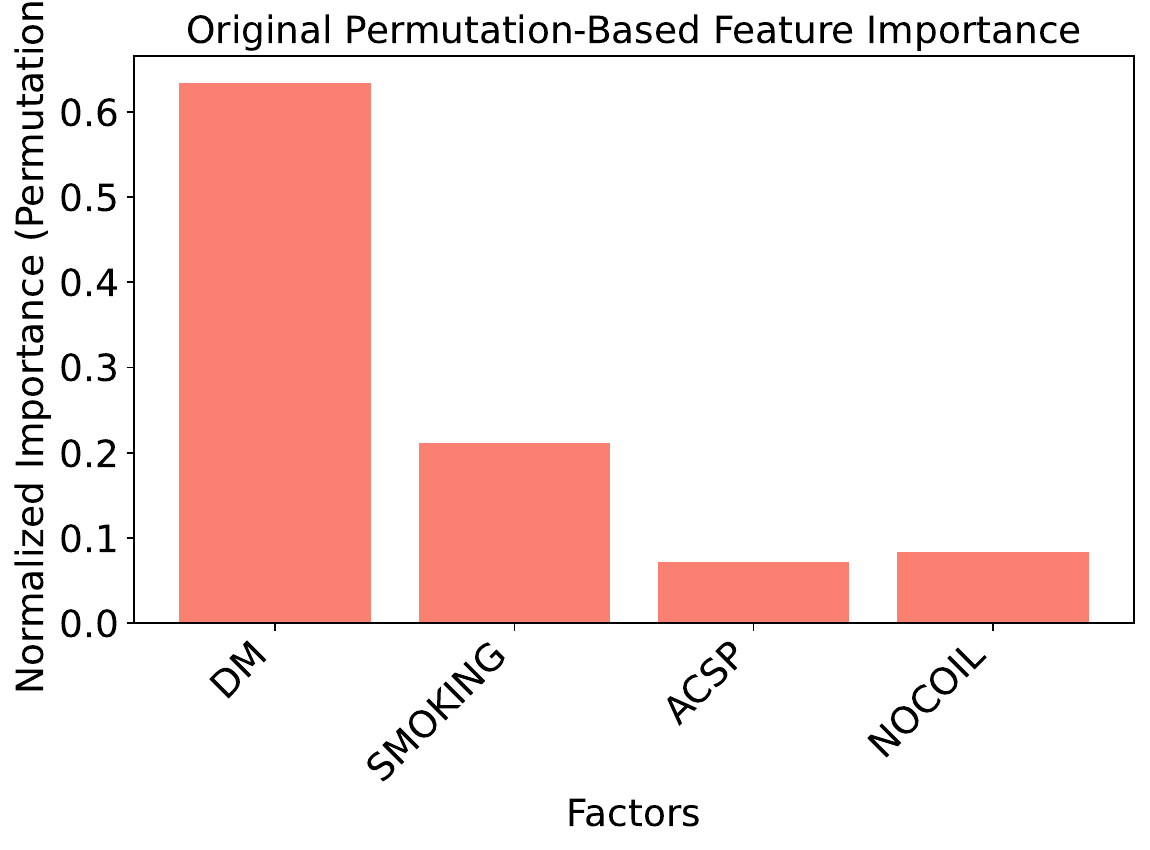}
\caption{Normalized feature importance scores from permutation-based analysis of the quantum neural network, highlighting the dominant influence of diabetes mellitus (DM) followed by smoking, NoCoil, and ACSP in predicting anastomotic leak.}
\label{fig:permutation_importance}
\end{figure}

The quantum neural network (QNN) feature importance analysis reveals distinct patterns when compared to the classical logistic regression approach, highlighting the complementary insights offered by quantum machine learning methodologies. The permutation-based importance analysis, shown in Figure~\ref{fig:permutation_importance}, demonstrates that diabetes mellitus (DM) emerges as the most influential predictor with a normalized importance score of 0.6373, substantially exceeding all other variables. This finding aligns with the logistic regression results, where DM showed a significant odds ratio of 3.16, but the quantum approach amplifies its relative importance within the feature space. Smoking exhibits moderate importance at 0.1890, which contrasts with its prominent role in the classical model where it demonstrated the second-highest odds ratio of 4.18. The non-use of a rectal tube (NoCoil) receives an importance score of 0.1042, while the absence of ACSP treatment shows the lowest contribution at 0.0695.

\begin{figure}
\centering
\includegraphics[width=1\linewidth]{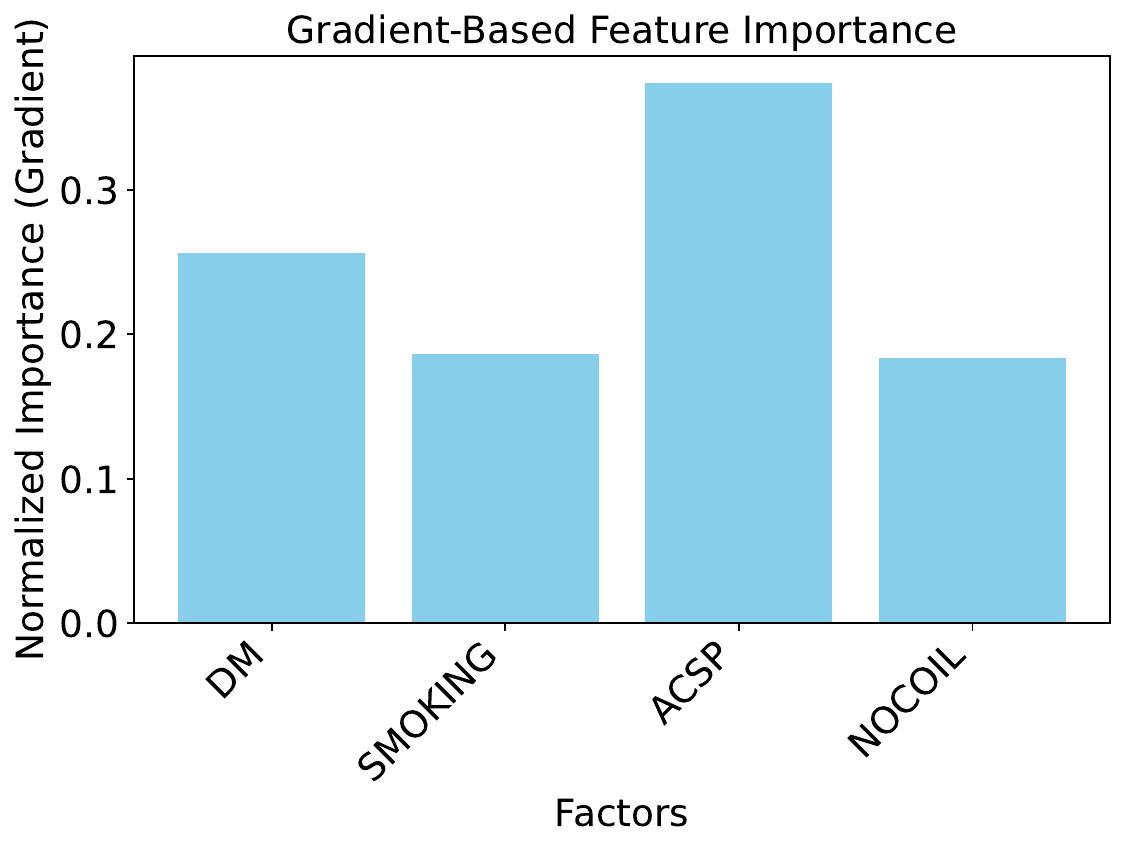}
\caption{Normalized feature importance scores from gradient-based analysis of the quantum neural network, emphasizing the critical role of ACSP, followed by DM, smoking, and NoCoil in predicting anastomotic leak.}
\label{fig:gradient_importance}
\end{figure}

The gradient-based importance analysis, illustrated in Figure~\ref{fig:gradient_importance}, provides a markedly different perspective on feature contributions, suggesting that the QNN captures complex, non-linear interactions that differ from both the permutation approach and classical regression. In this analysis, ACSP emerges as the most critical feature with an importance score of 0.3739, a dramatic shift from its minimal role in both the permutation analysis and logistic regression. Diabetes Mellitus (DM) maintains substantial influence with a score of 0.2564, though reduced compared to the permutation method. Smoking (0.1860) and NoCoil (0.1836) demonstrate nearly equivalent importance scores, indicating balanced contributions to the model's predictive capacity.

A key finding from our interpretability analysis is the striking divergence in feature importance rankings between the classical logistic regression model and the two QNN-based methods. While logistic regression, through its odds ratios, identifies Smoking and NoCoil as the most dominant predictors, the QNN analyses highlight DM (permutation-based) and ACSP (gradient-based). This divergence warrants deeper discussion, as it may point to the fundamentally different ways these models learn from data.

This discrepancy likely arises from the QNN's ability to capture complex, non-linear relationships that are inaccessible to a linear model like logistic regression. The logistic regression model assumes an additive contribution from each feature to the log-odds of the outcome. In contrast, the QNN, through its ZZFeatureMap and entangled circuit structure, can model higher-order interactions and non-linear dependencies. The dramatic rise of ACSP's importance in the gradient-based analysis, for example, could suggest that the model's decision boundary is highly sensitive to this feature, perhaps in specific conjunction with other factors—a non-linear effect the logistic model cannot represent.
The difference between the two quantum interpretation methods themselves is equally revealing and reflects their distinct methodological approaches. The permutation method, which assesses global impact by destroying a feature's predictive signal, suggests that DM is the most critical feature for the overall integrity of the QNN's learned correlations. Disrupting the DM feature may cause the most significant "collapse" of the learned entangled state, indicating that diabetes status serves as a central hub in the network of quantum correlations. This global sensitivity aligns with clinical understanding of diabetes as a systemic condition affecting multiple physiological pathways relevant to surgical healing.

Conversely, the gradient-based method acts as a local probe, measuring sensitivity to small perturbations around specific data points. Its emphasis on ACSP suggests this feature has the largest local influence on the output probability, potentially indicating that the quantum circuit's learned decision boundaries are most responsive to variations in this treatment variable. This could reflect the QNN's discovery of subtle, context-dependent relationships where ACSP's effectiveness varies based on specific combinations of patient characteristics—relationships that would be averaged out in a linear model.
The divergence between the two quantum-based importance measures reflects the fundamental differences in their methodological approaches and the QNN's ability to model complex feature interactions through entanglement. These quantum-derived insights complement the classical logistic regression findings while revealing additional layers of complexity in the risk factor relationships. The QNN's capacity to model non-linear interactions and quantum superposition states enables it to capture subtle interdependencies between risk factors that linear models cannot represent.

However, this divergence also underscores a central challenge in quantum machine learning: the potential to uncover novel, complex, and entangled patterns in clinical data comes with the difficulty of interpreting a model that operates on principles fundamentally different from classical statistics. While our experimental interpretability methods represent a necessary step towards understanding QNN behavior, this analysis highlights that QNNs may not simply learn "stronger" linear associations but may instead identify entirely different, and potentially more nuanced, risk structures within the data.
This analysis demonstrates the potential value of quantum machine learning approaches in clinical risk assessment, where complex biological systems may exhibit non-classical correlations that traditional statistical methods fail to detect. The complementary use of multiple interpretability techniques, including the established logistic regression approach and the experimental QNN methods, provides a more comprehensive understanding of feature contributions, essential for building clinician confidence in quantum-enhanced predictive models for critical surgical outcomes. Future work should focus on developing more sophisticated interpretability frameworks specifically designed for quantum models, potentially incorporating domain knowledge to better understand when quantum advantages represent genuine biological insights versus computational artifacts.


\subsection*{Discussion on Sample Size and Generalizability}
\label{sec:strengthened_limitations}

It is critical, however, to contextualize these promising findings within the primary limitation of this study: the small sample size (N=200, with 28 AL events). While we employed rigorous feature selection from the initial variable pool and extensive cross-validation to mitigate overfitting, the statistical power remains limited. This inherently constrains the generalizability of our findings and demands caution in their interpretation. The impressive performance metrics, particularly the high NPV and accuracy of the QNNs, must be considered preliminary. The small number of positive events (AL occurrences) presents a significant challenge for any machine learning model, especially for a QNN intended to learn complex, non-linear relationships and quantum correlations. Future work must prioritize the validation of these models on larger, independent, and ideally multi-center datasets. Only through such large-scale validation can the true clinical utility and potential advantages of QNNs shown here be definitively established and considered for any clinical application.

\section{Conclusion}

Our central finding, grounded in a robust evaluation of Quantum Neural Networks (QNNs) against hyperparameter-optimized classical models, is a distinct trade-off between classification power and probability calibration. At a fixed, clinically crucial sensitivity of 83\%, our results suggest that QNNs, simulated with realistic hardware noise, have the potential to be powerful classification tools, achieving higher accuracy and Negative Predictive Value on this dataset. Conversely, classical models, particularly the Multi-Layer Perceptron, excelled at probability calibration, making them better suited for nuanced risk stratification.

A significant contribution of this work is demonstrating the direct link between the efficacy of the chosen optimization algorithm in tuning the ansatz and the final predictive power of the QNN. Furthermore, by employing experimental perturbation-based techniques to interpret the QNN's decision-making, we found that it learns complex, non-linear feature interactions, assigning importance differently than traditional logistic regression models and hinting at its ability to uncover subtle, entangled patterns in clinical data.

It is critical, however, to contextualize these promising findings within the primary limitation of this study: the small sample size (N=200, with 28 AL events). While we employed rigorous feature selection from 76 initial variables and extensive cross-validation to mitigate overfitting, the results must be considered preliminary. The potential advantages of QNNs shown here require validation on larger, independent, multi-center datasets before any clinical application can be considered.

This study provides a robust blueprint for such future evaluations, from model benchmarking to interpretability analysis. Ultimately, our work underscores the necessity of a multifaceted evaluation to select the right tool for the right clinical task and paves the way for future research on quantum hardware and hybrid models that may one day combine the strengths of both computational paradigms.

\section*{CRediT authorship contribution statement}
\textbf{Vojtěch Novák}: Software, Investigation, Writing – original draft. \textbf{Ivan Zelinka}: Supervision, Validation, Reviewing. \textbf{Lenka Přibylová:} Conceptualization, Supervision. \textbf{Lubomír Martínek:} Conceptualization, Supervision. \textbf{Vladimír Benčurik:} Conceptualization, Supervision.

\section*{Acknowledgements}
The following grants are acknowledged for the financial support provided for this research: grant of SGS No. SP2025/072 and SP2024/017, VSB-Technical University of Ostrava, Czech Republic.

This work was supported by the Lithuania Research Council (LMTLT) under Agreement P-ITP-24-9.

\section*{Declaration of Competing Interest}

The authors declare no competing interests.

\section*{Data availability}
The datasets generated and analyzed during the current study, after appropriate anonymization to protect patient privacy, are available from the corresponding author upon reasonable request. The source code for simulating noise within the Variational Quantum Classifier (VQC) framework is publicly available at \href{https://github.com/VojtechNovak/qnn\_leak}{https://github.com/VojtechNovak/qnn\_leak}.

\appendix

\section{Individual Runs of Optimizers}
\label{sec:runs}

This appendix section provides the individual convergence curves for ten distinct optimization runs for each optimizer-ansatz combination. These plots visually demonstrate the stability and run-to-run variability of various optimizers when applied to the different quantum circuit ansatzes. For each set of runs, a mean convergence line is also included to facilitate comparison with the aggregated plots presented in Figure~\ref{fig:mean-convergence-all-methods}.

\FloatBarrier

\begin{figure}[htbp]
    \centering
    \includegraphics[width=0.49\textwidth]{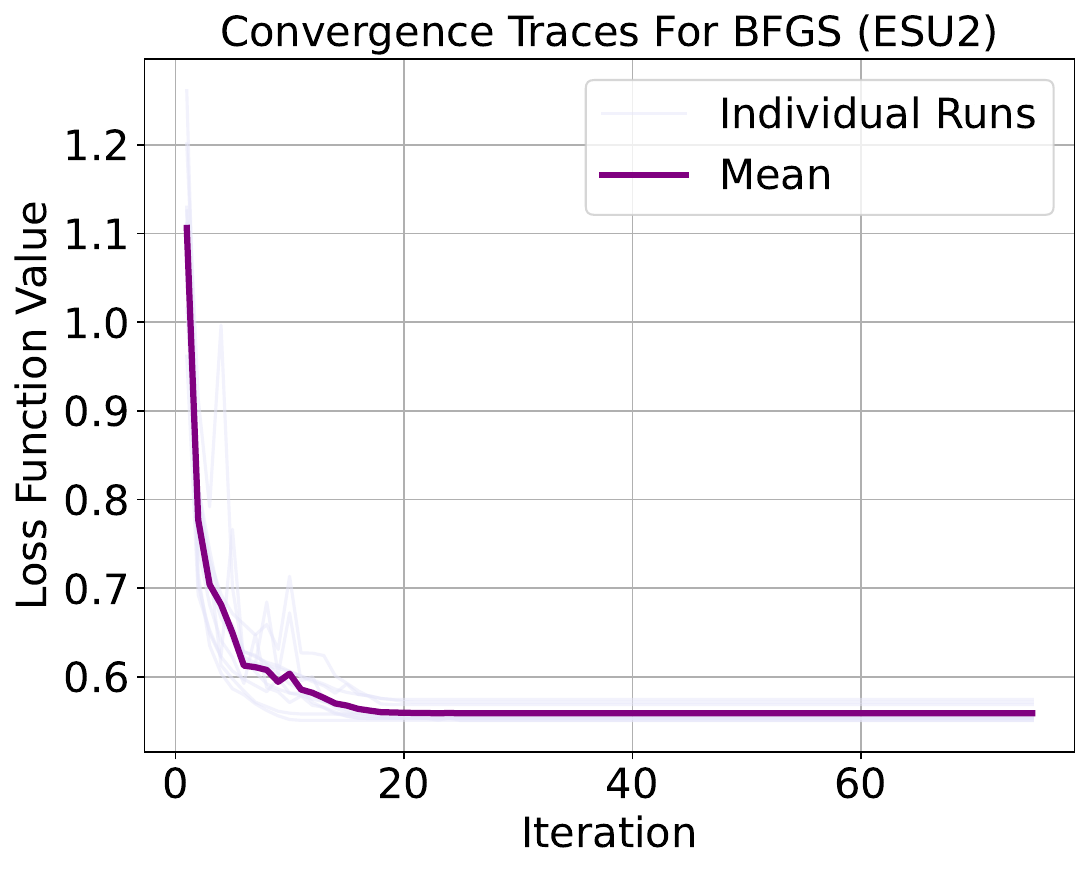}
    \caption{Convergence plot for BFGS with the ESU2 ansatz.} 
    \label{fig:bfgs_esu2}
\end{figure}

\begin{figure}[htbp]
    \centering
    \includegraphics[width=0.49\textwidth]{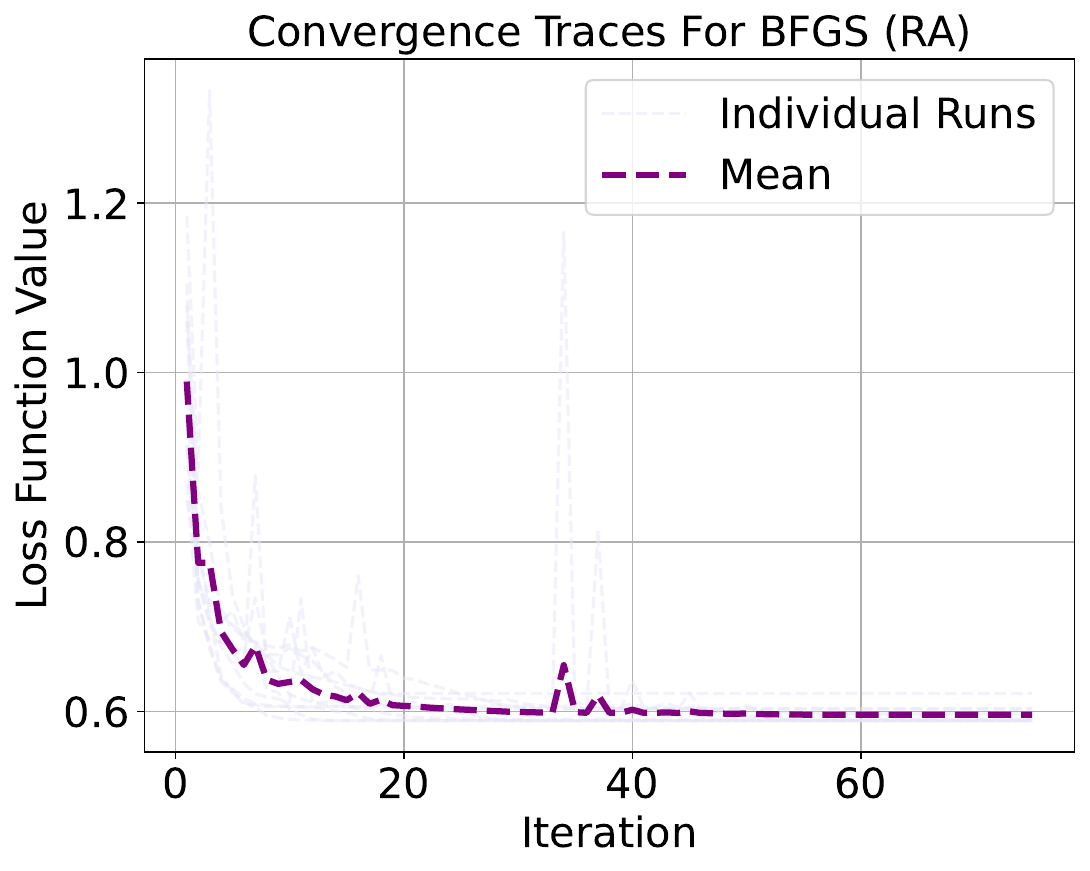}
    \caption{Convergence plot for BFGS with the RA ansatz.} 
    \label{fig:bfgs_ra}
\end{figure}

\begin{figure}[htbp]
    \centering
    \includegraphics[width=0.49\textwidth]{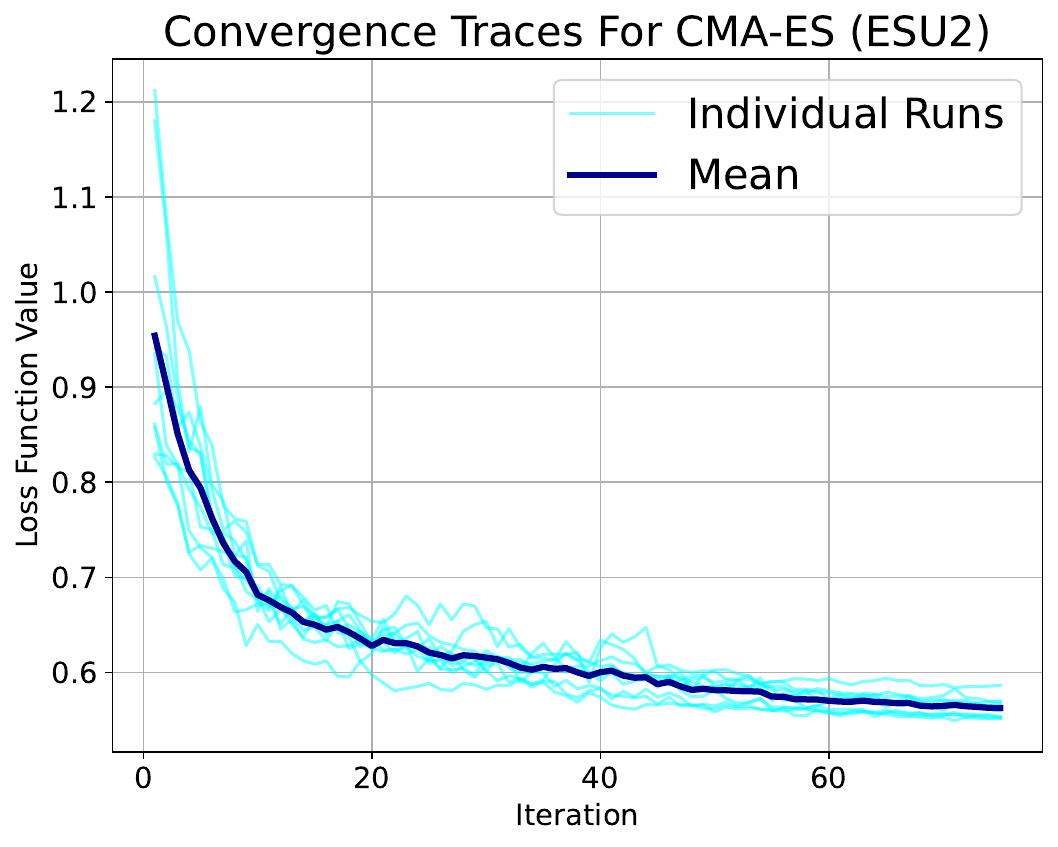}
    \caption{Convergence plot for CMA-ES with the ESU2 ansatz.} 
    \label{fig:cma_es_esu2}
\end{figure}

\begin{figure}[htbp]
    \centering
    \includegraphics[width=0.49\textwidth]{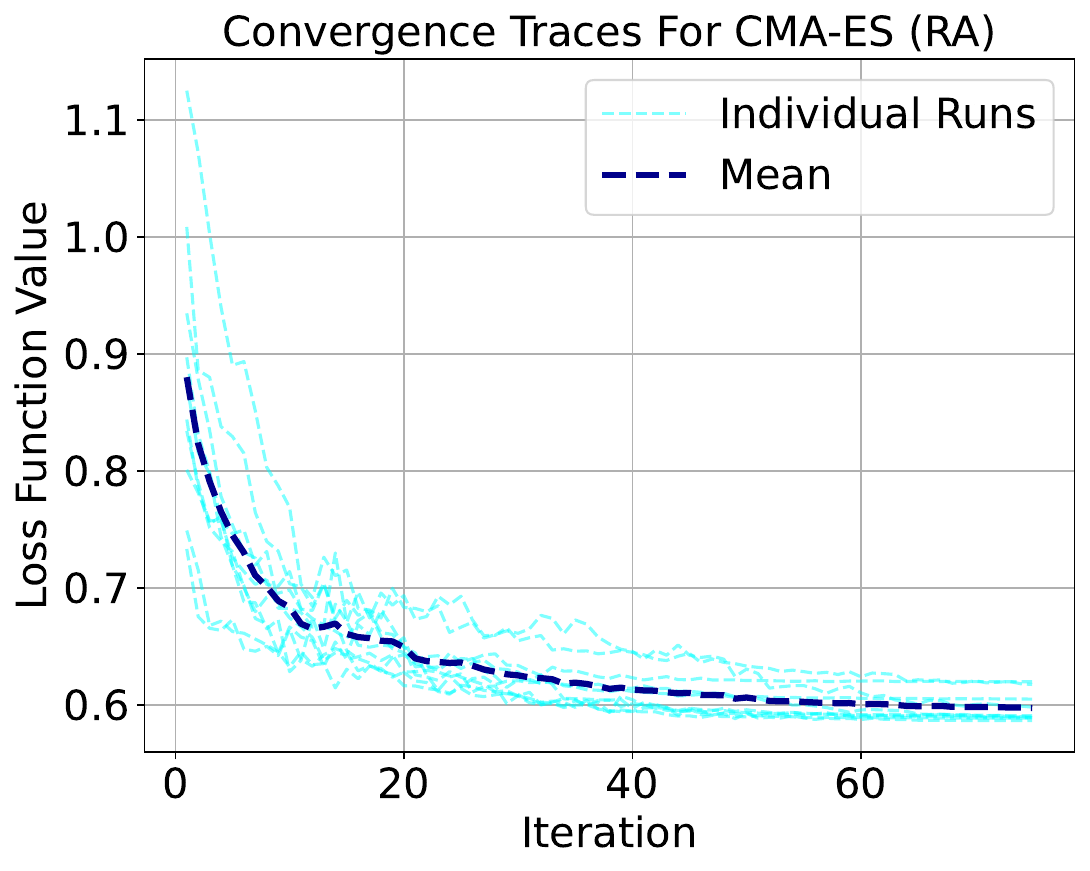}
    \caption{Convergence plot for CMA-ES with the RA ansatz.} 
    \label{fig:cma_es_ra}
\end{figure}

\begin{figure}[htbp]
    \centering
    \includegraphics[width=0.49\textwidth]{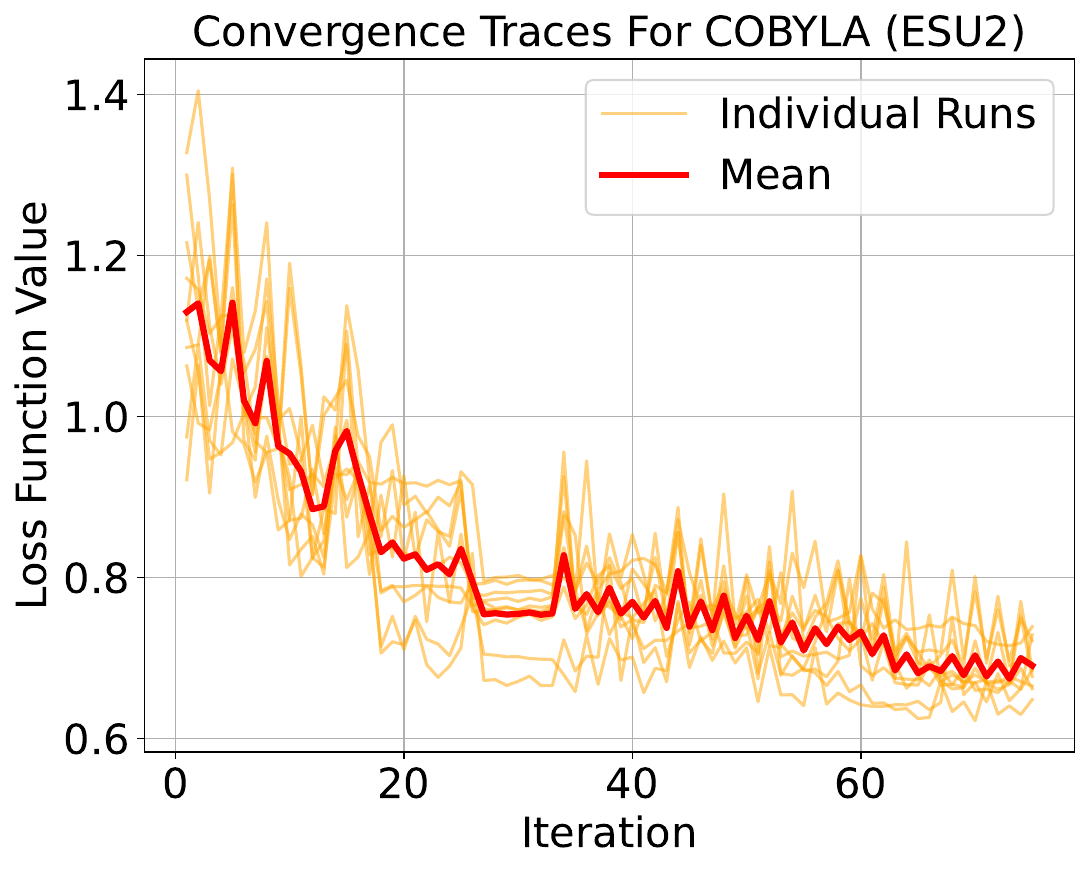}
    \caption{Convergence plot for COBYLA with the ESU2 ansatz.} 
    \label{fig:cobyla_esu2}
\end{figure}

\begin{figure}[htbp]
    \centering
    \includegraphics[width=0.49\textwidth]{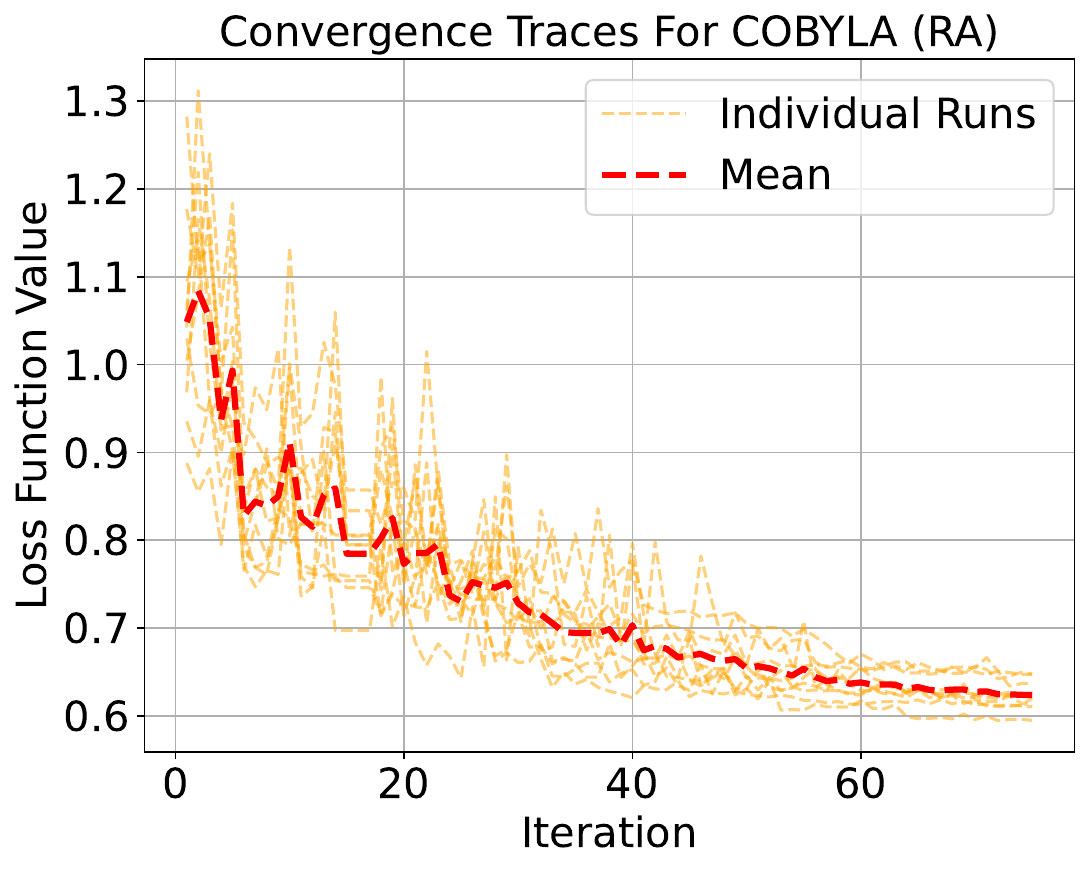}
    \caption{Convergence plot for COBYLA with the RA ansatz.} 
    \label{fig:cobyla_ra}
\end{figure}

\begin{figure}[htbp]
    \centering
    \includegraphics[width=0.49\textwidth]{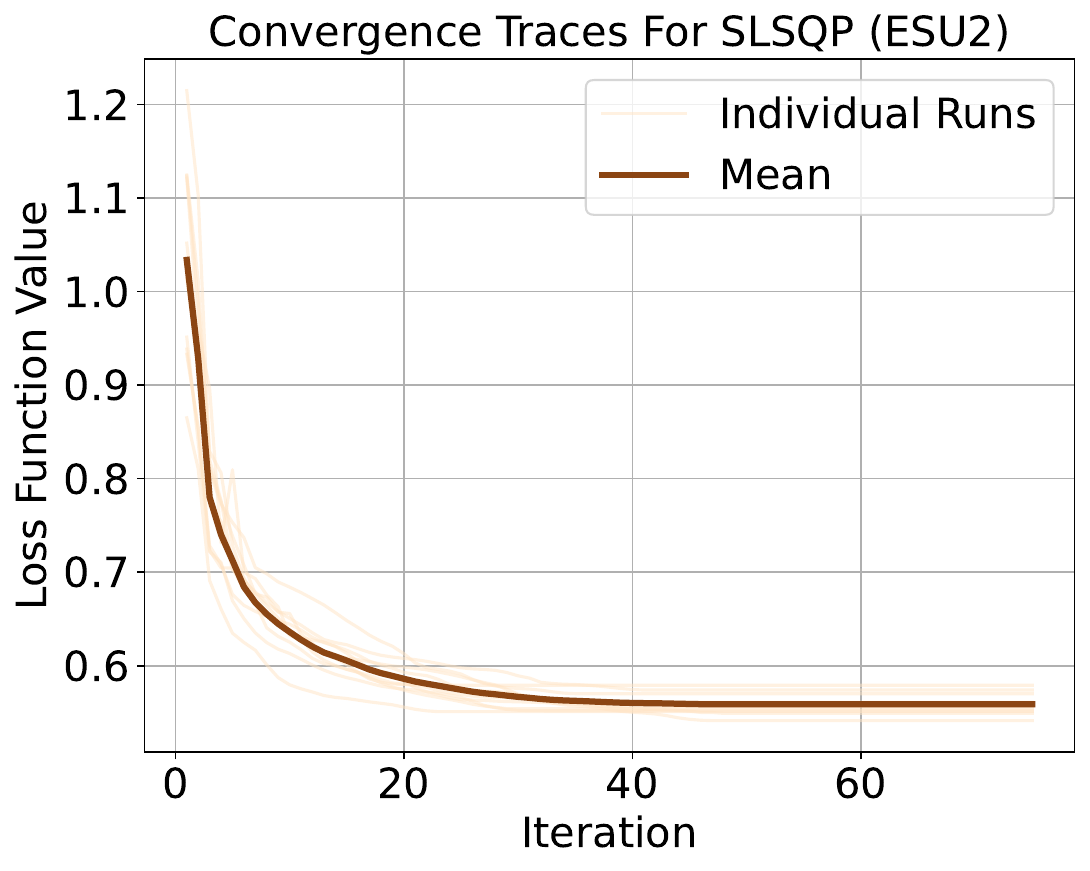}
    \caption{Convergence plot for SLSQP with the ESU2 ansatz.} 
    \label{fig:slsqp_esu2}
\end{figure}

\begin{figure}[htbp]
    \centering
    \includegraphics[width=0.49\textwidth]{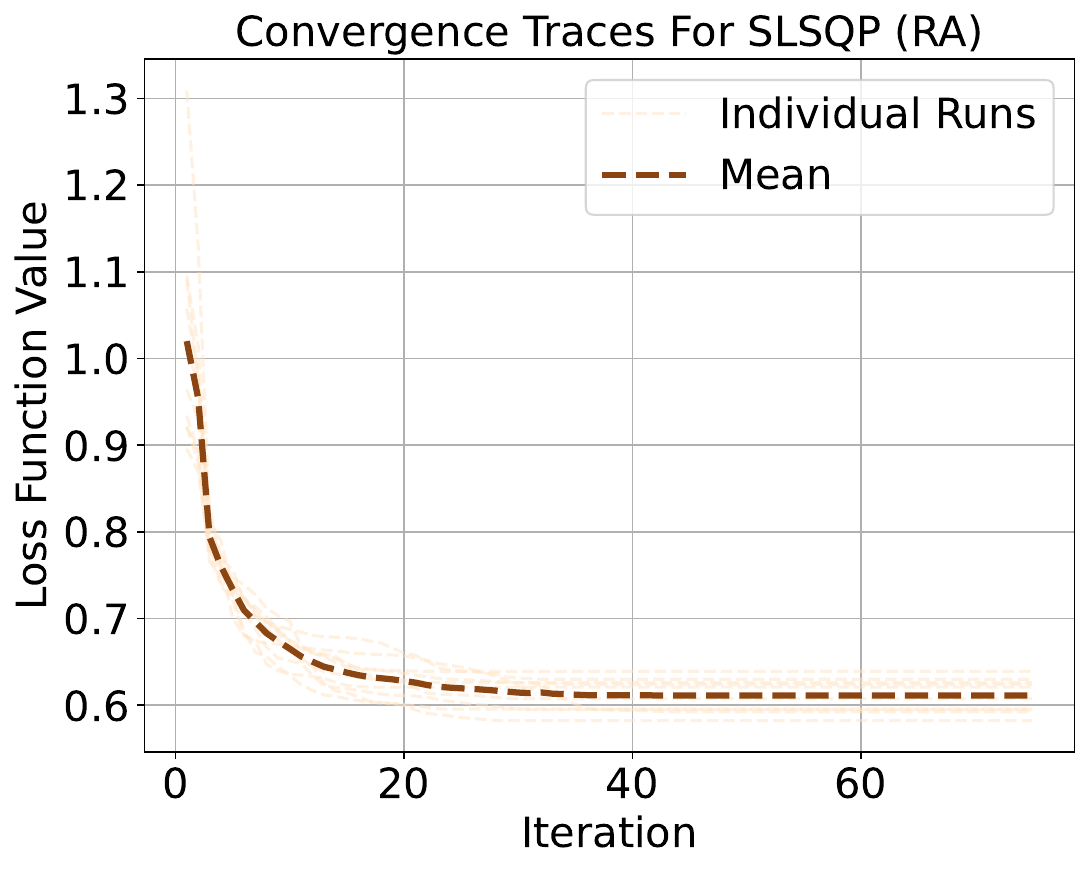}
    \caption{Convergence plot for SLSQP with the RA ansatz.} 
    \label{fig:slsqp_ra}
\end{figure}

\begin{figure}[htbp]
    \centering
    \includegraphics[width=0.49\textwidth]{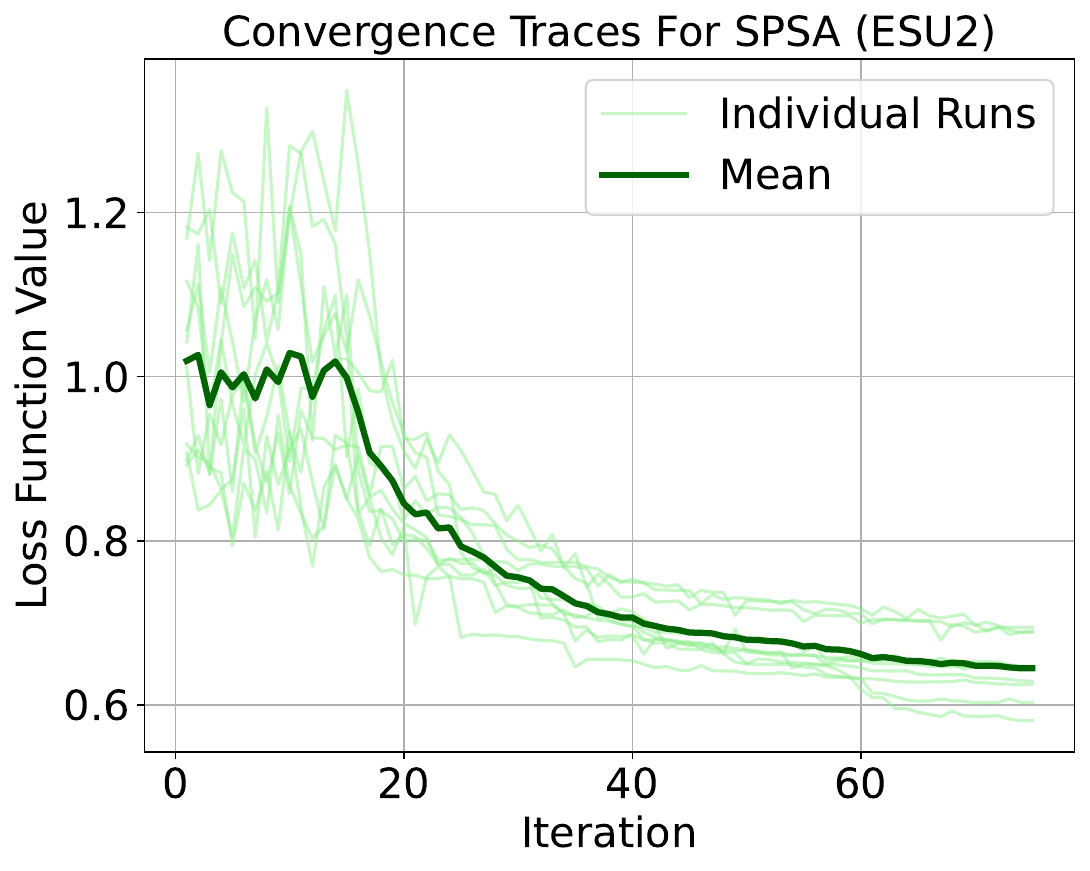}
    \caption{Convergence plot for SPSA with the ESU2 ansatz.} 
    \label{fig:spsa_esu2}
\end{figure}

\begin{figure}[htbp]
    \centering
    \includegraphics[width=0.49\textwidth]{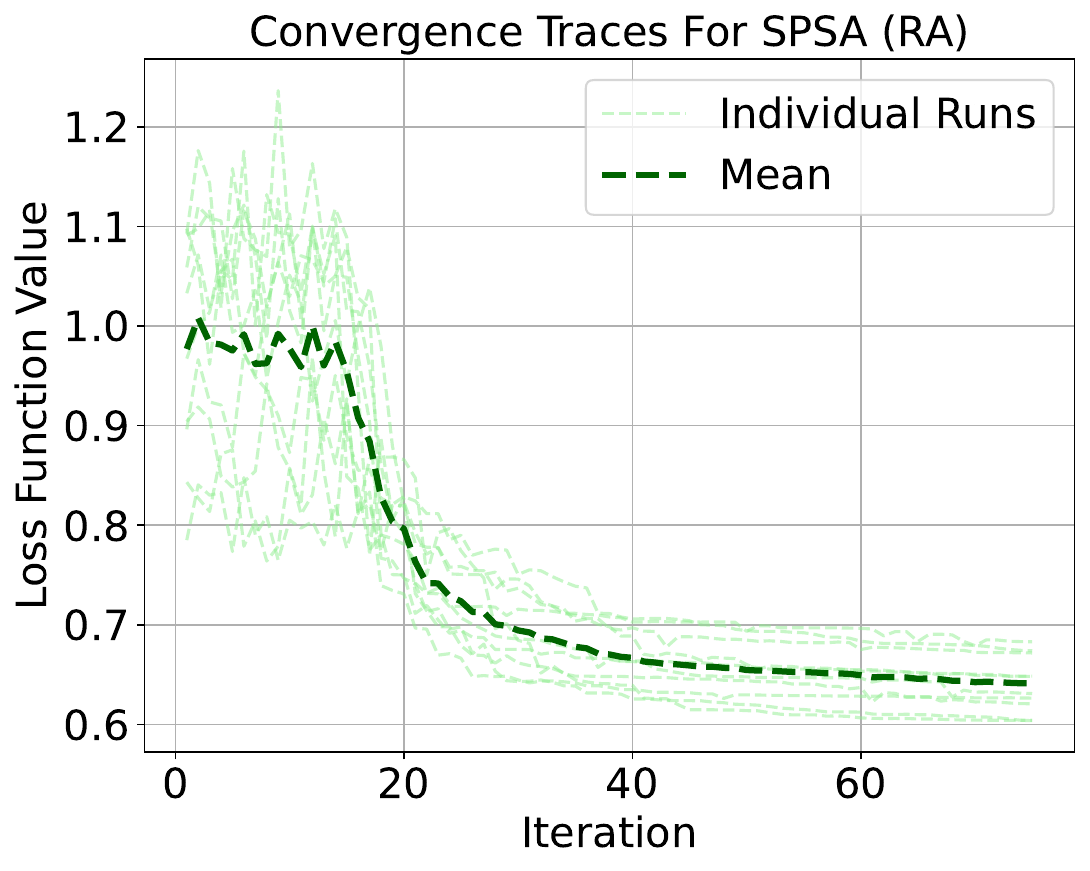}
    \caption{Convergence plot for SPSA with the RA ansatz.} 
    \label{fig:spsa_ra}
\end{figure}

\FloatBarrier

\section{Cross validation}

Our cross-validation approach involved partitioning the dataset into multiple folds. In each iteration, a different fold was held out as the test set, while the remaining folds were used for training. This process was repeated until each fold had served as the test set exactly once.  We quantified model performance in each fold using the Area Under the ROC Curve (AUC), a metric that reflects the overall discriminatory power of the model across various thresholds.

\begin{figure*}[htpb]
    \centering
    \includegraphics[width=1\linewidth]{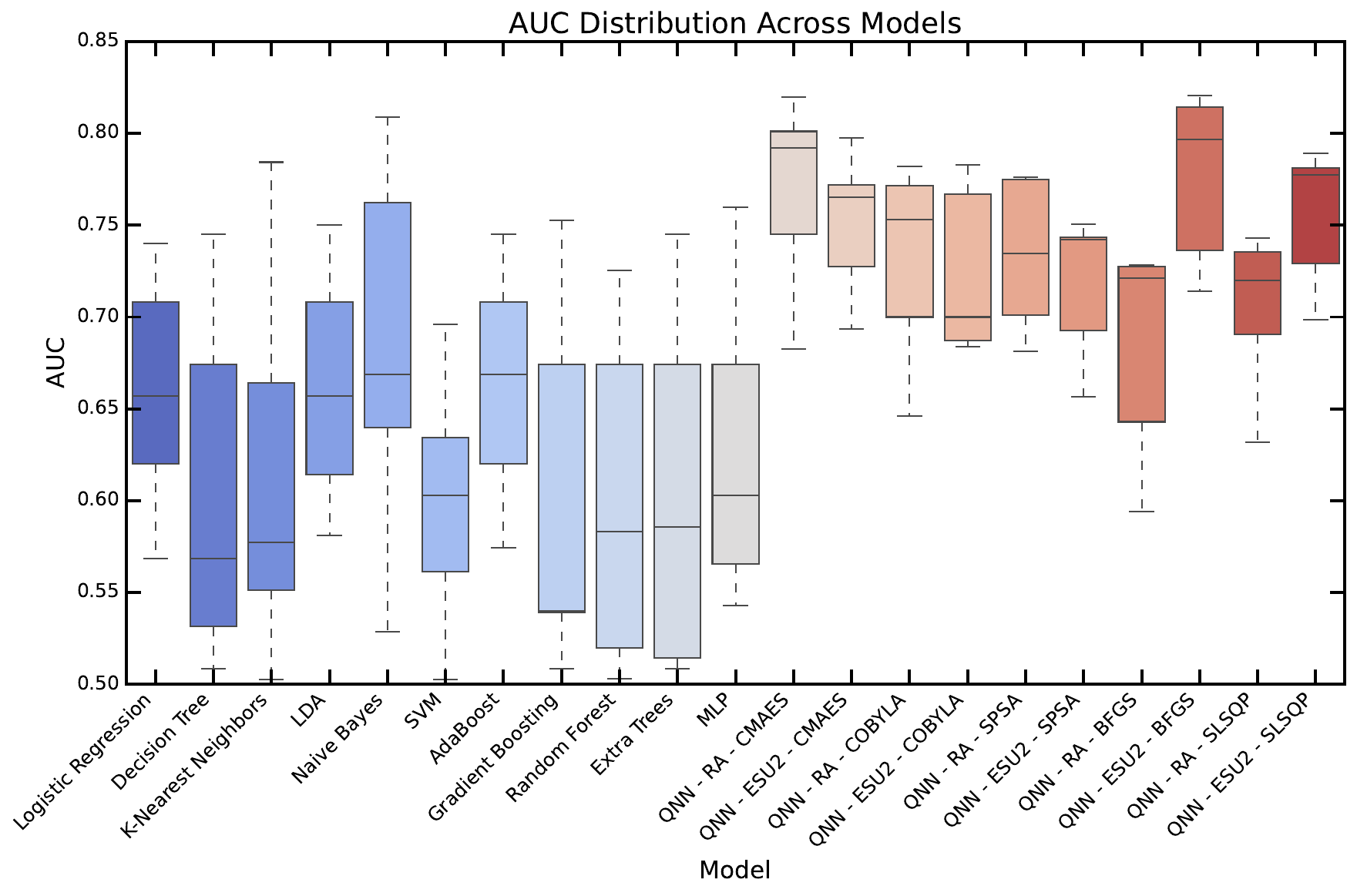}
    \caption{Boxplot of classification model performance across cross-validation folds. The distribution of AUC is shown for each model, illustrating variability and robustness to different training data subsets.}
    \label{fig:box}
\end{figure*}

\begin{figure*}[htpb]
    \centering
    \includegraphics[width=1\linewidth]{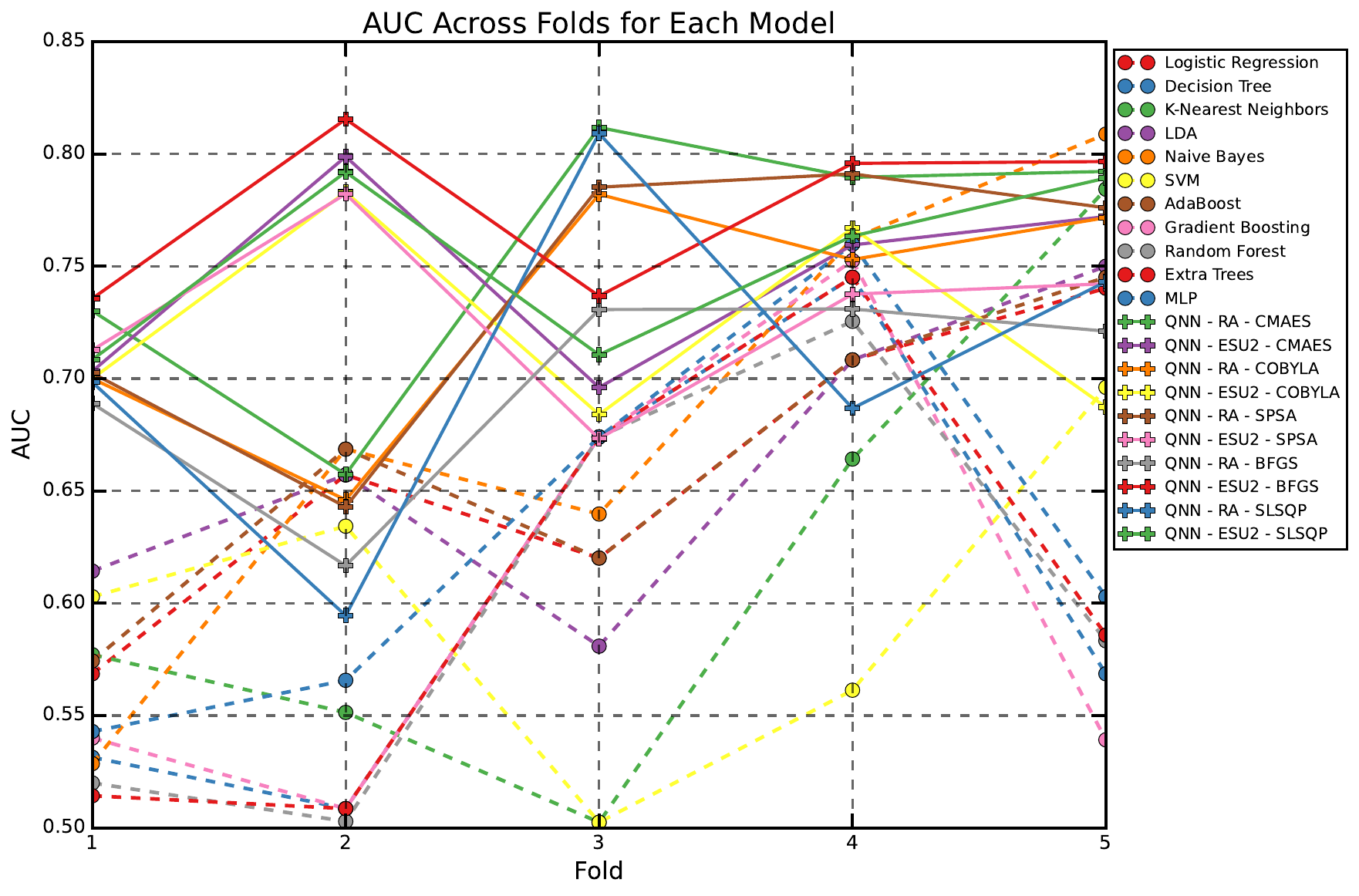}
    \caption{Line plot showing classification model performance across cross-validation folds. Each point represents a fold, and the trend lines indicate how models perform on different training subsets.}
    \label{fig:folds}
\end{figure*}

To visualize the cross-validation results, we present both a boxplot (Figure~\ref{fig:box}) and a line plot (Figure~\ref{fig:folds}). The boxplot (Figure~\ref{fig:box}) summarizes the distribution of AUC scores across all folds for each model. This visualization effectively illustrates the central tendency and variability of each model's performance across different training subsets. A tighter box in the boxplot indicates more consistent performance and greater robustness.

The line plot (Figure~\ref{fig:folds}) provides a complementary perspective by displaying the AUC score for each fold individually. This allows for the visualization of performance fluctuations across different training datasets and helps in identifying any potential instability or high variance in model performance.

The cross-validation results, as depicted in Figures~\ref{fig:box} and \ref{fig:folds}, demonstrate the consistent performance of the Quantum Neural Network (QNN) with the RealAmplitudes ansatz. The boxplot (Figure~\ref{fig:box}) clearly shows that the QNN consistently achieves higher median AUC scores compared to classical models such as Logistic Regression and Gradient Boosting, suggesting a superior average performance across different data partitions.  Furthermore, the line plot (Figure~\ref{fig:folds}) indicates that while some performance variation exists across folds for all models (attributable to the inherent variability in training data subsets), the QNN maintains a relatively stable and high-performing trajectory across these variations. This robustness across different cross-validation folds reinforces the generalizability of the QNN and its potential for reliable predictive performance on unseen patient data, setting it apart from classical models that exhibit either lower average performance or greater performance variability, as observed in the box and line plots.

\section{Future work}

This study provides a foundational step in exploring the application of Quantum Neural Networks (QNNs) for post-operative complication prediction using classical simulations. To rigorously validate the potential clinical utility of QNNs and to fully leverage the capabilities of quantum computation, future research will focus on several key directions:

\begin{itemize}
\item \textbf{Quantum Hardware Implementation:} The immediate next step is to transition from classical simulations to experiments on actual quantum computing hardware. This is crucial to assess the real-world performance of QNNs and to explore whether they can maintain or enhance their predictive power when executed on quantum devices. These experiments will involve deploying the developed QNN models on available quantum platforms and systematically evaluating their performance in practical quantum computing environments.
\item \textbf{Impact of Quantum Noise and Hardware Constraints:} While this study incorporated simulated noise models, including depolarizing errors and sampling noise, future work will extend these investigations to real quantum hardware. Experiments on quantum devices will inherently introduce additional noise sources, fidelity errors, limited coherence times, and connectivity constraints. Future studies will explicitly characterize how these hardware limitations affect QNN performance, with a focus on developing noise-resilient quantum circuits and error mitigation techniques tailored to biomedical prediction tasks.
\item \textbf{Exploration of Quantum Advantage:} As quantum hardware matures, future studies will aim to identify specific scenarios or model configurations where QNNs can demonstrably outperform state-of-the-art classical methods, establishing a tangible quantum advantage in biomedical classification. This will require careful benchmarking against optimized classical algorithms and architectures, alongside rigorous statistical validation to confirm any observed quantum speedup or accuracy improvements.

\item \textbf{Advanced QNN Architectures and Training Techniques:} Future research will explore more advanced QNN architectures, including alternative ansatzes (e.g., hardware-efficient or problem-inspired ansatzes) and feature maps (e.g., higher-order or data-adaptive feature maps), to further optimize performance for clinical prediction tasks. Additionally, hybrid quantum-classical training methodologies and techniques to improve the calibration of QNN probability estimates will be investigated, potentially bridging the gap between QNN classification strength and MLP calibration reliability.
\end{itemize}
These future directions are essential to move beyond classical simulations and towards a realistic assessment of QNNs' potential for transforming clinical diagnostics and predictive medicine, ultimately paving the way for practical quantum-enhanced tools in healthcare.

\section{Model Descriptions and Hyperparameter Optimization}
\label{sec:hyperparams}
To evaluate a range of classification algorithms, ten machine learning models were employed: Logistic Regression, Decision Tree, K-Nearest Neighbors (KNN), Linear Discriminant Analysis (LDA), Naive Bayes, Support Vector Machine (SVM) with a radial basis function (RBF) kernel, AdaBoost, Gradient Boosting, Random Forest, and Extra Trees. Each model was optimized using a grid search over a predefined set of hyperparameters to maximize the Area Under the Receiver Operating Characteristic Curve (AUC) score. The optimization was performed using 3-fold cross-validation within each of the five outer cross-validation folds of the dataset. Below, we describe each model, its possible hyperparameters, the optimization process, and the optimal hyperparameters found for each fold.

\subsection{Models and Hyperparameters}

\begin{itemize}
    \item \textbf{Logistic Regression}: A linear model that predicts the probability of a binary outcome using a logistic function. Regularization is applied to prevent overfitting.
    \begin{itemize}
        \item \textit{Possible Hyperparameters}: Regularization strength $C \in \{0.01, 0.1, 1, 10\}$, penalty type $\in \{\text{L1}, \text{L2}\}$.
    \end{itemize}

    \item \textbf{K-Nearest Neighbors (KNN)}: A non-parametric model that classifies instances based on the majority class of their $k$ nearest neighbors.
    \begin{itemize}
        \item \textit{Possible Hyperparameters}: Number of neighbors $k \in \{3, 5, 7, 9\}$, weighting scheme $\in \{\text{uniform}, \text{distance}\}$.
    \end{itemize}

    \item \textbf{Linear Discriminant Analysis (LDA)}: A linear model that finds a linear combination of features to separate classes.
    \begin{itemize}
        \item \textit{Possible Hyperparameters}: Solver $\in \{\text{svd}, \text{lsqr}\}$.
    \end{itemize}

    \item \textbf{Naive Bayes}: A probabilistic model based on Bayes' theorem, assuming feature independence.
    \begin{itemize}
        \item \textit{Possible Hyperparameters}: None (default Gaussian Naive Bayes parameters were used).
    \end{itemize}

    \item \textbf{Support Vector Machine (SVM)}: A model that finds the optimal hyperplane to separate classes, using an RBF kernel for non-linear boundaries.
    \begin{itemize}
        \item \textit{Possible Hyperparameters}: Regularization parameter $C \in \{0.1, 1, 10\}$, kernel scale $\gamma \in \{0.01, 0.1, 1\}$, kernel type $\in \{\text{rbf}\}$.
    \end{itemize}

    \item \textbf{AdaBoost}: An ensemble method that combines weak learners (decision stumps) with adaptive weighting.
    \begin{itemize}
        \item \textit{Possible Hyperparameters}: Number of estimators $\in \{50, 100, 200\}$, learning rate $\in \{0.01, 0.1, 1\}$. The SAMME \cite{zhu2009multi} algorithm was used to avoid deprecation warnings.
    \end{itemize}

    \item \textbf{Gradient Boosting}: An ensemble method that builds trees sequentially, minimizing a loss function.
    \begin{itemize}
        \item \textit{Possible Hyperparameters}: Number of estimators $\in \{100, 200\}$, learning rate $\in \{0.01, 0.1\}$, maximum tree depth $\in \{3, 5\}$.
    \end{itemize}
\end{itemize}

The dataset was split into five folds for outer cross-validation to ensure robust evaluation. For each fold, the training data was used to perform a 3-fold inner cross-validation grid search to identify the optimal hyperparameters for each classifier, maximizing the AUC score. The target variable, originally encoded as $\{-1, 1\}$, was transformed to $\{0, 1\}$ using a label encoder to ensure compatibility with scikit-learn's metrics and models. Features were standardized using a standard scaler to have zero mean and unit variance. After optimization, the best model for each classifier was evaluated on the test set of each fold. The AUC score was checked, and if it was below 0.5, the predicted probabilities were flipped (i.e., $p \to 1-p$) to correct for potential label misalignment, and the AUC was recomputed. The optimal threshold for classification was determined by maximizing the F1-score on the test set.

\subsection{Optimal Hyperparameters}

The optimal hyperparameters for each classifier across the five folds are summarized in Table~\ref{tab:optimal_params}. These parameters were selected based on the highest AUC score during the inner 3-fold cross-validation.

\begin{table*}[htpb]
\centering
\caption{Optimal Hyperparameters for Each Classifier Across Five Folds}
\label{tab:optimal_params}
\resizebox{\textwidth}{!}{
\scriptsize
\begin{tabular}{|l|l|}
\hline
\textbf{Classifier} & \textbf{Optimal Hyperparameters per Fold} \\
\hline
Logistic Regression &
\begin{tabular}[c]{@{}l@{}}
Fold 1: $\{C=1, \text{penalty}=\text{l1}\}$ \\
Fold 2: $\{C=1, \text{penalty}=\text{l1}\}$ \\
Fold 3: $\{C=1, \text{penalty}=\text{l1}\}$ \\
Fold 4: $\{C=1, \text{penalty}=\text{l1}\}$ \\
Fold 5: $\{C=1, \text{penalty}=\text{l2}\}$
\end{tabular} \\
\hline
K-Nearest Neighbors &
\begin{tabular}[c]{@{}l@{}}
Fold 1: $\{\text{n\_neighbors}=3, \text{weights}=\text{distance}\}$ \\
Fold 2: $\{\text{n\_neighbors}=5, \text{weights}=\text{distance}\}$ \\
Fold 3: $\{\text{n\_neighbors}=9, \text{weights}=\text{uniform}\}$ \\
Fold 4: $\{\text{n\_neighbors}=5, \text{weights}=\text{distance}\}$ \\
Fold 5: $\{\text{n\_neighbors}=5, \text{weights}=\text{distance}\}$
\end{tabular} \\
\hline
LDA &
\begin{tabular}[c]{@{}l@{}}
Fold 1: $\{\text{solver}=\text{svd}\}$ \\
Fold 2: $\{\text{solver}=\text{svd}\}$ \\
Fold 3: $\{\text{solver}=\text{svd}\}$ \\
Fold 4: $\{\text{solver}=\text{svd}\}$ \\
Fold 5: $\{\text{solver}=\text{svd}\}$
\end{tabular} \\
\hline
Naive Bayes &
\begin{tabular}[c]{@{}l@{}}
Fold 1: $\{\}$ \\
Fold 2: $\{\}$ \\
Fold 3: $\{\}$ \\
Fold 4: $\{\}$ \\
Fold 5: $\{\}$
\end{tabular} \\
\hline
SVM &
\begin{tabular}[c]{@{}l@{}}
Fold 1: $\{C=10, \gamma=0.1, \text{kernel}=\text{rbf}\}$ \\
Fold 2: $\{C=0.1, \gamma=0.01, \text{kernel}=\text{rbf}\}$ \\
Fold 3: $\{C=0.1, \gamma=0.1, \text{kernel}=\text{rbf}\}$ \\
Fold 4: $\{C=1, \gamma=1, \text{kernel}=\text{rbf}\}$ \\
Fold 5: $\{C=10, \gamma=1, \text{kernel}=\text{rbf}\}$
\end{tabular} \\
\hline
AdaBoost &
\begin{tabular}[c]{@{}l@{}}
Fold 1: $\{\text{learning\_rate}=1, \text{n\_estimators}=100\}$ \\
Fold 2: $\{\text{learning\_rate}=0.1, \text{n\_estimators}=50\}$ \\
Fold 3: $\{\text{learning\_rate}=1, \text{n\_estimators}=50\}$ \\
Fold 4: $\{\text{learning\_rate}=0.01, \text{n\_estimators}=100\}$ \\
Fold 5: $\{\text{learning\_rate}=0.1, \text{n\_estimators}=200\}$
\end{tabular} \\
\hline
Gradient Boosting &
\begin{tabular}[c]{@{}l@{}}
Fold 1: $\{\text{learning\_rate}=0.1, \text{max\_depth}=3, \text{n\_estimators}=100\}$ \\
Fold 2: $\{\text{learning\_rate}=0.01, \text{max\_depth}=5, \text{n\_estimators}=100\}$ \\
Fold 3: $\{\text{learning\_rate}=0.01, \text{max\_depth}=3, \text{n\_estimators}=100\}$ \\
Fold 4: $\{\text{learning\_rate}=0.01, \text{max\_depth}=3, \text{n\_estimators}=100\}$ \\
Fold 5: $\{\text{learning\_rate}=0.01, \text{max\_depth}=5, \text{n\_estimators}=100\}$
\end{tabular} \\
\hline
\end{tabular}
}
\end{table*}

\section{Quantum Measurement Operators}
\label{sec:measurement_operators}

The quantum neural network classification is performed through projective measurements on the quantum circuit output. This section describes the measurement operators employed in our implementation using the Qiskit Sampler primitive.

Our quantum classifier operates on a 4-qubit system, where measurements are performed in the computational basis $\{|0\rangle, |1\rangle\}^{\otimes 4}$. The measurement operators are defined as the set of projection operators:

\begin{equation*}
\hat{M}_s = |s\rangle\langle s|
\end{equation*}

where $s \in \{0, 1\}^4$ represents all possible 4-bit computational basis states, and $|s\rangle = |s_0 s_1 s_2 s_3\rangle$ with $s_i \in \{0, 1\}$.

The complete set of 16 measurement operators for our 4-qubit system is:

\begin{align*}
\hat{M}_{0000} &= |0000\rangle\langle 0000| \\
\hat{M}_{0001} &= |0001\rangle\langle 0001| \\
\hat{M}_{0010} &= |0010\rangle\langle 0010| \\
&\vdots \\
\hat{M}_{1111} &= |1111\rangle\langle 1111|
\end{align*}

These operators satisfy the completeness relation:
\begin{equation*}
\sum_{s=0}^{15} \hat{M}_s = \hat{I}^{\otimes 4}
\end{equation*}

where $\hat{I}$ is the single-qubit identity operator.

For a quantum state $|\psi(\boldsymbol{\theta}, \mathbf{x})\rangle$ prepared by the parameterized quantum circuit with parameters $\boldsymbol{\theta}$ and input features $\mathbf{x}$, the measurement probabilities are:

\begin{equation*}
p_s = \langle\psi(\boldsymbol{\theta}, \mathbf{x})|\hat{M}_s|\psi(\boldsymbol{\theta}, \mathbf{x})\rangle = |\langle s|\psi(\boldsymbol{\theta}, \mathbf{x})\rangle|^2
\end{equation*}

The Qiskit Sampler primitive performs shot-based sampling to estimate these probabilities through repeated measurements. The classification probabilities are computed as:

\begin{align*}
P(\text{class 0}) &= \sum_{s: \text{parity}(s) = 0} p_s \\
P(\text{class 1}) &= \sum_{s: \text{parity}(s) = 1} p_s
\end{align*}

where $\text{parity}(s) = s_0 \oplus s_1 \oplus s_2 \oplus s_3$ is the parity of the bit string $s$.

In our implementation using the Variational Quantum Classifier (VQC) from Qiskit Machine Learning:

\begin{itemize}
\item The \texttt{Sampler} primitive handles the measurement process automatically
\item Default shot count is used for probability estimation (typically 1024 shots)
\item The \texttt{predict\_proba} method returns the two-class probabilities $[P(\text{class 0}), P(\text{class 1})]$
\item No explicit measurement operators need to be defined in code as they are handled internally by the VQC framework
\end{itemize}

The shot-based sampling introduces statistical uncertainty in probability estimates. For a probability $p$ estimated with $N$ shots, the standard error is approximately:

\begin{equation*}
\sigma_p \approx \sqrt{\frac{p(1-p)}{N}}
\end{equation*}

This measurement uncertainty contributes to the overall variance in the quantum neural network performance and is accounted for in our cross-validation methodology.

\section*{ROC curves}

Understanding ROC Curves

The Receiver Operating Characteristic (ROC) curve illustrates a classification model's diagnostic ability by plotting the True Positive Rate (TPR) against the False Positive Rate (FPR) at various threshold settings. The Area Under the Curve (AUC) provides a single value summarizing the model's overall discriminative performance across all possible thresholds; an AUC of 1.0 is perfect, while 0.5 is no better than random. Higher AUC values indicate superior model performance.

\begin{figure}[htpb]
    \centering
    \includegraphics[width=0.48\textwidth]{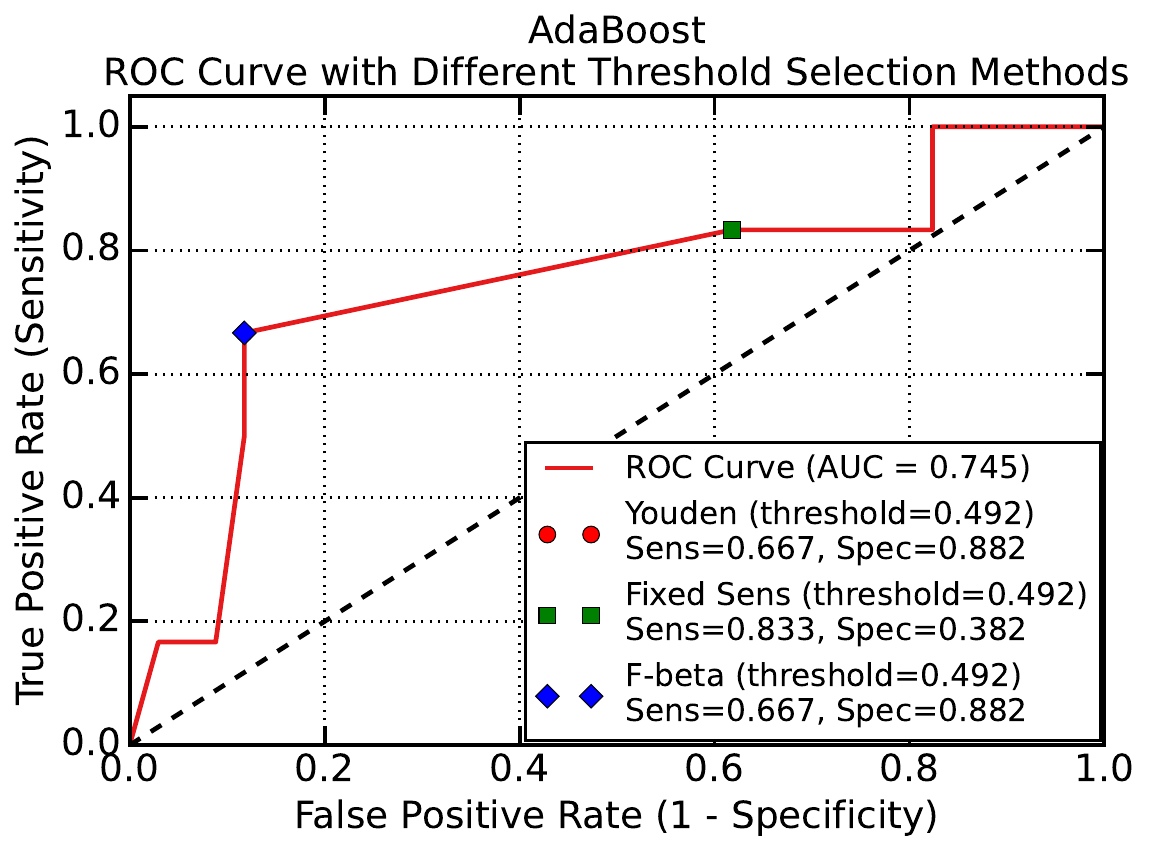}
    \caption{ROC curve for the AdaBoost model.}
    \label{fig:roc_adaboost}
\end{figure}

\begin{figure}[htpb]
    \centering
    \includegraphics[width=0.48\textwidth]{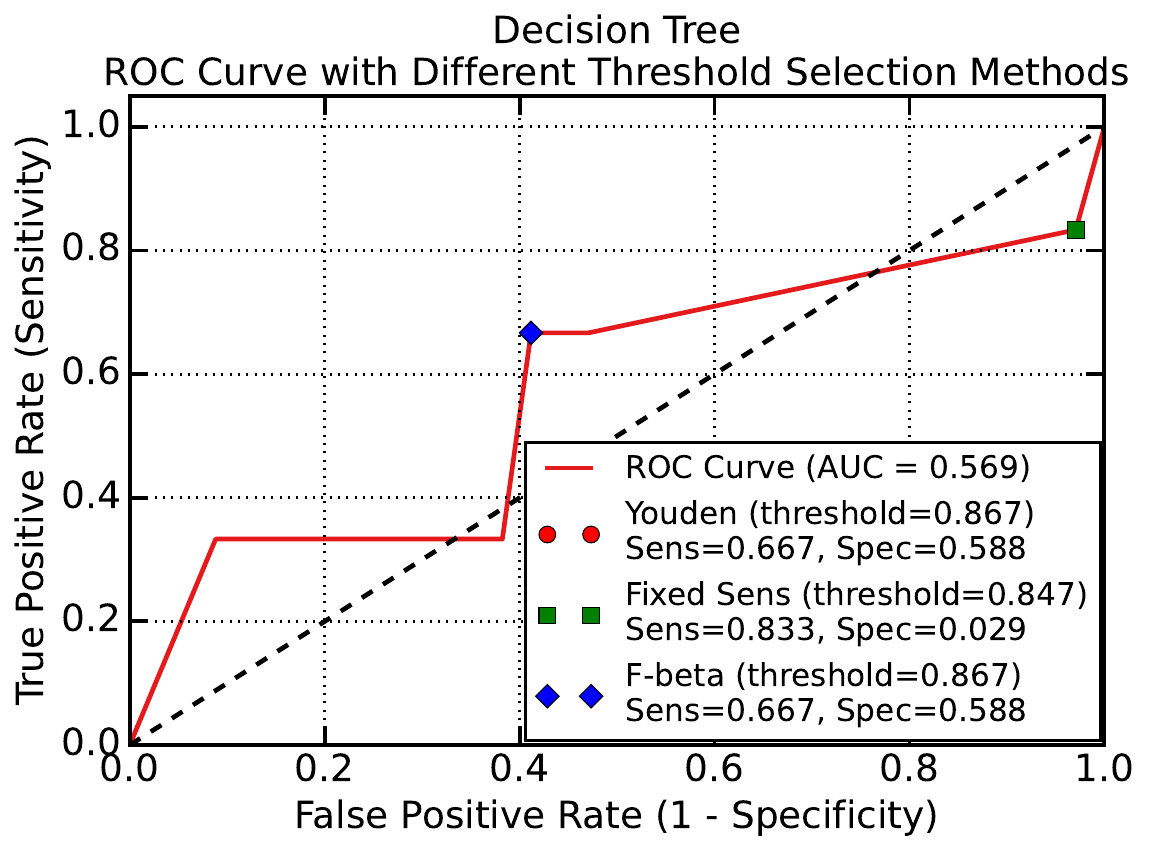}
    \caption{ROC curve for the Decision Tree model.}
    \label{fig:roc_decision_tree}
\end{figure}

\begin{figure}[htpb]
    \centering
    \includegraphics[width=0.48\textwidth]{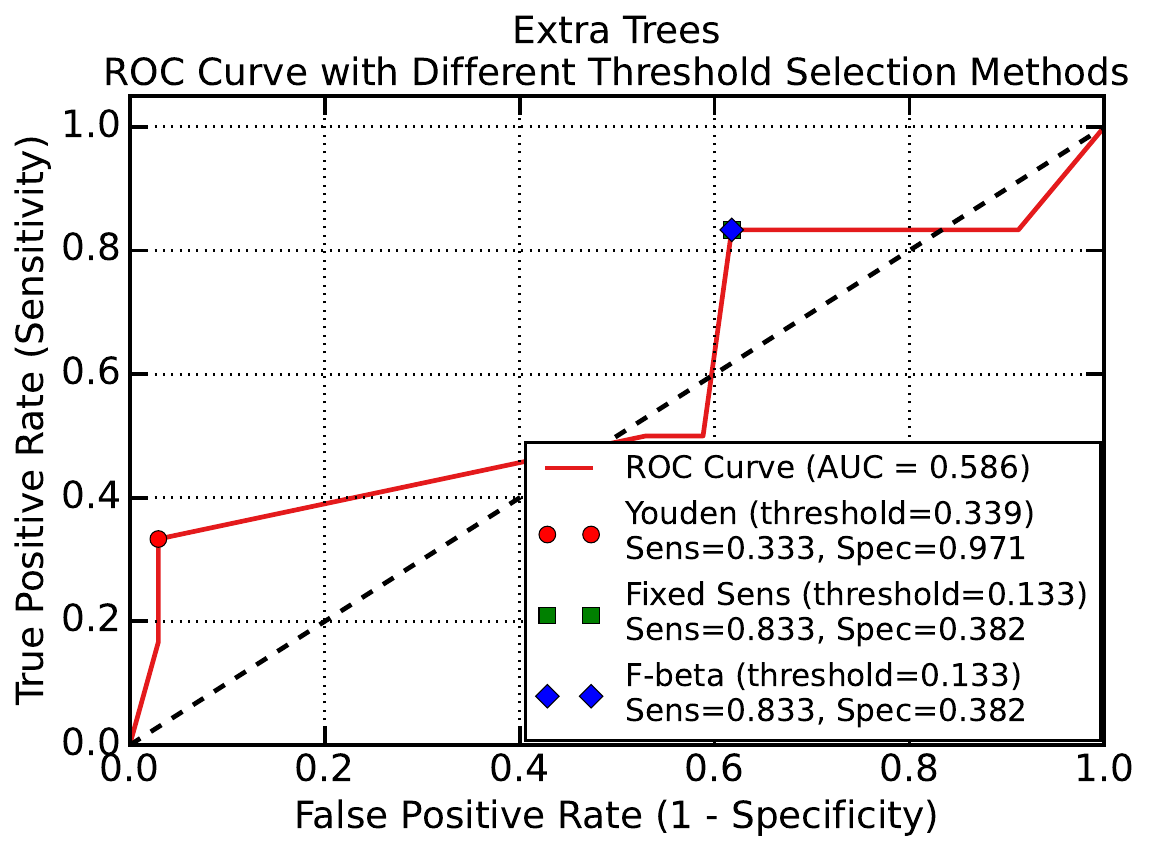}
    \caption{ROC curve for the Extra Trees model.}
    \label{fig:roc_extra_trees}
\end{figure}

\begin{figure}[htpb]
    \centering
    \includegraphics[width=0.48\textwidth]{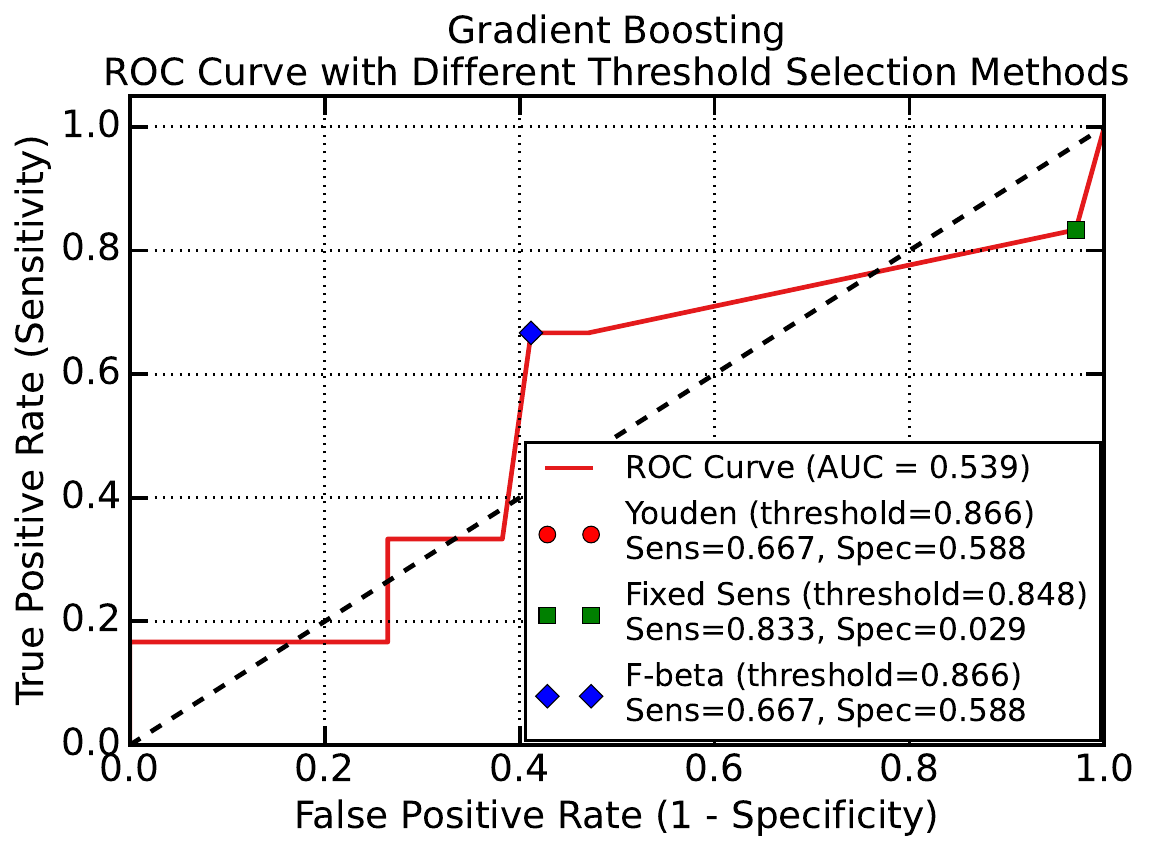}
    \caption{ROC curve for the Gradient Boosting model.}
    \label{fig:roc_gradient_boosting}
\end{figure}

\begin{figure}[htpb]
    \centering
    \includegraphics[width=0.48\textwidth]{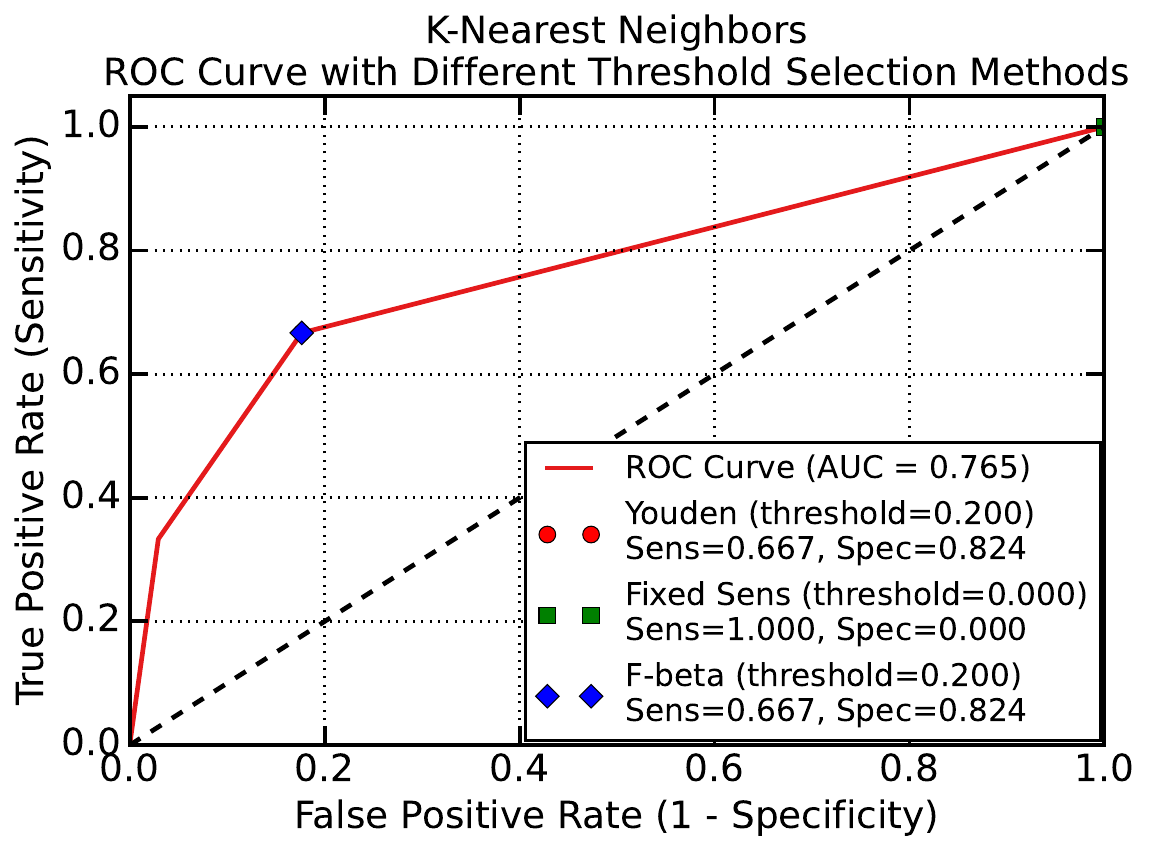}
    \caption{ROC curve for the K-Nearest Neighbors model.}
    \label{fig:roc_knn}
\end{figure}

\begin{figure}[htpb]
    \centering
    \includegraphics[width=0.48\textwidth]{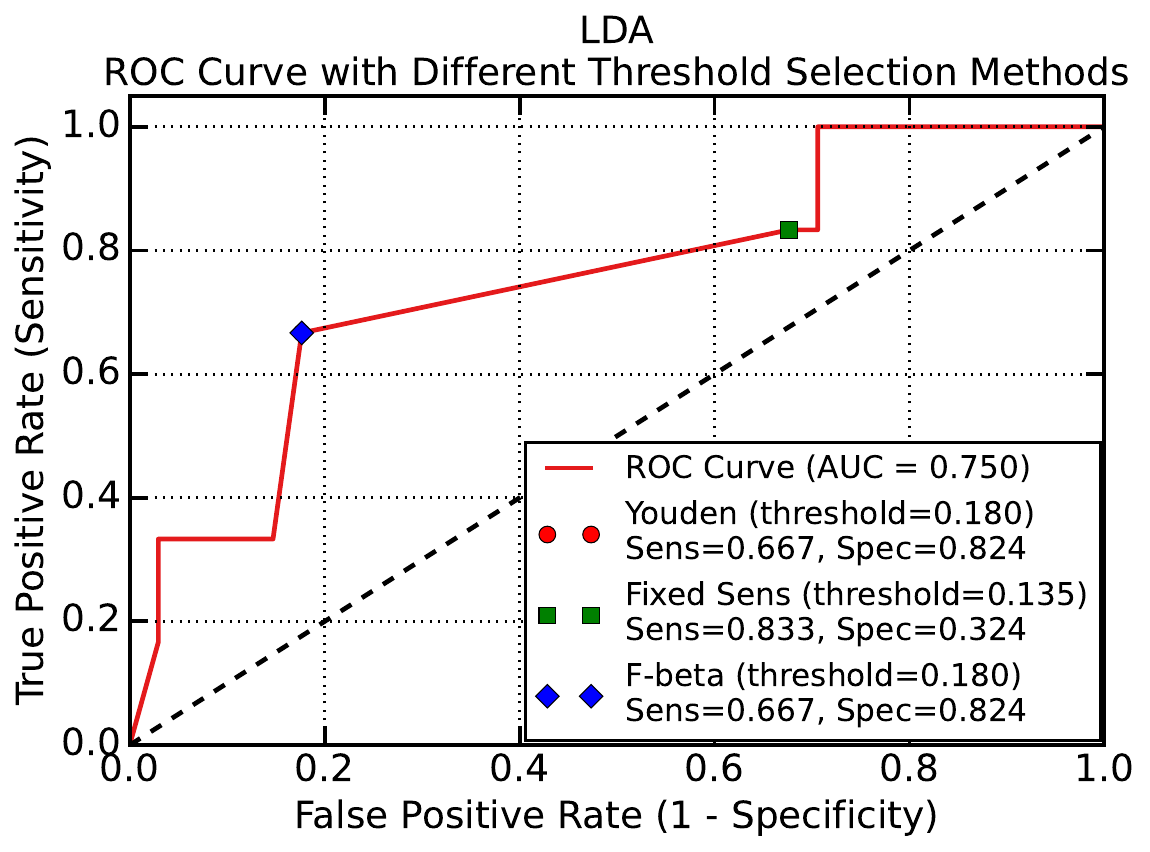}
    \caption{ROC curve for the Linear Discriminant Analysis (LDA) model.}
    \label{fig:roc_lda}
\end{figure}

\begin{figure}[htpb]
    \centering
    \includegraphics[width=0.48\textwidth]{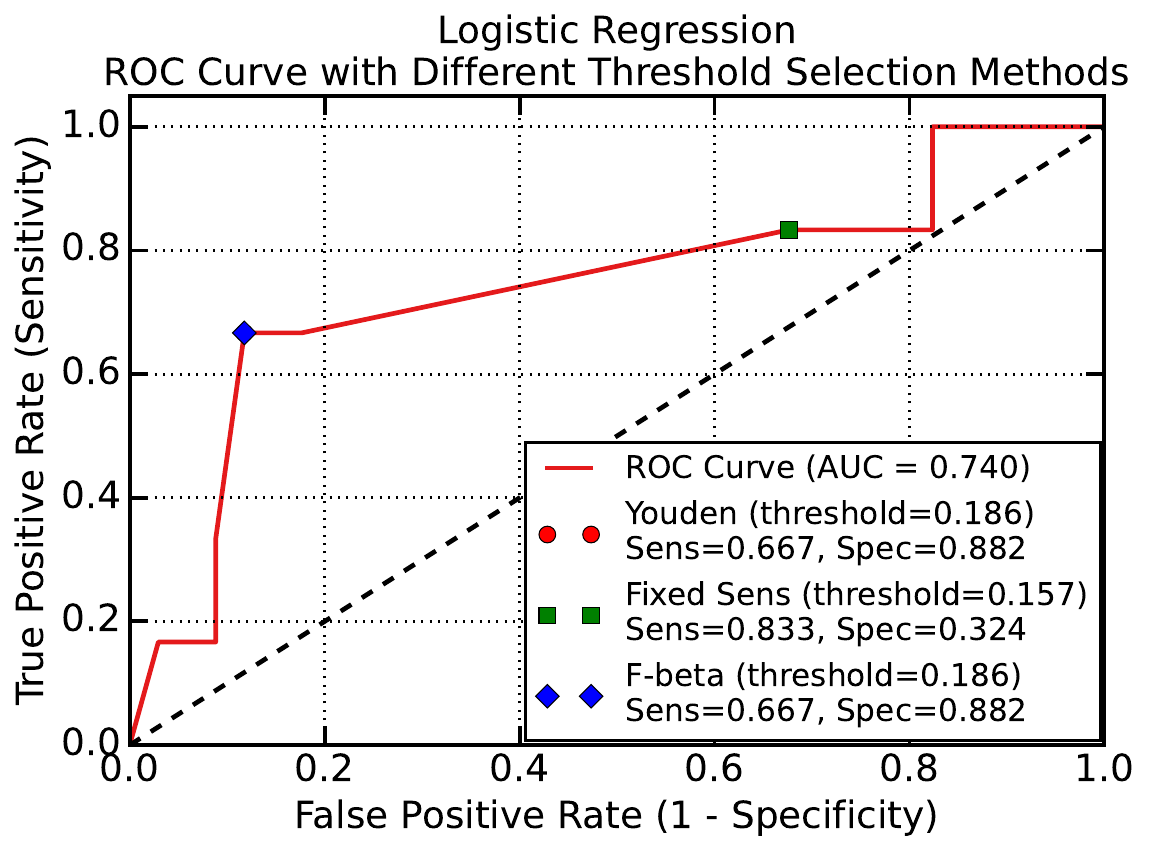}
    \caption{ROC curve for the Logistic Regression model.}
    \label{fig:roc_logistic_regression}
\end{figure}

\begin{figure}[htpb]
    \centering
    \includegraphics[width=0.48\textwidth]{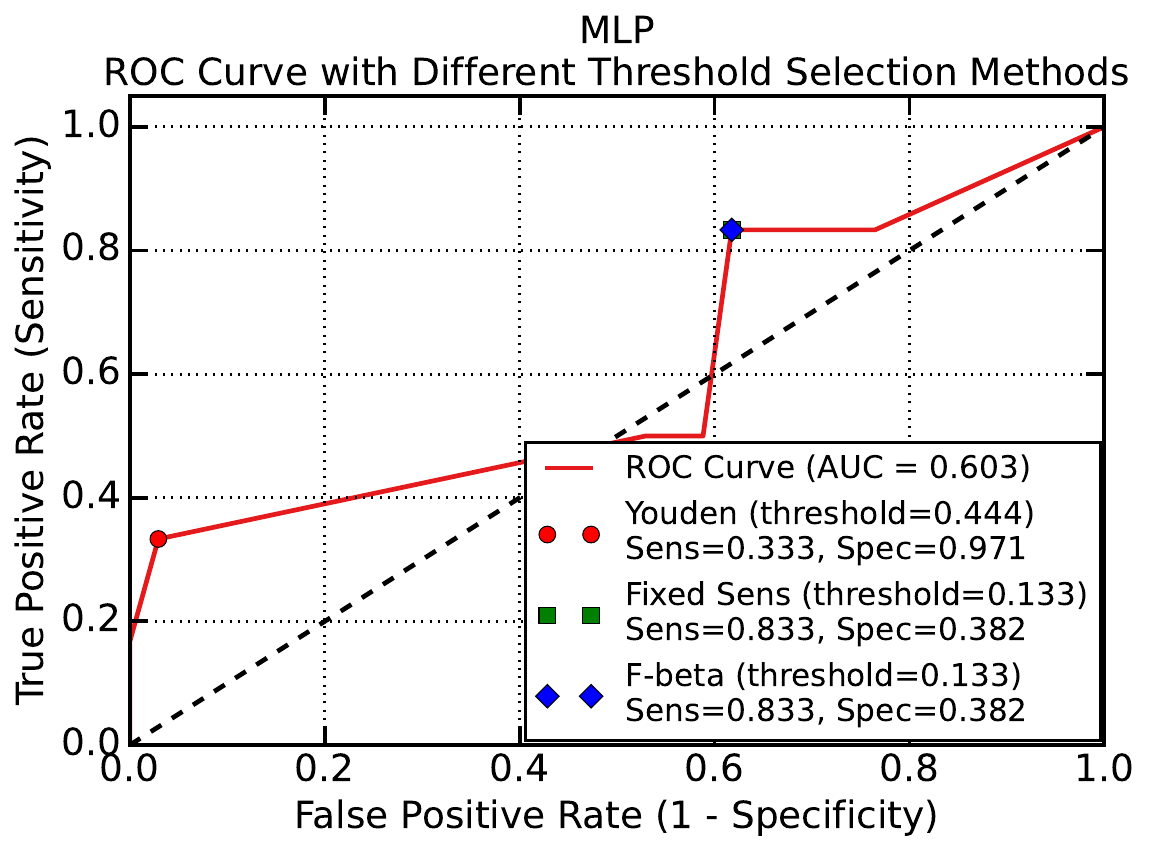}
    \caption{ROC curve for the Multi-Layer Perceptron (MLP) model.}
    \label{fig:roc_mlp}
\end{figure}

\begin{figure}[htpb]
    \centering
    \includegraphics[width=0.48\textwidth]{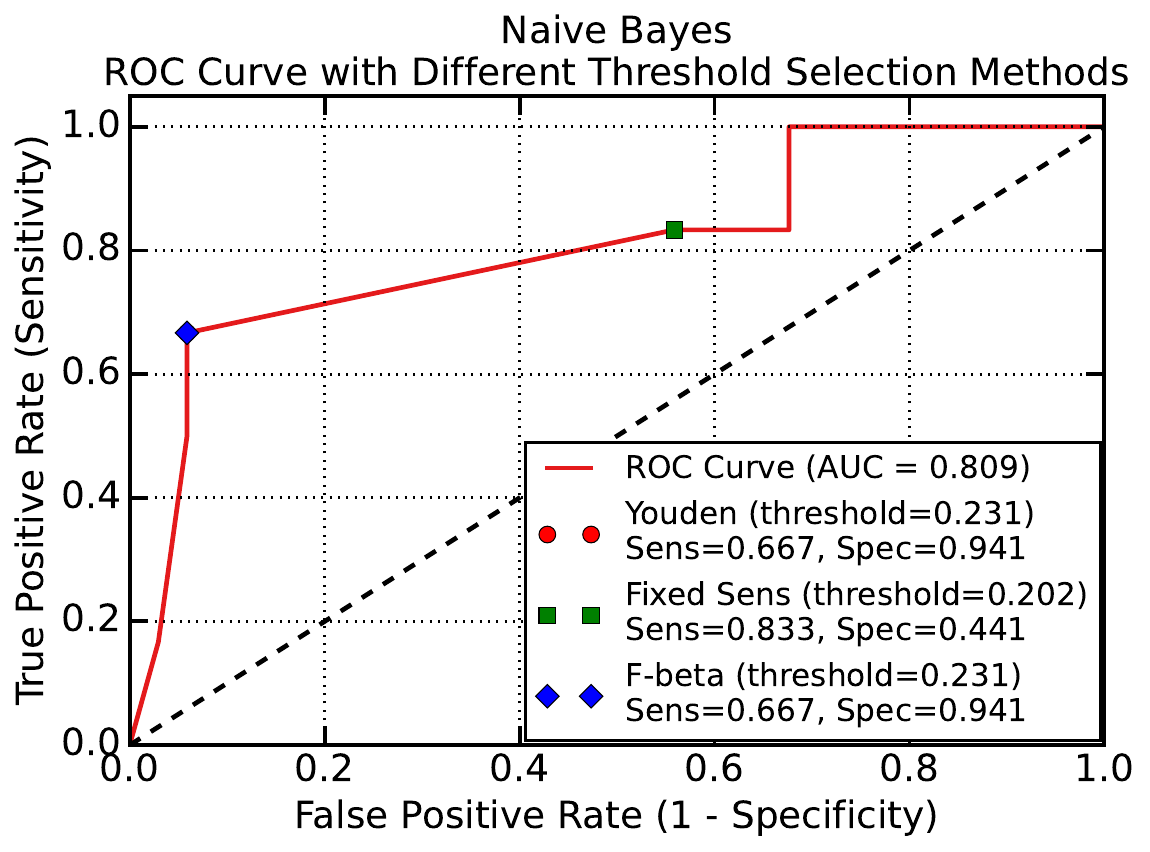}
    \caption{ROC curve for the Naive Bayes model.}
    \label{fig:roc_naive_bayes}
\end{figure}

\begin{figure}[htpb]
    \centering
    \includegraphics[width=0.48\textwidth]{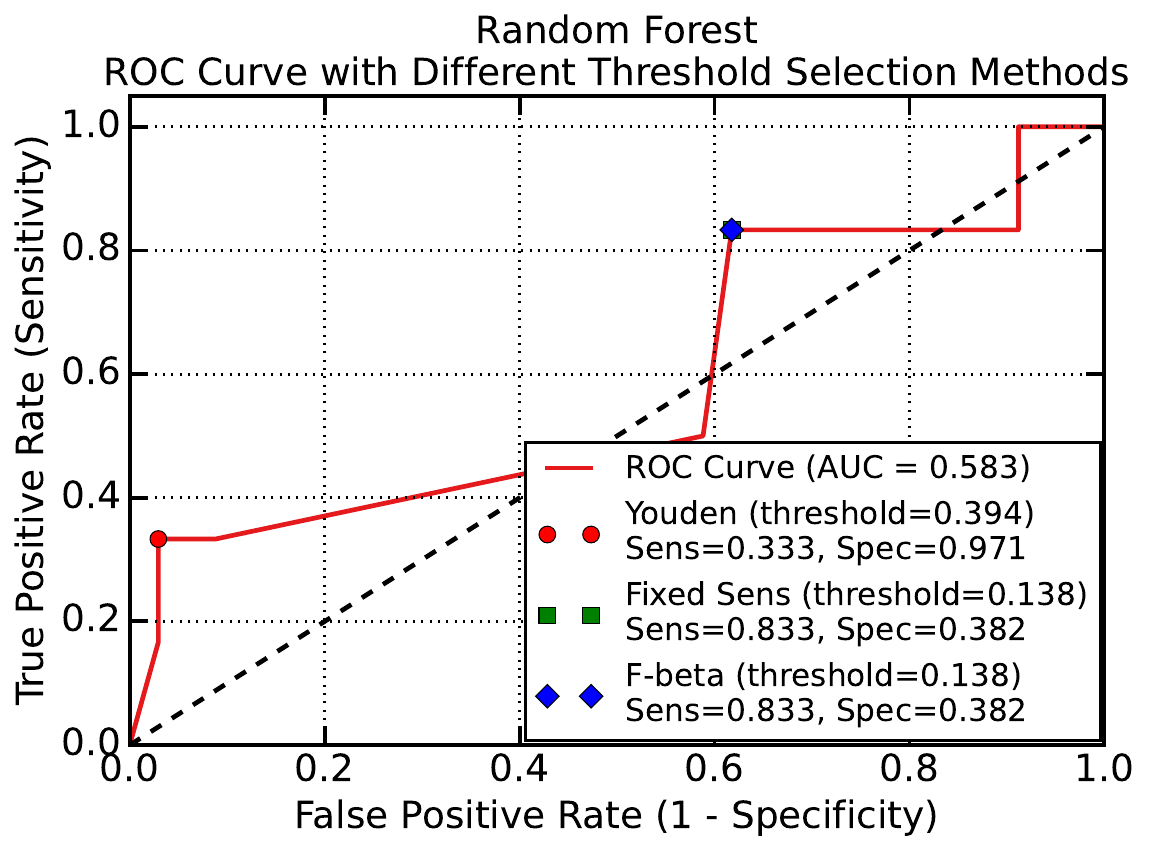}
    \caption{ROC curve for the Random Forest model.}
    \label{fig:roc_random_forest}
\end{figure}

\begin{figure}[htpb]
    \centering
    \includegraphics[width=0.48\textwidth]{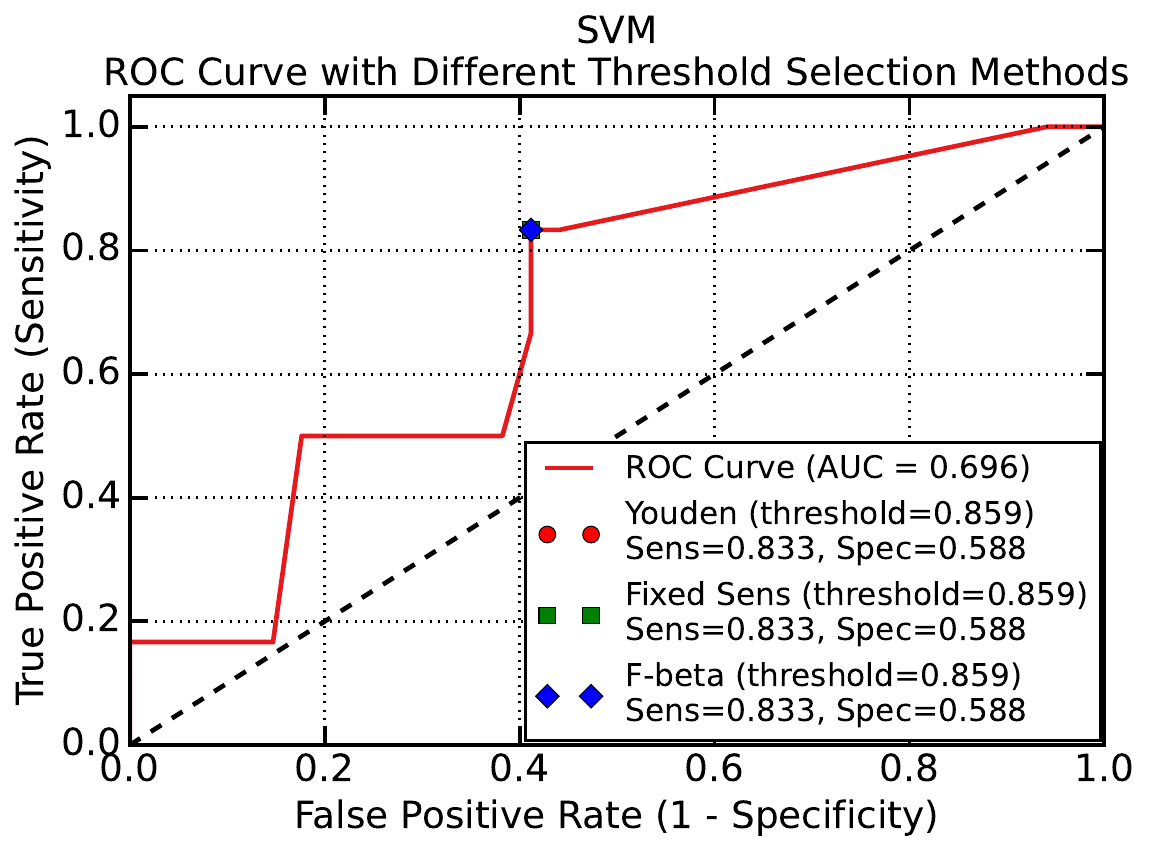}
    \caption{ROC curve for the Support Vector Machine (SVM) model.}
    \label{fig:roc_svm}
\end{figure}

\begin{figure}[htpb]
    \centering
    \includegraphics[width=0.48\textwidth]{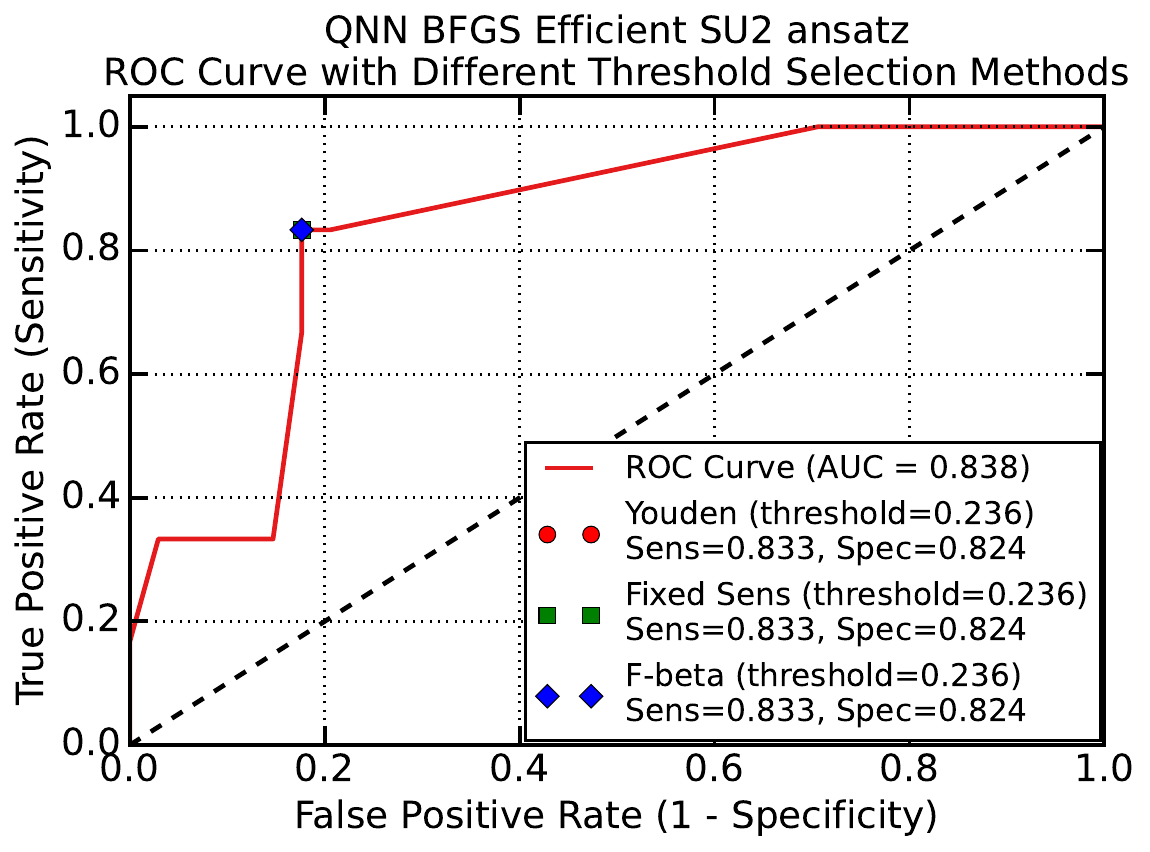}
    \caption{ROC curve for the QNN model with BFGS optimizer and Efficient SU2 ansatz.}
    \label{fig:roc_qnn_bfgs_esu2}
\end{figure}

\begin{figure}[htpb]
    \centering
    \includegraphics[width=0.48\textwidth]{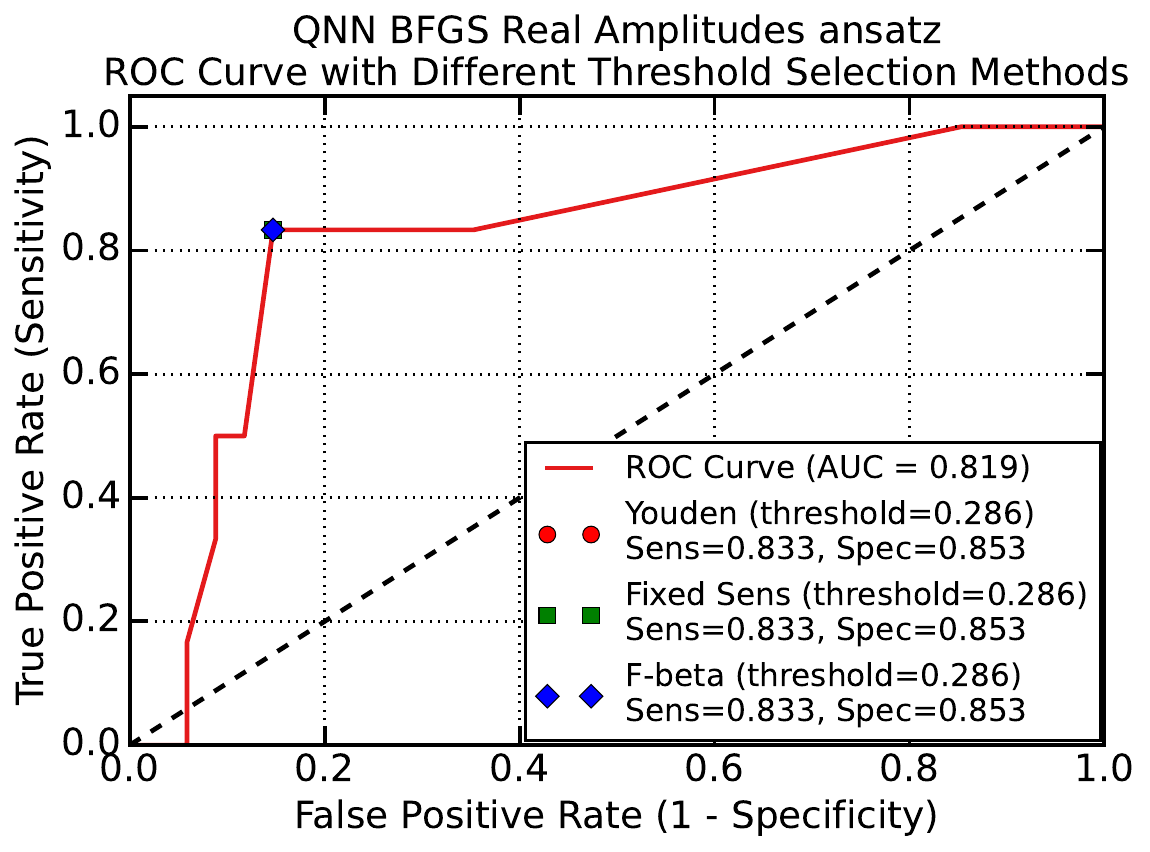}
    \caption{ROC curve for the QNN model with BFGS optimizer and Real Amplitudes ansatz.}
    \label{fig:roc_qnn_bfgs_ra}
\end{figure}

\begin{figure}[htpb]
    \centering
    \includegraphics[width=0.48\textwidth]{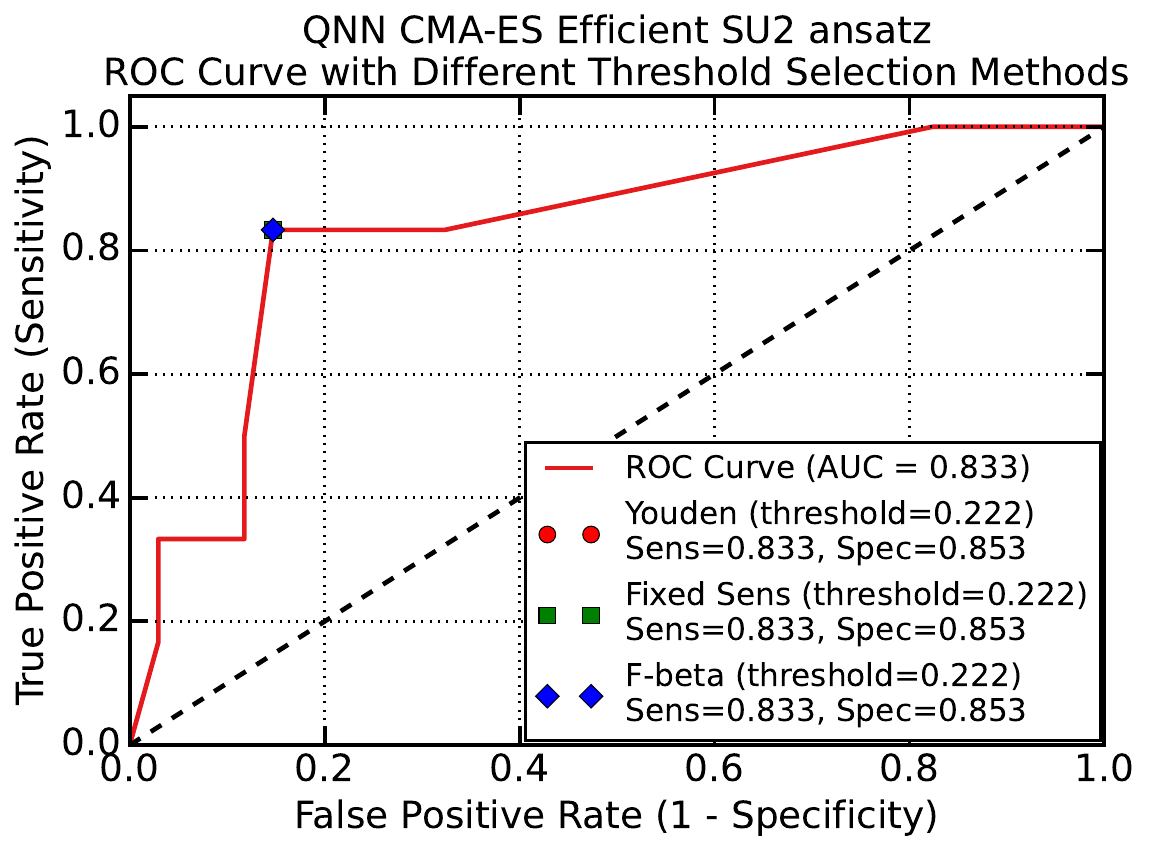}
    \caption{ROC curve for the QNN model with CMA-ES optimizer and Efficient SU2 ansatz.}
    \label{fig:roc_qnn_cmaes_esu2}
\end{figure}

\begin{figure}[htpb]
    \centering
    \includegraphics[width=0.48\textwidth]{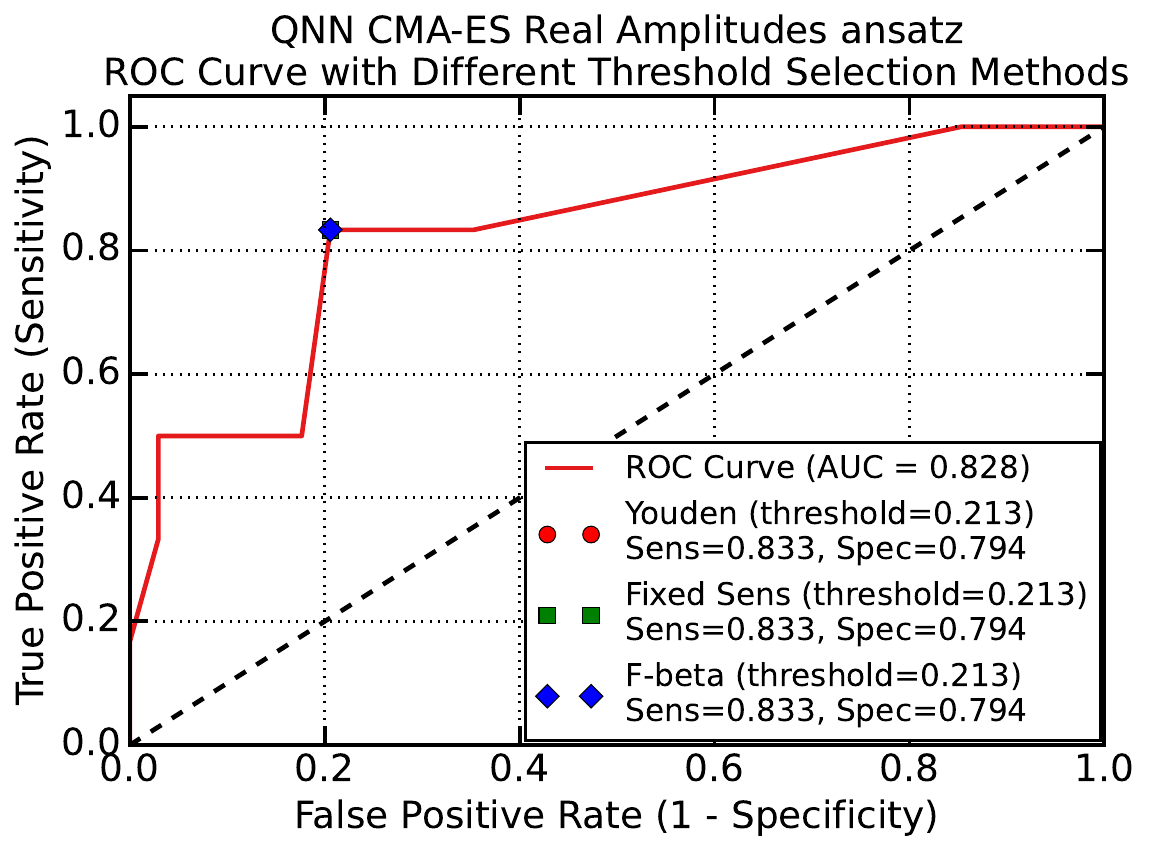}
    \caption{ROC curve for the QNN model with CMA-ES optimizer and Real Amplitudes ansatz.}
    \label{fig:roc_qnn_cmaes_ra}
\end{figure}

\begin{figure}[htpb]
    \centering
    \includegraphics[width=0.48\textwidth]{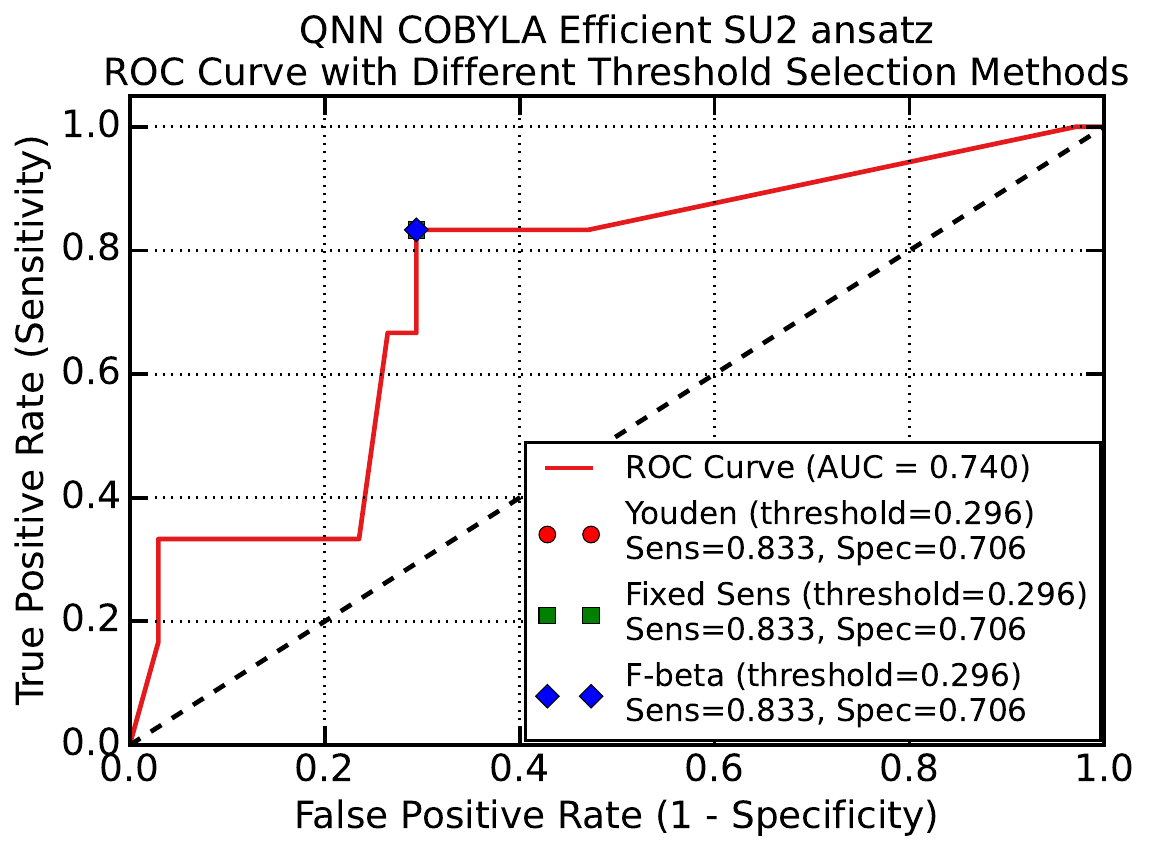}
    \caption{ROC curve for the QNN model with COBYLA optimizer and Efficient SU2 ansatz.}
    \label{fig:roc_qnn_cobyla_esu2}
\end{figure}

\begin{figure}[htpb]
    \centering
    \includegraphics[width=0.48\textwidth]{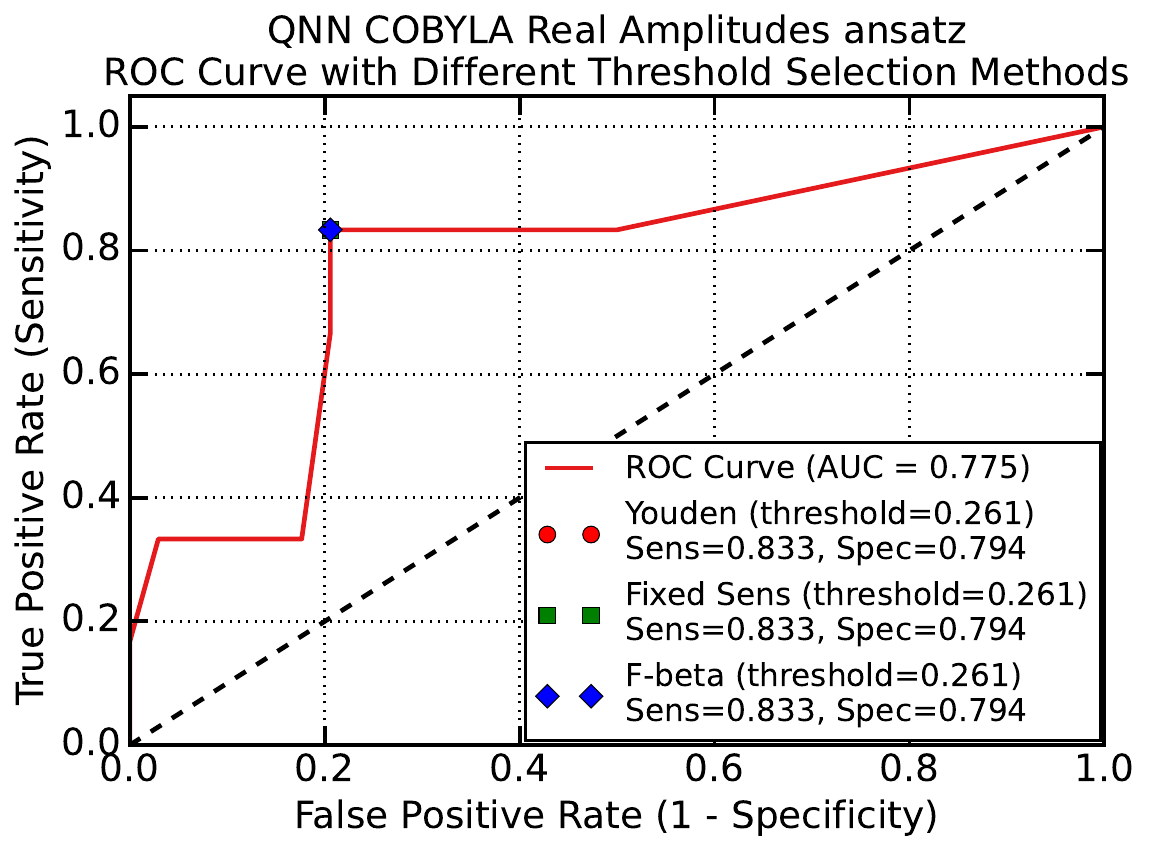}
    \caption{ROC curve for the QNN model with COBYLA optimizer and Real Amplitudes ansatz.}
    \label{fig:roc_qnn_cobyla_ra}
\end{figure}

\begin{figure}[htpb]
    \centering
    \includegraphics[width=0.48\textwidth]{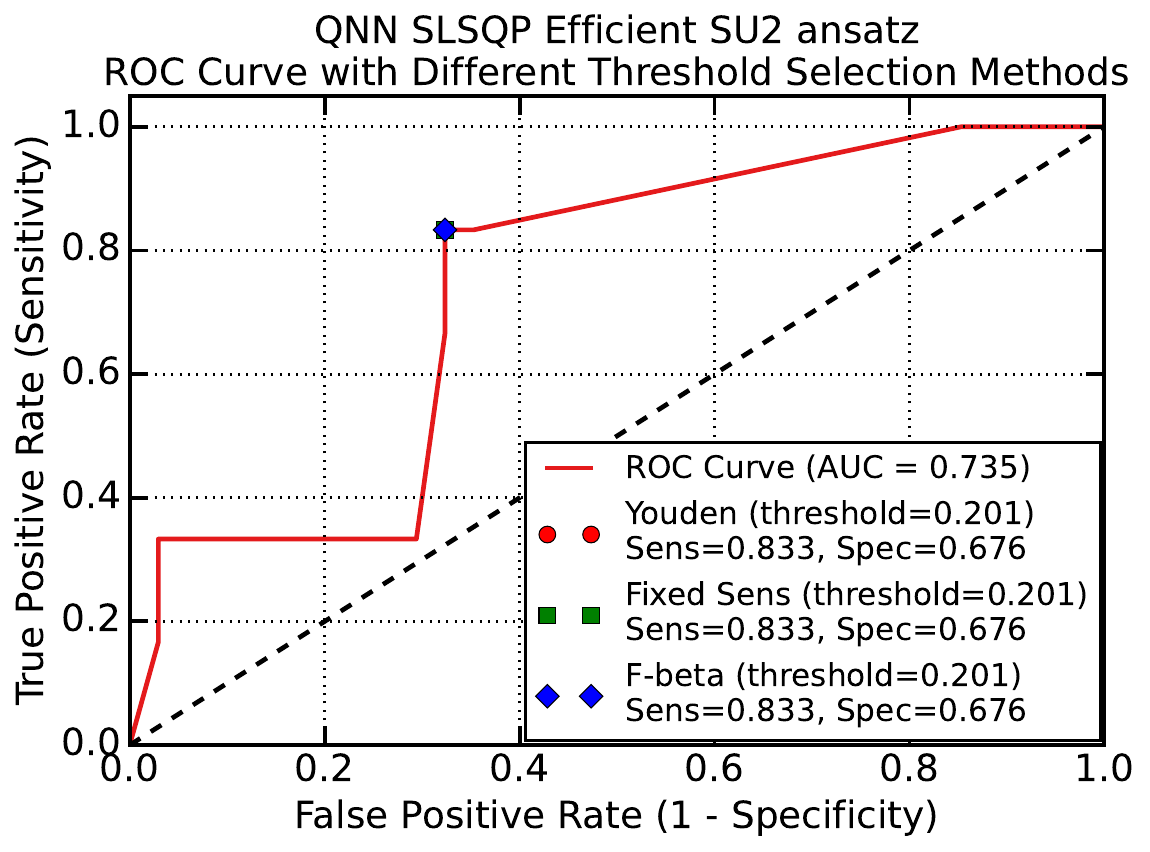}
    \caption{ROC curve for the QNN model with SLSQP optimizer and Efficient SU2 ansatz.}
    \label{fig:roc_qnn_slsqp_esu2}
\end{figure}

\begin{figure}[htpb]
    \centering
    \includegraphics[width=0.48\textwidth]{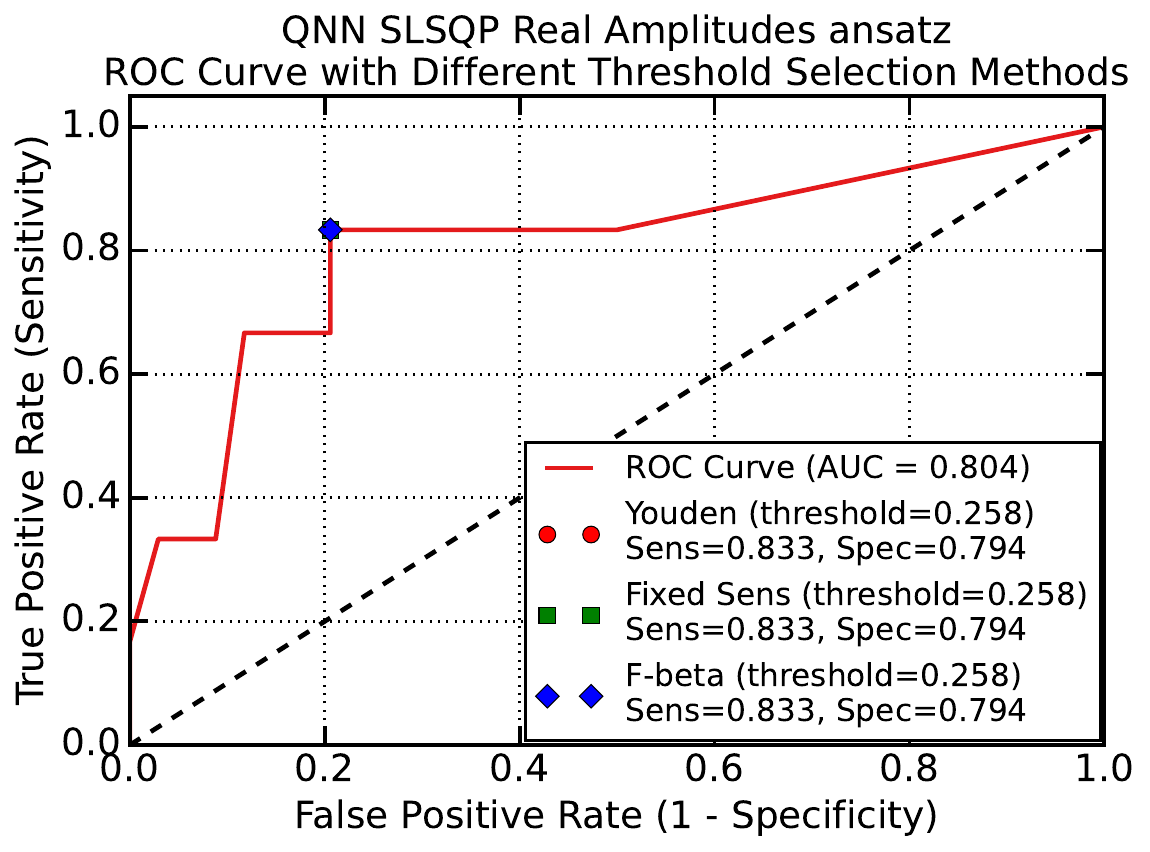}
    \caption{ROC curve for the QNN model with SLSQP optimizer and Real Amplitudes ansatz.}
    \label{fig:roc_qnn_slsqp_ra}
\end{figure}

\begin{figure}[htpb]
    \centering
    \includegraphics[width=0.48\textwidth]{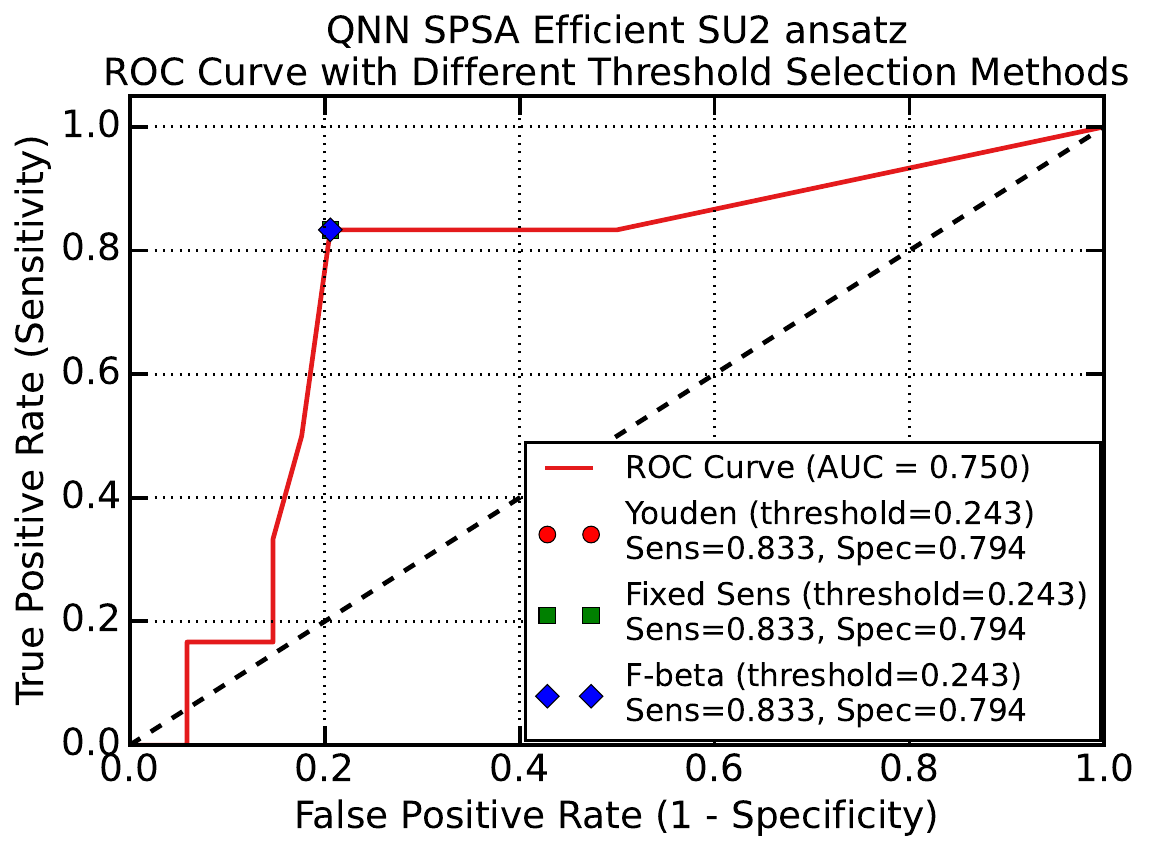}
    \caption{ROC curve for the QNN model with SPSA optimizer and Efficient SU2 ansatz.}
    \label{fig:roc_qnn_spsa_esu2}
\end{figure}

\begin{figure}[htpb]
    \centering
    \includegraphics[width=0.48\textwidth]{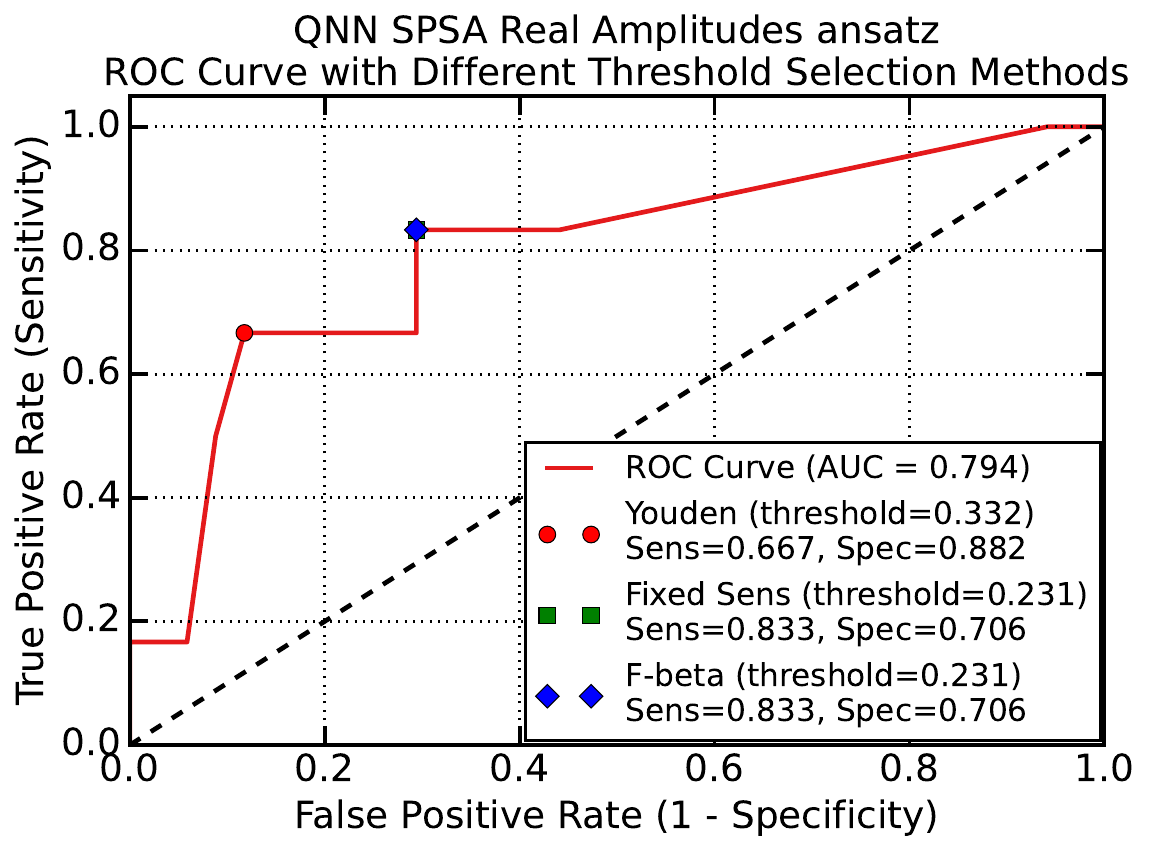}
    \caption{ROC curve for the QNN model with SPSA optimizer and Real Amplitudes ansatz.}
    \label{fig:roc_qnn_spsa_ra}
\end{figure}

The figures in this appendix (e.g., Figure \ref{fig:roc_adaboost} through Figure \ref{fig:roc_qnn_spsa_ra}) display the individual ROC curves for each evaluated model. These curves are derived from models trained on the fifth fold of the cross-validation process for consistent comparison. Each figure prominently features the AUC value for its respective model. Additionally, optimal threshold points, determined by the methods detailed below, are visually marked on these curves, allowing for a direct comparison of how different thresholds impact sensitivity and specificity trade-offs across models and methods.

We generated ROC curves for all evaluated models, derived from models trained on a representative fold (fold five) of the cross-validation process to ensure comparability. To select clinically relevant thresholds from these ROC curves, we employed three distinct methods:

\begin{itemize}
    \item Youden Index: This method aims to maximize diagnostic accuracy by identifying the threshold that maximizes the sum of sensitivity and specificity.
    \item Fixed Sensitivity: Recognizing the paramount importance of sensitivity in detecting rare post-surgical complications, we selected thresholds corresponding to a fixed sensitivity of 83\%. This approach ensures a fair comparison of other performance metrics at a clinically relevant sensitivity level.
    \item F-beta Score Optimization: To balance sensitivity and specificity based on clinical priorities, we utilized the F-beta score, allowing for adjustable weighting of sensitivity.
\end{itemize}
These thresholds are visually indicated on the ROC curves in figures using distinct markers, enabling a direct comparison of threshold locations and associated sensitivity-specificity trade-offs across models and methods.

The following figures display the Receiver Operating Characteristic (ROC) curves for various classical and quantum neural network (QNN) models evaluated on the fifth cross-validation fold.
\FloatBarrier

\FloatBarrier

\bibliography{apssamp}

\end{document}